\documentclass[journal]{IEEEtran}
\usepackage{bm}
\usepackage{latexsym}
\usepackage{amsfonts,amssymb,amsmath}
\usepackage{array}
\usepackage{xcolor}
\usepackage{balance}
\usepackage[caption=false,font=footnotesize]{subfig}
\usepackage{fixltx2e}
\usepackage{textcomp}
\usepackage{stfloats}
\usepackage{url}
\usepackage{verbatim}
\usepackage{cite}
\usepackage{multirow,multicol}
\usepackage{booktabs}
\usepackage{graphicx}
\let\oldfootnote\footnote
\def\footnote{\ifhmode\unskip\fi\oldfootnote}
\usepackage{pifont}
\DeclareMathOperator*{\argmax}{arg\,max}

\DeclareMathOperator{\EX}{\mathbb{E}}% expected value
\DeclareMathOperator{\tr}{tr} % Trace symbol
\DeclareMathAlphabet\mathbfcal{OMS}{cmsy}{b}{n} % bold mathcal font
\newcommand{\RNum}[1]{\lowercase\expandafter{\romannumeral #1\relax}}
\usepackage[linesnumbered,ruled,vlined]{algorithm2e} %for algorithms
\makeatletter
\newcommand{\nosemic}{\renewcommand{\@endalgocfline}{\relax}}% Drop semi-colon ;
\newcommand{\dosemic}{\renewcommand{\@endalgocfline}{\algocf@endline}}% Reinstate semi-colon ;
\let\oldnl\nl% Store \nl in \oldnl
\newcommand{\nonl}{\renewcommand{\nl}{\let\nl\oldnl}}% Remove line number for one line
\makeatother
\ifCLASSINFOpdf
\else
\fi
\newcommand*\squeezespaces[1]{% %% <- #1 is a number between 0 and 1
	\thickmuskip=\scalemuskip{\thickmuskip}{#1}%
	\medmuskip=\scalemuskip{\medmuskip}{#1}%
	\thinmuskip=\scalemuskip{\thinmuskip}{#1}%
	\nulldelimiterspace=#1\nulldelimiterspace
	\scriptspace=#1\scriptspace
} 
\newcommand*\scalemuskip[2]{%
	\muexpr #1*\numexpr\dimexpr#2pt\relax\relax/65536\relax
} %% <- based on  https://tex.stackexchange.com/a/198966/156366

\begin{document}
	\bstctlcite{IEEEexample:BSTcontrol}
	\title{Deep Learning Meets Swarm Intelligence for UAV-Assisted IoT Coverage in Massive MIMO}
%	\title{Swarm Intelligence Meets Deep Learning: UAV-Assisted Relaying for Enhanced IoT Network Coverage and Capacity in Massive MIMO Systems}
	\author{Mobeen Mahmood, MohammadMahdi Ghadaksaz, Asil Koc, Tho Le-Ngoc % <-this % stops a space
	\thanks{This work was supported in part by Huawei Technologies Canada and in part by the Natural Sciences and Engineering Research Council of Canada. An earlier version of this work is accepted in part for publication in IEEE 96th Vehicular Technology Conference (VTC2022-Fall \cite{Mobeen_UAV1}).}% <-this % stops a space
	\thanks{The authors are with the Department of Electrical and Computer Engineering, McGill University, Montreal, QC H3A 0G4, Canada (e-mail: mobeen.mahmood@mail.mcgill.ca, mm.ghadaksaz2000@gmail.com, asil.koc@mail.mcgill.ca, tho.le-ngoc@mcgill.ca)}	
\vspace{-7ex}}
\maketitle

\begin{abstract}
This study considers a UAV-assisted multi-user massive multiple-input multiple-output (MU-mMIMO) systems, where a decode-and-forward (DF) relay in the form of an unmanned aerial vehicle (UAV) facilitates the transmission of multiple data streams from a base station (BS) to multiple Internet-of-Things (IoT) users. A joint optimization problem of hybrid beamforming (HBF), UAV relay positioning, and power allocation (PA) to multiple IoT users to maximize the total achievable rate (AR) is investigated. The study adopts a geometry-based millimeter-wave (mmWave) channel model for both links and proposes three different swarm intelligence (SI)-based algorithmic solutions to optimize: 1) UAV location with equal PA; 2) PA with fixed UAV location; and 3) joint PA with UAV deployment. The radio frequency (RF) stages are designed to reduce the number of RF chains based on the slow time-varying angular information, while the baseband (BB) stages are designed using the reduced-dimension effective channel matrices. Then, a novel deep learning (DL)-based low-complexity joint hybrid beamforming, UAV location and power allocation optimization scheme (J-HBF-DLLPA) is proposed via fully-connected deep neural network (DNN), consisting of an offline training phase, and an online prediction of UAV location and optimal power values for maximizing the AR. The illustrative results show that the proposed algorithmic solutions can attain higher capacity and reduce average delay for delay-constrained transmissions in a UAV-assisted MU-mMIMO IoT systems. Additionally, the proposed J-HBF-DLLPA can closely approach the optimal capacity while significantly reducing the runtime by $99\%$, which makes the DL-based solution a promising implementation for real-time online applications in UAV-assisted MU-mMIMO IoT systems. 
\end{abstract}

\begin{IEEEkeywords}
Decode-and-forward (DF) relay, deep learning, hybrid beamforming,  massive MIMO, millimeter wave communications, power allocation (PA), unmanned aerial vehicles (UAVs). \vspace{-1em} 
\end{IEEEkeywords}

\IEEEpeerreviewmaketitle
\vspace{-2ex}
\section{Introduction}
\IEEEPARstart{T}{he} advent of advanced wireless communications and networking technologies has heralded a new age of innovation with the Internet-of-Things (IoT) at its forefront. This pioneering concept has captured the imagination of the industry and promises to transform the way we interact with the world around us. The potential applications of IoT are vast, ranging from healthcare and urban environments to households \cite{IoT_5G}. However, deploying IoT effectively and extensively still poses significant challenges, including efficient information transfer between wireless nodes and gateways. To address this issue, various routing schemes have been proposed, including direct transmission or relay structures. Nonetheless, when the distance between the IoT end node and the gateway is substantial, direct transmission may not be feasible or may consume excessive power. In such cases, communication through relay can be a more power-efficient alternative. Moreover, deploying cellular stations in urban areas can be a costly and challenging task, which can further complicate the communications coverage issue in the IoT framework \cite{IoT_UAV}. \par 
Unmanned aerial vehicles (UAVs), commonly referred to as drones, are viewed as a key component of the next generation of wireless communications networks. UAV as a relay offers several advantages over traditional static relays. Specifically, the ability to deploy on-demand, mobile relaying systems at a relatively low cost and in a timely manner, makes them particularly well-suited for unforeseen or short-term events, such as emergency situations or network offloading \cite{ mozaffari2019tutorial}. Furthermore, the high mobility of UAVs allows for the dynamic adjustment of their locations to optimize communications conditions, a technique particularly promising for delay-tolerant applications, such as periodic sensing and the transfer of large data \cite{delay_UAV, UAV_delay_1, UAV_delay_2}. UAVs' capability to reach inaccessible locations makes them a viable option for future IoT applications, as they can fly close to IoT devices, sequentially collect sensing data, address coverage issues, and reduce IoT communications networks' overhead \cite{UAV_IoT_5G_2}.  \par
The incorporation of UAVs as relay nodes in wireless sensor networks (WSNs) has the potential to augment communications capacity by connecting remote sensor gateways and addressing the escalating data-rate demands in applications such as virtual reality, device-to-device communications, and smart cities. UAVs can be deployed at high altitudes to increase the likelihood of line-of-sight (LoS) dominated air-to-ground communications channels, thereby supporting high-rate communications. However, the severely congested sub-6 GHz bands can be inadequate to meet the rising data rate requirements. In contrast, millimeter-wave (mmWave) communications, with their abundant spectrum resources, can potentially support the high-throughput and low-latency requirements of various UAV application scenarios \cite{mmWave_5G}. Nonetheless, mmWave signals suffer from high propagation loss, including free-space path loss, atmospheric and molecular absorption, and rain attenuation. This challenge can be surmounted by leveraging massive multiple-input multiple-output (mMIMO) technology with large array structures to generate high beam gains, which can improve the transmission range and simultaneously suppress interference among IoT nodes by utilizing the advanced capabilities of three-dimensional (3D) beamforming. In the realm of mmWave mMIMO systems, fully-digital beamforming (FDBF) and hybrid beamforming (HBF) are two common approaches for mitigating interference. However, FDBF becomes infeasible in UAV-assisted IoT systems due to the prohibitively high cost, complexity, and limited power supply of UAVs \cite{mahmood2021energy}. Conversely, HBF, which involves the design of both the radio frequency (RF)-stage and baseband (BB)-stage, can approach the performance of FDBF by reducing the number of energy-intensive RF chains, thereby minimizing power consumption \cite{koc2020Access, mobeen_3D,sohrabi2016hybrid, HBF_PE-AltMin, mobeen2020}.
\vspace{-1em}
\subsection{Related Works}
\subsubsection{UAV-Assisted Relaying}
The deployment of UAVs as a relay has garnered significant research attention in recent years, with the objective of designing UAV-assisted systems that maximize throughput or minimize transmit power \cite{UAV_relay_1,UAV_relay,UAV_relay_IoT_1,UAV_relay_IoT_2,UAV_relay_MU,UAV_relay_6G,UAV_relay_Cognitive, UAV_Buffer_1, UAV_Buffer_2,UAV_Buffer_3}. In particular, the authors in \cite{UAV_relay_1} propose a UAV relay networks optimization method that minimizes outage probability through joint trajectory design and power control. Their proposed solution reduces outage probability in comparison to a fixed power relay and a circular UAV trajectory. In \cite{UAV_relay}, the optimal altitude for a UAV relay in communications systems is analyzed, and the authors demonstrate that different reliability measures have slightly different optimum altitudes. Additionally, they show that decode-and-forward (DF) relay is superior to amplify-and-forward (AF) relay. In \cite{UAV_relay_IoT_1}, a single UAV is used as a mobile relay for wireless communications networks in harsh environments. The source/relay power allocation (PA) and UAV trajectory are optimized to maximize the end-to-end throughput. Similarly, in \cite{UAV_relay_IoT_2}, a UAV serves as a mobile relay in cases of disrupted communications lines during emergencies, and a variable-rate relaying strategy is proposed to reduce the risk of outages and improve data rate. The authors in \cite{UAV_relay_MU} propose a multi-user (MU) UAV-relaying system that maximizes the sum rate of ground users while satisfying cellular user rate requirements through the optimization of UAV placement and transmit PA. The non-convex problem is solved using block coordinate descent (BCD) and successive convex optimization. In \cite{UAV_relay_6G}, an interference coordination approach using UAV relaying is proposed to enhance data transmission reliability and alleviate mutual interference among ground users by jointly optimizing throughput and UAV energy consumption using the BCD method. A UAV relay is used in an uplink MIMO cognitive radio system in \cite{UAV_relay_Cognitive} to enable communications between a primary user and secondary user. The optimal PA is derived to maximize the achievable rate of the secondary user while respecting power budget, interference, and relay power constraints. In \cite{UAV_Buffer_1}, a UAV-assisted free-space optical/radio frequency (FSO/RF) relaying system is studied to maximize the data throughput with a finite buffer size of UAV relay. A store-then-amplify-and-forward protocol is introduced in \cite{UAV_Buffer_2}, where the received signal from the source node is stored at UAV buffer and amplified and forwarded when the UAV flies close to the destination node. Similarly, the UAV trajectory and user association are optimized in \cite{UAV_Buffer_3} for delay-tolerant communications using Lyapunov optimization technique.
\subsubsection{UAV-Assisted HBF, PA and Positioning}
In \cite{UAV_mmWave_BF}, an optimization problem for UAV location, user clustering, and HBF design is presented to maximize capacity under a minimum rate constraint for each user. This is achieved using an alternating optimization, successive convex optimization, and combinatorial optimization scheme. Similarly, \cite{uav_du2020} proposes an optimal PA and HBF design for UAV-assisted mMIMO communications and derives a closed-form expression of rate for LoS channels. The authors show that the total transmit power should be inversely proportional to the number of UAV antennas to meet each user's rate requirement. In \cite{UAV_mmWave_NOMA}, the energy efficiency (EE) of mmWave-enabled non-orthogonal multiple access (NOMA) is maximized by jointly optimizing UAV location, HBF, and PA through decomposition of the non-convex problem into several sub-problems. In \cite{UAV_NOMA_1}, the joint optimization of UAV's 3D placement, HBF, and resource allocation is addressed to maximize the sum-rate while satisfying the constraints of energy harvesting and coverage using a polyhedral annexation method and deep deterministic policy gradient algorithm. Reference \cite{UAV_mmWave_HBF} considers UAV position and HBF optimization by employing UAVs as flying base stations (BSs), where the non-convex problem is solved using a two-stage optimization strategy. First, the UAV is deployed based on an ideal beampattern assumption, and then the optimal location is used for HBF design by considering the UAV jitter. In \cite{UAV_mmWave_HBF_2}, BCD-based algorithmic solutions are proposed for joint PA, analog beamforming, and UAV positioning to maximize EE while satisfying constraints such as maximum transmitting power, deployment span, and minimum data rate from the ground users. Similarly, in \cite{UAV_mmWave_Trajectory}, the joint optimization problem of UAV 3D placement, beampattern, and transmit power is considered. The non-convex problem is solved using a multi-objective evolutionary algorithm-based on decomposition (MOEA/D)-based algorithm. In \cite{UAV_mmWave_BS}, a UAV-BS is used, and the joint optimization problem of UAV deployment and analog beamformer design is solved to maximize the achievable sum-rate (ASR) subject to minimum rate constraint for each user, deployment span for UAV, and a constant modulus (CM) constraint using an artificial bee colony algorithm.
\subsubsection{Deep-Learning in UAV-Assisted Communications}
DL has emerged as a significant contributor to the progress of artificial intelligence (AI) and has already impacted various domains such as computer vision, speech recognition, and natural language processing \cite{Deep_learning_2006}. The remarkable success of DL in these fields has motivated researchers to explore its potential applications in wireless communications systems \cite{AI_5G, DL_5G, DL_Wireless_Survey, DL_PHY}. Although DL-based wireless communications is in its early stages of development, its superior performance has been verified in certain applications, such as UAV-assisted wireless communications \cite{DL_UAV_2, DL_UAV_1, DL_UAV_3, UAV_DL_4, UAV_RL}. In this regard, researchers have introduced a hybrid DL-based online offloading framework in \cite{DL_UAV_2}, which predicts the optimal positions of ground vehicles and UAVs in a mobile edge computing platform. Additionally, a DL algorithm based on gated recurrent units and autoencoders has been proposed for UAV trajectory prediction in \cite{DL_UAV_1}. Similarly, \cite{DL_UAV_3} employs a DL solution based on Genetic Algorithm-based dataset for multi-UAV path planning that can predict the UAV trajectories with reduced computational complexity. The authors in \cite{UAV_DL_4} have developed a long short-term memory-based recurrent neural network to predict the UAV location and beam angle between BS and UAV for fast beam alignment. Reference \cite{UAV_RL} has designed a deep Q network to optimize UAV navigation based on reinforcement learning.
\subsection{Motivations and Our Contributions}
\vspace{-1ex}
Most prior research on the use of UAVs to assist in communications has overlooked the potential benefits of beamforming solutions, as evidenced by studies such as \cite{UAV_relay_1,UAV_relay,UAV_relay_IoT_1,UAV_relay_IoT_2,UAV_relay_MU,UAV_relay_6G,UAV_relay_Cognitive, UAV_Buffer_1, UAV_Buffer_2,UAV_Buffer_3,DL_UAV_2, DL_UAV_3,UAV_RL}. While some studies have explored beamforming solutions, they have been limited to single-hop communications scenarios and have only examined its application in single links, such as in studies \cite{UAV_mmWave_BF,uav_du2020,UAV_mmWave_NOMA,UAV_NOMA_1,UAV_mmWave_HBF,UAV_mmWave_HBF_2,UAV_mmWave_Trajectory,UAV_mmWave_BS}, where UAVs act as flying BSs rather than relays, making the beamforming solutions inapplicable for dual-hop or relaying structures. To the best of our knowledge, the joint optimization of UAV location, PA, and HBF design for a dual-hop mmWave MU-mMIMO IoT communications networks is an unaddressed problem, presenting a significant opportunity to advance the field of UAV-based wireless communications. To address this gap in literature, this research aims to highlight the full potential of beamforming, particularly HBF, and its ability to enhance performance in dual-hop UAV communications networks. It is worth noting that, unlike prior literature that employs a single antenna in UAV-assisted systems, this study considers large antenna arrays deployed in the mmWave band to consider beamforming issues in fading scenarios. In non-fading situations, for instance, LoS channels, the proposed HBF solutions can further improve performance by providing higher directional gain and reducing interference. Our motivation is to jointly optimize UAV location, PA to multiple IoT nodes, and design hybrid RF and BB stages of BS and UAV for enhanced capacity. Table \ref{tab:tab0} provides a detailed comparison of this work with respect to the state-of-the-art. The main contributions of this work are summarized as follows:
\begin{table*}[t!]
	\caption{A brief comparison of the related literature. \vspace{-1em}} %title of the table
	\centering % centering table
	\resizebox{1\textwidth}{!}{	
		\begin{tabular}{|c||c|c||c|c||c|c||c|c||c|c|c||c|}
			\hline
			\multirow{3}{*}{Literature} & \multicolumn{12}{c|}{Contents}  \\
			\cline{2-13}
			&  \multicolumn{2}{c||}{UAV Operation Mode} & \multicolumn{2}{c||}{Propagation Environment} & \multicolumn{2}{c||}{MIMO} & \multicolumn{2}{c||}{Relay Transmission} & \multicolumn{3}{c||}{Radio Resource Management} & \multirow{2}{*}{Deep Learning}  \\
			\cline{2-12}
			& As BS & As Relay & No Fading (LoS) & Fading (LoS + NLoS) & SU-MIMO & MU-MIMO & With Buffer & Without Buffer & UAV Positioning/Trajectory & Power Allocation & Beamforming & \\
			\hline
			\cite{UAV_relay_1,UAV_relay} &  & \ding{51} & \ding{51} & & \ding{51} & & & \ding{51} & \ding{51} & & & \\ \hline
			\cite{UAV_relay_IoT_1} &  & \ding{51} & \ding{51} &  &\ding{51} & & &  \ding{51} & \ding{51} &\ding{51} & &  \\\hline
			\cite{UAV_relay_MU} &  & \ding{51} & \ding{51} & & & \ding{51} & &  \ding{51} & \ding{51} & \ding{51} & & \\\hline
			\cite{UAV_relay_6G} & \ding{51} & &\ding{51} & & & \ding{51} & &  \ding{51} &\ding{51} & & & \\\hline
			\cite{UAV_relay_Cognitive} &  & \ding{51} & & \ding{51} & & \ding{51} & & \ding{51} & & \ding{51} & & \\\hline
			\cite{UAV_Buffer_1} &  & \ding{51} & & \ding{51} &\ding{51} & & \ding{51} & & \ding{51} & & & \\ \hline
			\cite{UAV_Buffer_2} &  & \ding{51} & \ding{51} & & \ding{51} & & \ding{51} & & \ding{51} & & &\\ \hline
			\cite{UAV_Buffer_3} & \ding{51} & & & \ding{51} & & \ding{51} & \ding{51} & & \ding{51} & & &\\ \hline
			\cite{UAV_mmWave_BF, UAV_mmWave_HBF, UAV_mmWave_Trajectory} & \ding{51} &  &  & \ding{51} & & \ding{51} & & \ding{51} & \ding{51} & & \ding{51} &  \\\hline
			\cite{uav_du2020, UAV_mmWave_NOMA,UAV_NOMA_1} & \ding{51} & & \ding{51} & & & \ding{51} & & \ding{51} & \ding{51} & \ding{51} & \ding{51} & \\\hline
			\cite{UAV_mmWave_HBF_2} & \ding{51} & & &\ding{51}  & & \ding{51} & & \ding{51} & \ding{51} &  \ding{51} & \ding{51} & \\ \hline 
			\cite{UAV_mmWave_Trajectory} & \ding{51} & & \ding{51} & & & \ding{51} & & \ding{51} & \ding{51} & \ding{51} & &\\ \hline
			\textbf{This work} &  & \ding{51} & & \ding{51} & & \ding{51} & \ding{51} & & \ding{51} & \ding{51} & \ding{51} & \ding{51} \\ \hline
	\end{tabular}}
	\label{tab:tab0}
	\vspace{-2em}
\end{table*}
\begin{enumerate}
	\item We propose three novel optimization schemes for maximizing total capacity in UAV-assisted MU-mMIMO IoT systems: 1) joint HBF and optimal PA for fixed UAV location (J-HBF-PSOPA-FL); 2) joint HBF and UAV location optimization for equal PA (J-HBF-PSOL-EQPA); and 3) joint HBF, UAV location optimization and PA (J-HBF-PSOLPA). Using swarm intelligence-based particle swarm optimization (PSO), we tackle the challenging non-convex problem with high-dimensional variable matrices and fractional programming variables, while adhering to constraints like UAV deployment span, total transmit power, PA, and CM. In particular, the RF stages are formulated using slow time-varying angle-of-departure (AoD)/angle-of-arrival (AoA) information, while the BB stages use the reduced-dimension effective channel information with regularized zero-forcing (RZF). In J-HBF-PSOPA-FL, we allocate optimal power for multiple IoT users with fixed UAV location, and design the RF and BB stages for maximum capacity. In J-HBF-PSOL-EQPA, we optimize UAV location for equal PA, whereas, in J-HBF-PSOLPA, we jointly optimize both UAV location and PA, and design HBF stages for both UAV and BS. The illustrative results show that J-HBF-PSOLPA can achieve higher total rate compared to J-HBF-PSOL-EQPA and J-HBF-PSOPA-FL schemes. 
	\item To overcome the high computational complexity of J-HBF-PSOLPA, we propose a novel low-complexity DL-based joint HBF, UAV location optimization and PA (J-HBF-DLLPA) algorithm for a UAV-assisted MU-mMIMO IoT systems. The proposed J-HBF-DLLPA-based solution can achieve AR of J-HBF-PSOLPA, while reducing the runtime by $98-99$ $\%$. In particular, the proposed J-HBF-DLLPA is built via a fully-connected deep neural network (DNN) consisting of two phases: 1) offline supervised learning via the optimal allocated powers and UAV locations calculated with J-HBF-PSOLPA; and 2) real-time prediction of optimal power values and UAV location via the trained DNN.
	\item We analyze the performance of UAV DF relaying with and without buffering. Most existing studies have explored the option of forwarding the received signal without buffering the data, which fails to fully leverage the UAV's mobility \cite{UAV_relay_1,UAV_relay,UAV_relay_IoT_1,UAV_relay_IoT_2,UAV_relay_MU,UAV_relay_6G,UAV_relay_Cognitive,UAV_mmWave_BF,uav_du2020,UAV_mmWave_NOMA,UAV_NOMA_1,UAV_mmWave_HBF,UAV_mmWave_HBF_2,UAV_mmWave_Trajectory,UAV_mmWave_BS}. The proposed solutions incorporate signal buffering at the UAV, enhancing the relaying performance significantly. Particularly, we analyze the average delay of delay-unconstrained and delay-constrained transmissions, which can be significantly reduced by J-HBF-PSOLPA compared to fixed UAV deployment and equal PA (FL-EQPA).
\end{enumerate}
\color{black}
\vspace{-1em}
\subsection{Organization}
The rest of this paper is organized as follows. Section II presents the system and channel model of UAV-assisted mmWave MU-mMIMO systems followed by the proposed joint HBF, PA and UAV location optimization schemes in Section III. The low-complexity DL-based joint HBF, PA and UAV location optimization scheme is presented in Section IV followed by the illustrative results in Section V. Finally, Section VI concludes the paper.\newline
\indent \textit{Notation:} The following notations are used throughout this paper. Boldface lower-case and upper-case letters denote column vectors and matrices, respectively. $(·)^T$, $(·)^H$, $\left\lVert.\right\rVert_2$ and $\left\lVert . \right\rVert_F^2$ represent the transpose, complex-transpose, the 2-norm and the Frobenius norm of a vector or matrix, respectively. $\mathbf{I}_k$, $\EX[.]$ and $\tr(.)$ denote $k \times k$ identity matrix, the expectation operator and the trace operator, respectively. $\mathbf{X} \otimes \mathbf{Y}$ denotes the Kronecker product of two matrices $\mathbf{X}$ and $\mathbf{Y}$. We use $x_k$ $\sim$ $\mathcal{CN} (0,\sigma^2)$ when $x_k$ is a complex Gaussian random variable with zero-mean and variance $\sigma^2$.
\vspace{-1em}
\section{System and Channel Model}
In this section, we present the system and channel models of proposed UAV-assisted mmWave MU-mMIMO IoT systems.
\vspace{-1em}
\subsection{System Model}
The present study delves into a challenging scenario where multiple IoT devices are connected to an IoT gateway through either wired or wireless links. This setup is situated in a remote area that is difficult to access directly by BS/eNodeB due to various obstacles such as buildings, mountains, etc. Then, a UAV is introduced as a dual-hop DF relay to access the IoT users, as depicted in Fig. 1. We assume that there is no direct link between BS and IoT devices due to severe blockage. In contrast to traditional static relaying, which relies on fixed relay locations, we investigate the potential of using UAVs as DF relays for future mMIMO-enabled IoT systems where direct link communication between BS and IoT node ends is not feasible. It is important to note that the algorithms proposed in Sections III and IV are not only limited to fixed/static node locations but can be applicable to movable nodes (i.e., dynamic environment). Let $\left(x_b, y_b, z_b\right)$, $\left(x_u, y_u, z_u\right)$ and $\left(x_k, y_k, z_k \right)$ denote the locations of BS, UAV relay and $k^{th}$ IoT node, respectively. We define the 3D distances for a UAV-assisted mmWave MU-mMIMO IoT system as follows: 
\begin{equation}
	\begin{split}
\tau_1 &= \sqrt{(x_u - x_b)^2 + (y_u - y_b)^2 + (z_u - z_b)^2}  \\ 
\tau_{2,k} &= \sqrt{(x_u - x_k)^2 + (y_u - y_k)^2 + (z_u - z_k)^2} \\ 
\tau_{k} &= \sqrt{(x_b - x_k)^2 + (y_b - y_k)^2 + (z_b - z_k)^2} \label{eq: 3D_distance}
	\end{split}
\end{equation}
where $\tau_1$, $\tau_{2,k}$ and $\tau_{k}$ are the 3D distance between UAV \& BS, between UAV and $k^{th}$ IoT node, and between BS and $k^{th}$ IoT node, respectively. \par 
\begin{figure}[!t]
	\centering
	\captionsetup{justification=centering}
	\includegraphics[height= 4.8cm, width=1\columnwidth]{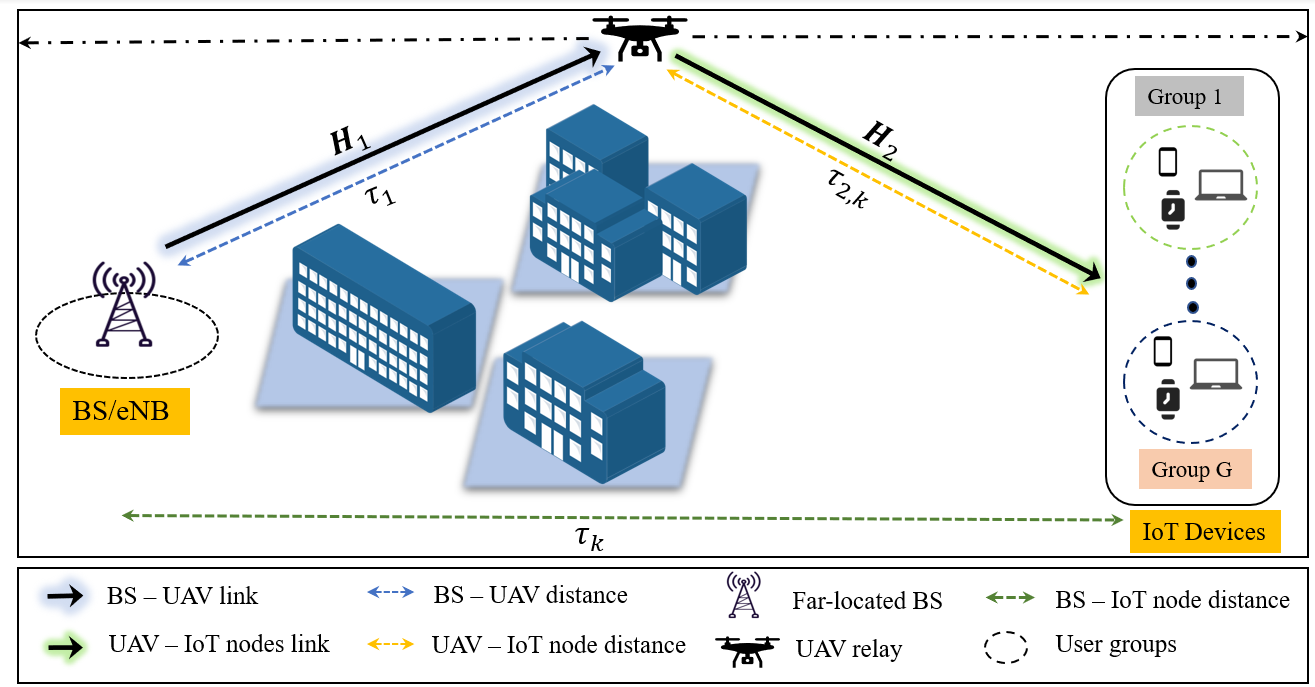} 
	\caption{System model of UAV-assisted relaying in mmWave MU-mMIMO IoT communications.}
	\label{fig:fig1}
	\vspace{-2em}
\end{figure} 
\begin{figure*}[!t]
	\captionsetup{justification=centering}
	\centering
	\includegraphics[height= 4.8cm, width=2.1\columnwidth]{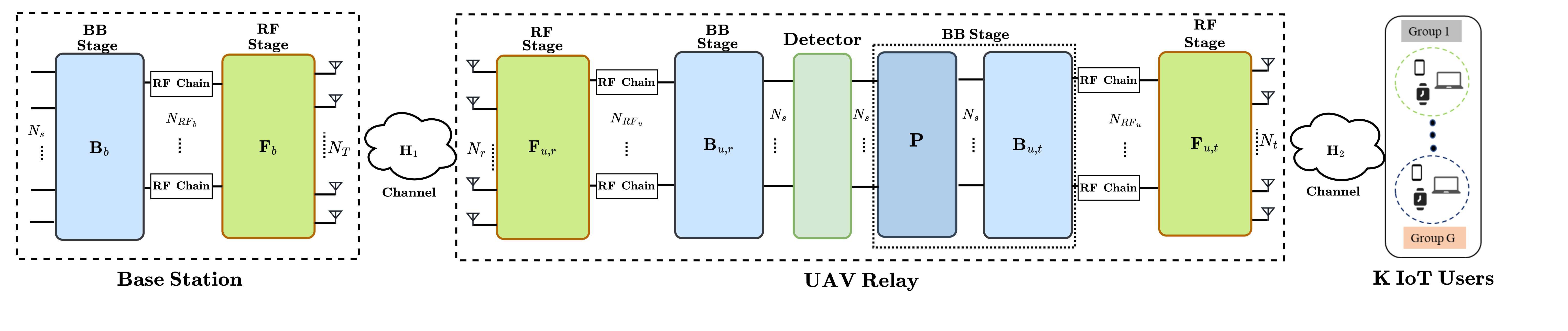} 
	\vspace{-3em}
	\caption{System model of UAV-assisted DF relaying in MU-mMIMO IoT networks.}
	\label{fig:fig2}
	\vspace{-2em}
\end{figure*} 
In the system model shown in Fig. 2, we consider BS equipped with $N_T$ antennas, UAV relay with $N_{r}$ antennas for receiving and $N_{t}$ antennas to serve $K$ single-antenna IoT nodes clustered in $G$ groups, where $g^{th}$ group has $K_g$ IoT nodes such that $K = \sum_{g=1}^{G}K_g$. For the downlink transmission of $N_S = K$ data streams, we consider HBF for BS and UAV, where BS consists of a RF beamforming stage $\mathbf{F}_b$ $\in$ $\mathbb{C}^{N_T \times N_{{RF}_b}}$ and BB stage $\mathbf{B}_b$ $\in$ $\mathbb{C}^{N_{{RF}_b} \times K}$. Here, $N_{{RF}_b}$ is the RF chains such that $N_S \leq N_{{RF}_b} \leq N_T$ to guarantee multi-stream transmission. We consider half-duplex (HD) DF relaying, whereas the use of full-duplex (FD) UAV relaying is left as our future work. During the first time slot, $K$ data streams are transmitted through channel $\mathbf{H}_1$ $\in$ $\mathbb{C}^{N_{r} \times N_T}$. Using $N_{r}$ antennas, UAV receives signals with RF stage $\mathbf{F}_{u,r}$ $\in$ $\mathbb{C}^{N_{{RF}_{u}} \times N_{r}}$ and BB stage $\mathbf{B}_{u,r}$ $\in$ $\mathbb{C}^{K \times N_{{RF}_{u}}}$. We assume UAV transmits the data in the second time slot using RF beamformer $\mathbf{F}_{u,t} = \left[\mathbf{f}_{{u,t,1}},\cdots,\mathbf{f}_{{u,t, N_{{RF}_{u}}}}\right]$ $\in$ $\mathbb{C}^{N_{t} \times N_{{RF}_{u}}}$, BB stage $\mathbf{B}_{u,t} = \left[\mathbf{b}_{{u,t,1}},\cdots,\mathbf{b}_{{u,t,K}} \right]$ $\in$ $\mathbb{C}^{N_{{RF}_{u}} \times K}$, and MU PA matrix $\mathbf{P} = \text{diag}(\sqrt{p_1}, \hdots, \sqrt{p_K})$ $\in$ $\mathbb{C}^{K \times K}$ through channel $\mathbf{H}_2$ $\in$ $\mathbb{C}^{K \times N_{t}}$, where $p_k$ reflects the allocated power to $k^{th}$ user. The implementation of all RF beamforming/combining stages involves the use of PSs and thus, impose a CM constraint, i.e., $|\mathbf{F}_b (i,j)| = \frac{1}{\sqrt{N_T}}$, $|\mathbf{F}_{u,r} (i,j)| = \frac{1}{\sqrt{N_{r}}}$, $|\mathbf{F}_{u,t} (i,j)| = \frac{1}{\sqrt{N_{t}}} \hspace{1ex} \forall i,j$. The design of HBF greatly reduces the number of RF chains, for instance, from $N_T$ RF chains to $N_{{RF}_{b}}$ for BS, from $N_t(N_r)$ RF chains to $N_{{RF}_{u}}$ for UAV whilst satisfying: 1) $K \leq N_{{RF}_{b}} \ll N_T$; and 2) $K \leq N_{{RF}_{u}} \ll  N_{r}(N_{t})$. For the data signal $\mathbf{d} = [d_1, d_2, \hdots, d_{K}]^T$ with $\EX \{\mathbf{d}\mathbf{d}^H\} = \mathbf{I}_{K}$ $\in$ $\mathbb{C}^{K \times K}$, the signal transmitted by BS is given as follows:
\begin{equation}
	\mathbf{s}_1 = \mathbf{F}_b\mathbf{B}_b \mathbf{d}. \label{eq:1} \vspace{-1ex}
\end{equation}
Let $P_T$ be the total transmit power of BS. The design of $\mathbf{F}_b$ and $\mathbf{B}_b$ satisfies the power constraint $\lVert \mathbf{F}_b\mathbf{B}_b \rVert^2_F$ = $P_T$. Then, the received signal at UAV is: 
\begin{equation}
	\begin{split}
		\mathbf{y}_1 &= \mathbf{H}_1 \mathbf{s}_1 + \mathbf{n}_1, \\
		&= \mathbf{H}_1 \mathbf{F}_b\mathbf{B}_b \mathbf{d} + \mathbf{n}_1, 
	\end{split}	\label{eq:2}  \vspace{-1ex}
\end{equation}
where $n_1$ $\in$ $\mathbb{C}^{N_r}$ denotes the zero-mean complex circularly symmetric Gaussian noise vector at UAV relay with covariance matrix $\EX \{\mathbf{n}_1\mathbf{n}_1^H\} = \sigma^2_n \mathbf{I}_{N_r}$ \hspace{-2ex} $\in$ $\mathbb{C}^{N_r \times N_r}$. Subsequently, the received signal after baseband processing at UAV relay is written as follows: \vspace{-0.5ex}
\begin{equation}
	\tilde{\mathbf{y}}_1 = \mathbf{B}_{u,r}\mathbf{F}_{u,r}\mathbf{H}_1 \mathbf{F}_b\mathbf{B}_b \mathbf{d} + \mathbf{B}_{u,r}\mathbf{F}_{u,r}\mathbf{n}_1. \label{eq:received_UAV}   \vspace{-1ex}
\end{equation}
UAV as DF relay uses $\tilde{\mathbf{y}}_1$ to decode the information and re-encodes it for transmission in second time slot. Then, the signal transmitted by UAV is: \vspace{-0.5ex}
\begin{equation}
	\mathbf{s}_2 = \mathbf{F}_{u,t}\mathbf{B}_{u,t}\mathbf{P} \hat{\mathbf{d}}, \label{eq:s1} \vspace{-1ex}
\end{equation} 
where $\mathbf{\hat{d}}$ is the re-encoded signal at UAV relay. After transmitting via channel $\mathbf{H}_2$, the received signal at $k^{th}$ IoT node in $g^{th}$ group can be written as:
\begin{equation}\label{eq:received_User}
	\begin{split}
		y_{g_k} &= \mathbf{h}_{2,g_k}^{T}\mathbf{s}_2 + n_{2,g_k} = \mathbf{h}_{2,g_k}^{T} \mathbf{F}_{u,t}\mathbf{B}_{u,t}\mathbf{P} \mathbf{\hat{d}} + n_{g_k},  \\ \vspace{-4ex}
		&= \mbox{$\squeezespaces{0.5}\underbrace{\sqrt{p_{g_k}}\mathbf{h}_{2,g_k}^T \mathbf{F}_{u,t} \mathbf{b}_{{u,t,g_k}} \hat{d}_{g_k}}_{\text{Desired Signal}} \hspace{-0.5ex}+ \hspace{-0.5ex}\underbrace{\sum_{\hat{k} \neq k}^{K_g} \sqrt{p_{g_{\hat{k}}}} \mathbf{h}_{2,g_k}^T \mathbf{F}_{u,t} \mathbf{b}_{{u,t,g_{\hat{k}}}} \hat{d}_{g_{\hat{k}}}}_{\text{Intra-group interference}}$} \vspace{-1ex} \\ 
		& + \mbox{$\squeezespaces{0.5}\underbrace{\sum_{q\neq g}^G \sum_{\hat{k}=1}^{K_g} \sqrt{p_{q_{\hat{k}}}} \mathbf{h}_{2,g_k}^T \mathbf{F}_{u,t} \mathbf{b}_{{u,t,q_{\hat{k}}}} \hat{d}_{q_{\hat{k}}}}_{\text{Inter-group interference}} + \underbrace{n_{g_k}}_{\text{Noise}}$}, \raisetag{1.5\baselineskip} 
	\end{split}	
\end{equation}
where $g_k = k+ \sum_{g'=1}^{g-1} K_{g'}$ is the IoT node index, $\mathbf{h}_{2,g_k}$ $\in$ $\mathbb{C}^{N_t}$ is the channel vector between UAV and corresponding IoT node, and $\mathbf{n}_{g_k}$ denotes the complex circular symmetric Gaussian noise distributed as $\mathcal{CN} (\mathbf{0}, \sigma^2)$. Then, the achievable rate of first link (between BS and UAV) is: 
\begin{equation}
\mbox{$\squeezespaces{0.05}	\mathrm{R}_1 \hspace{-0.1ex}\left(\mathbf{F}_b,\mathbf{B}_b, \mathbf{F}_{u,r},\hspace{-0.1ex} \mathbf{B}_{u,r}\hspace{-0.2ex}\right)\hspace{-0.1ex} =\hspace{-0.1ex} \log_2 \left|\hspace{-0.1ex}\mathbf{I}_{K} \hspace{-0.1ex} +  \hspace{-0.1ex}\mathbf{Q}_1^\mathrm{-1}\hspace{-0.1ex}\mathbf{B}_{u,r}\mathbfcal{H}_1 \mathbf{B}_b \mathbf{B}_b^H \mathbfcal{H}_1^H \mathbf{B}_{u,r}^H \hspace{-0.1ex}\right|$}, 
\end{equation} 
where $\mathbf{Q}_1^\mathrm{-1}\hspace{-1ex} = \hspace{-1ex}(\sigma_n^2 \mathbf{B}_{u,r} \mathbf{F}_{u,r})^\mathrm{-1}  \mathbf{F}_{u,r}^H \mathbf{B}_{u,r}^H$ and $\mathbfcal{H}_1 \hspace{-1ex}=\hspace{-1ex} \mathbf{F}_{u,r}\mathbf{H}_1\mathbf{F}_b$. Similarly, the total AR for the second link (between UAV and multiple IoT nodes) is based on the instantaneous signal-to-interference-plus-noise ratio (SINR), which is given as:
\begin{equation}
\mbox{$\squeezespaces{0.5}	\text{SINR}_{\hspace{-0.25ex}g_k} \hspace{-0.95ex}=\hspace{-0.5ex} \frac{p_{g_k}|\mathbf{h}_{2,k}^{H} \mathbf{F}_{u,t}\mathbf{b}_{u,t,g_k}|^2}{ \hspace{-0.25ex}\sum_{\hat{k} \neq k}^{K_g} \hspace{-0.25ex} p_{g_{\hat{k}}}\hspace{-0.25ex}|\hspace{-0.3ex}\mathbf{h}_{2,k}^{H}\hspace{-0.2ex} \mathbf{F}_{u,t}\hspace{-0.25ex}\mathbf{b}_{u,t,g_{\hat{k}}}\hspace{-0.25ex}|^2 \hspace{-0.25ex} + \hspace{-0.25ex}\sum_{q \neq g}^G\hspace{-0.25ex}\sum_{\hat{k} \neq k}^{K_g}\hspace{-0.25ex} p_{q_{\hat{k}}} \hspace{-0.25ex}|\hspace{-0.25ex}\mathbf{h}_{2,k}^{H}\hspace{-0.25ex} \mathbf{F}_{u,t}\hspace{-0.25ex}\mathbf{b}_{u,t,q_{\hat{k}}}\hspace{-0.25ex}|^2 \hspace{-0.25ex}+ \hspace{-0.25ex}\sigma^2}.$} \label{eq:sinr}	
\end{equation}
By using the instantaneous SINR, the ergodic sum-rate capacity of the second link $R_2$ for the UAV-assisted mmWave MU-mMIMO IoT systems can be written as:
\begin{equation}
\mbox{$\squeezespaces{0.2}	\mathrm{R}_2 \left(\hspace{-0.2ex}\mathbf{F}_{u,t},\hspace{-0.2ex}\mathbf{B}_{u,t}, \hspace{-0.2ex}\mathbf{P},\hspace{-0.1ex} x_u, y_u\hspace{-0.1ex}\right) \hspace{-0.2ex}= \hspace{-0.2ex}\EX \biggl\{\hspace{-0.2ex} \sum\limits_{g = 1}^{G}\hspace{-0.2ex}\sum\limits_{k = 1}^{K_g}\hspace{-0.2ex} \EX[\log_2 \hspace{-0.2ex}(1 + \text{SINR}_{g_k}\hspace{-0.2ex})]\hspace{-0.2ex}\biggr\}.$} \label{eq:R_sum} 
\end{equation}
\vspace{-2em}
\subsection{Channel Model}
While LoS channel models can be useful for simple scenarios, they can be limited in their ability to capture the channel complexities (e.g., multi-path fading and shadowing). On the other hand, mmWave channel models can provide a more accurate representation of the channel characteristics, including the impact of non-LoS (NLoS) paths and obstacles on the signal propagation in UAV-assisted communications. Therefore, we consider mmWave channels for both links. The channel between BS and UAV is modeled based on the Saleh-Valenzuela channel model \cite{mmWave_Roi2016}, and is given as:
\begin{equation}
\begin{split}
\mathbf{H}_1 &= \sum\limits_{c = 1}^{C} \sum\limits_{l = 1}^{L} z_{1_{cl}} \tau_{1_{cl}}^{-\eta} \mathbf{a}_{1}^{(r)} ( \hspace{-0.25ex}{{\theta^{(r)}_{cl}}},{{\phi^{(r)}_{cl}}} ) \mathbf{a}_{1}^{(t) T}\hspace{-0.2ex} (\hspace{-0.25ex} {{\theta_{cl}^{(t)}}},{{\phi_{cl}^{(t)}}}\hspace{-0.25ex} )\hspace{-0.1ex} \\ 
&= \mathbf{A}_{1}^{(r)}\mathbf{Z}_1\mathbf{A}_{1}^{(t)}, \label{eq:first_channel} 
\end{split}
\end{equation}
where $C$ is the total number of clusters, $L$ is the total number of paths, $\eta$ is the path loss exponent, and $z_{1_{cl}}$ is the complex gain of $l^{th}$ path in $c^{th}$ cluster. Here, ${\bf{a}}_1^{(j)}(\cdot,\cdot)$ is the corresponding transmit or receive array steering vector for uniform rectangular array (URA), which is given as \cite{koc2020Access}: 
\begin{equation}
	\begin{split}
	\hspace{-0.5ex}{\bf{a}}_1^{(j)}\hspace{-0.5ex}\left( \hspace{-0.2ex}{{\theta, \phi}}\hspace{-0.2ex} \right) \hspace{-0.5ex}&=\hspace{-0.2ex}\mbox{$\squeezespaces{0.1}\hspace{-0.75ex} \big[ {1,{e^{- j2\pi d  {{\sin (\theta) \cos (\phi)}} }}, \cdots,{e^{ -j2\pi d\left( {N_x - 1} \right) {{\sin (\theta) \cos (\phi)}} }}} \big]$}\\
		 & \hspace{-1em}\mbox{$\squeezespaces{0.1}\otimes \hspace{-0.5ex}\big[\hspace{-0.2ex} {1,{e^{ -j2\pi d  {{\sin (\theta) \sin (\phi)}}}}, \hspace{-0.2ex}\cdots\hspace{-0.2ex},{e^{ -j2\pi d\left( {N_y - 1} \right) {{\sin (\theta) \sin (\phi)}} }}} \hspace{-0.2ex}\big]$}, \raisetag{1\baselineskip}
	\end{split}  \label{eq_phase_vector} 
\end{equation} 
where $j = \{t,r\}$, $d$ is the inter-element spacing, $N_x(N_y)$ is the horizontal (vertical) size of corresponding antenna array at BS and UAV, $\mathbf{Z}_1 = \text{diag} ( z_{1,1} \tau_{1,1}^{-\eta},\hdots, z_{1,L} \tau_{1,L}^{-\eta})$ $\in$ $\mathbb{C}^{L \times L}$ is the diagonal gain matrix, $\mathbf{A}_{1}^{(r)}$ $\in$ $\mathbb{C}^{N_r \times L}$ and $\mathbf{A}_{1}^{(t)}$ $\in$ $\mathbb{C}^{L \times N_T}$ are the receive and transmit phase response matrices, respectively. Here, the angles ${\theta _{cl}^{(t)}} \in \big[ {{\theta}_c^{(t)} - {\delta_c^{^{\theta (t)}}}}, {{\theta}_c^{(t)} + {\delta_c^{^{\theta (t)}}}} \big]$ and   
${\phi _{cl}^{(t)}} \in \big[ {{\phi}_c^{(t)} - {\delta_c^{\phi (t)}}}, {{\phi}_c^{(t)} + {\delta_c^{\phi (t)}}} \big]$ are the elevation AoD (EAoD) and azimuth AoD (AAoD) for $l^{th}$ path in channel $\mathbf{H}_1$, respectively. ${\theta}_c^{(t)}$ is the mean EAoD and ${\delta_c^{^{\theta (t)}}}$ represents the EAoD spread, whereas ${\phi}_c^{(t)}$ is mean AAoD with spread ${\delta_c^{\phi (t)}}$. Similarly, the angles ${\theta_{cl}^{(r)}} \in \big[ {{\theta}_c^{(r)} - {\delta_c^{\theta{(r)}}}}, {{\theta}_c^{(r)} + {\delta_c^{\theta{(r)}}}} \big]$ and ${\phi_{cl}^{(r)}} \in \big[ {{\phi}_c^{(r)} - {\delta_c^{\phi (r)}}}, {{\phi}_c^{(r)} + {\delta_c^{\phi (r)}}} \big]$ are the elevation AoA (EAoA) and azimuth AoA (AAoA), where ${\theta}_c^{(r)}$ and ${\phi}_c^{(r)}$ are the mean EAoA and AAoA with angular spread ${\delta_c^{\theta (r)}}$ and $ {\delta_c^{\phi (r)}}$, respectively. Then, the channel vector between the UAV and the $k^{th}$ IoT user can be written as follows:
\begin{equation}
	\begin{split}
	\mathbf{h}_{2,k}^T \hspace{-0.3ex} &=\hspace{-0.3ex}  \sum\nolimits_{q = 1}^{Q} z_{2,k_q}\tau_{2,k_q}^{-\eta}  \mathbf{a}(\theta_{k_q}, \phi_{k_q})\hspace{-0.3ex} \\
	 &= \hspace{-0.3ex} \mathbf{z}_{2,k}^T \mathbf{A}_{2,k} \in \mathbb{C}^{N_{t}}, \label{eq:second_channel} 
	\end{split}
\end{equation}
where $Q$ is the total number of downlink paths from UAV to IoT nodes, $z_{2,k_{q}}$ $\sim$ $\mathcal{C N}(0, \frac{1}{Q})$ is the complex path gain of $q^{th}$ path in second link, $\mathbf{a}(\cdot, \cdot)$ $\in$ $\mathbb{C}^{N_{t}}$ is the UAV downlink array phase response vector. As expressed in (\ref{eq:second_channel}), the intended downlink channel constitute two parts: 1) fast time-varying path gain vector $\mathbf{z}_{2, k}=[z_{2, k_{1}} \tau_{2, k_{1}}^{-\eta}, \cdots, z_{2, k_{Q}}\tau_{2, k_{Q}}^{-\eta}]^{T} \in \mathbb{C}^{Q}$; and 2) slow time-varying downlink array phase response matrix $\mathbf{A}_{2,k} \in \mathbb{C}^{Q \times N_{t}}$ with the rows of $\mathbf{a}(\theta_{k_l}, \phi_{k_l})$. Then, the channel matrix for second link is written as:
\begin{equation}
\mathbf{H}_{2}=[\mathbf{h}_{2, 1}, \cdots, \mathbf{h}_{2, K}]^{T}=\mathbf{Z}_{2} \mathbf{A}_2 \in \mathbb{C}^{K \times N_{t}},
\end{equation}
where $\mathbf{Z}_{2}=[\mathbf{z}_{2, 1}, \cdots, \mathbf{z}_{2, K}]^{T} \in \mathbb{C}^{K \times Q}$ is the complete path gain matrix for all downlink IoT nodes.
\vspace{-1em}
\subsection{Problem Formulation}
Considering the UAV is positioned at a fixed height $z_u$, and operates as DF relay, the total AR can be maximized by the joint optimization of the beamforming stages $\mathbf{F}_b$, $\mathbf{B}_b$, $\mathbf{F}_{u,t}$, $\mathbf{F}_{u,r}$, $\mathbf{B}_{u,t}$ and $\mathbf{B}_{u,r}$ with optimal PA matrix $\mathbf{P}$ and UAV positioning $\mathbf{x}_o$ $=$ $[x_o, y_o]^T$ $\in$ $\mathbb{R}^{2}$. Here, $\mathbf{x}_o$ represents the 2-D UAV deployment in a given flying span. Then, we can formulate the optimization problem as follows:
\begin{gather}
	\begin{aligned} 
		&\max_{\left\{ \mathbf{F}_b, \mathbf{B}_b, \mathbf{F}_{u,t}, \mathbf{B}_{u,t}, \mathbf{F}_{u,r}, \mathbf{B}_{u,r}, \mathbf{P}, \mathbf{x}_o \right\}} \quad \textrm{R}_T \\
		&\textrm{s.t.} \hspace{2ex} C_1: \hspace{1ex}  |\mathbf{F}_{u,t} (i,j)|\hspace{-0.5ex} = \hspace{-0.5ex} \frac{1}{\sqrt{N_t}}, |\mathbf{F}_{u,r} (i,j)| \hspace{-0.5ex}= \hspace{-0.5ex} \frac{1}{\sqrt{N_r}}, \hspace{1ex} \forall i,j, \\
		& \quad \quad C_2: \hspace{1ex}  |\mathbf{F}_{b} (i,j)| = \frac{1}{\sqrt{N_T}}, \hspace{1ex} \forall i,j, \\ 
		&\quad \quad C_3: \hspace{1ex} \EX \{ \left\lVert \mathbf{s}_1 \right\rVert_2^2\} \leq P_T, \\
		&\quad \quad C_4: \hspace{1ex} \EX \{ \left\lVert \mathbf{s}_2 \right\rVert_2^2\} = \sum\nolimits_{k=1}^{K}p_k \mathbf{b}_{u,t,k}^H\mathbf{F}_{u,t}^H\mathbf{F}_{u,t}\mathbf{b}_{u,t,k}\leq P_T, \\
		&\quad \quad C_5: \hspace{1ex} p_k \geq 0, \hspace{1ex} \forall k, \\
		&\quad \quad C_6: \hspace{1ex} \mathbf{x}_\mathrm{min} \leq \mathbf{x}_o \leq \mathbf{x}_\mathrm{max},
	\end{aligned}  \raisetag{0.7\baselineskip} \label{eq:optimization} 
\end{gather}
where $\textrm{R}_T = \frac{1}{2} \min( \mathrm{R}_1, \mathrm{R}_2)$ is the total transmission rate from BS to multiple IoT devices under DF protocol, $C_1$ and $C_2$ refers to the CM constraint due to the use of PSs for UAV and BS, respectively, $C_3$ and $C_4$ represents the transmit power constraint for BS and UAV, respectively, $C_5$ is the non-negative allocated power to each IoT node, and $C_6$ implies UAV deployment within the given flying span. Here, $[\mathbf{x}_\mathrm{min}, \mathbf{x}_\mathrm{max}] = [(x_\mathrm{min}, y_\mathrm{min}), (x_\mathrm{max}, y_\mathrm{max})]$ represents the UAV deployment range in 2-D space. The optimization problem defined in (\ref{eq:optimization}) is non-convex and intractable due to the following reason: 1) the CM constraint at each RF stage; and 2) fractional programming variables are entangled with each other. To solve this challenging problem, we propose three different PSO-based algorithmic solutions in Section III, which can achieve a near-optimal solution in finding optimal PA $\mathbf{P}$ and UAV deployment $\mathbf{x}_o$. Then, in Section IV, we introduce a novel low-complexity DL-based solution, which can reduce the runtime while providing similar performance to proposed PSO-based solutions.
\vspace{-2ex}
\section{Joint HBF, PA and UAV Location Optimization}
In this section, our objectives are to reduce the channel state information (CSI) overhead and the number of RF chains, while mitigating the inter-user interference to maximize the total achievable rate of a dual-hop UAV-assisted MU-mMIMO IoT systems. In this regard, we consider the joint optimization of the UAV location, PA to multiple IoT users, and HBF design for BS and UAV. Since the optimization problem is intractable, we first design $\mathbf{F}_b, \mathbf{B}_b, \mathbf{F}_{u,t}, \mathbf{F}_{u,r}, \mathbf{B}_{u,t}, \mathbf{B}_{u,r}$ based on some fixed UAV location, and then re-formulate RF and BB stages for the optimal UAV location as well as adjusting the allocated power in the MU PA block $\mathbf{P}$ by using three different algorithmic schemes: 1) optimal PA for fixed UAV location; 2) UAV location optimization for equal PA; and 3) joint UAV location optimization and optimal PA.
\vspace{-2ex} 
\subsection{RF-Beamformer Design}
The design of RF stages $\mathbf{F}_b,\mathbf{F}_{u,r},\mathbf{F}_{u,t}$ intends to maximize the beamforming gain for the desired signals expressed in (\ref{eq:received_UAV}) and (\ref{eq:received_User}). We design the RF stages $\mathbf{F}_b,\mathbf{F}_{u,r},\mathbf{F}_{u,t}$ based on the slow time-varying AoD and AoA. By using (\ref{eq:first_channel}), the effective channel for first link can be written as follows:\vspace{-0.5ex}
\begin{equation}
	\mathbfcal{H}_1=\mathbf{F}_{u,r} \mathbf{H}_1 \mathbf{F}_{b} = \mathbf{F}_{u, r} \boldsymbol{A}_{1}^{(r)} \mathbf{Z}_1 \boldsymbol{A}_{1}^{(t)} \mathbf{F}_{b} \label{eq:effective_first_channel} \vspace{-0.5ex}
\end{equation}
To maximize the transmit (receive) beamforming gain and exploit all degrees of freedom (DoF) provided by the first channel, the columns of $\mathbf{F}_{b}(\mathbf{F}_{u, r})$ should belong to the subspace spanned by $\mathbf{A}_{1}^{(t)}(\mathbf{A}_{1}^{(r)})$. Thus, we should satisfy $\operatorname{Span}\left(\mathbf{F}_{b}\right) \subset \operatorname{Span}(\mathbf{A}_{\mathrm{1}}^{(t)})$ and $\operatorname{Span}\left(\mathbf{F}_{\mathrm{u}, \mathrm{r}}\right) \subset \operatorname{Span}(\mathbf{A}_{\mathrm{1}}^{(r)})$. Here, it is worthwhile to mention that the transmit (receive) phase response matrix $\mathbf{A}_{1}^{(t)}(\mathbf{A}_{1}^{(r)})$ is a function of slow time-varying $\mathrm{AoD}(\mathrm{AOA})$ information. Thus, the $\mathrm{AoD}$ and $\mathrm{AoA}$ supports for first channel are defined as follows:
\begin{align}
	\textrm{AoD} = \big\lbrace {\sin ( \theta  ){{\left[ {\cos ( \phi  ),\sin ( \phi )} \right]}}} \big| {\theta} \in \bm{\theta}_1^{(t)}, {\phi} \in \bm{\phi}_1^{(t)} \big\rbrace	,\label{eq_AoD_Supp} \\
	\textrm{AoA} = \big\lbrace {\sin ( \theta  ){{\left[ {\cos ( \phi  ),\sin ( \phi )} \right]}}} \big| 	\theta \in{\bm {\theta }}_1^{(r)}, \phi \in{\bm {\phi }}_1^{(r)}\big\rbrace, \label{eq_AoA_Supp} 
\end{align}
where $\bm{\theta}_1^{(t)}=\cup_{c=1}^C[\theta_{1, c}^{(t)}-\delta_{1, c}^{(t) \theta}, \theta_{1, c}^{(t)}+\delta_{1, c}^{(t) \theta}]$ and $\bm{\phi}_1^{(t)}=$ $\cup_{c=1}^C[\phi_{1, c}^{(t)}-\delta_{1, c}^{(t) \phi}, \phi_{1, c}^{(t)}+\delta_{1, c}^{(t) \phi}]$ denote the elevation and azimuth $\mathrm{AOD}$ supports of first channel, respectively. Similarly, $\bm{\theta}_1^{(r)}=\cup_{c=1}^C\left[\theta_{1, c}^{(r)}-\delta_{1, c}^{(r) \theta}, \theta_{1, c}^{(r)}+\delta_{1, c}^{(r) \theta}\right]$ and $\bm{\phi}_1^{(r)}=$ $\cup_{c=1}^C\left[\phi_{1, c}^{(r)}-\delta_{1, c}^{(r)\phi}, \phi_{1, c}^{(r)}+\delta_{1, c}^{(r) \phi}\right]$ represent the elevation and azimuth AoA supports of the channel between BS and UAV, respectively. Then, the transmit RF beamformer $\mathbf{F}_b$ is constructed via transmit steering vector as follows: 
\begin{align}
 &{\bf{e}}_{b}^{(t)}\hspace{-0.5ex}\left( {{\theta, \phi}} \right) \hspace{-0.5ex} = \notag \\ &\hspace{-0.75ex} \frac{1}{\sqrt{N_{x,T}}}\big[\hspace{-0.25ex} {1,{e^{j2\pi d  {{\sin (\theta) \cos (\phi)}} }}\hspace{-0.25ex},\hspace{-0.25ex} \cdots\hspace{-0.25ex},\hspace{-0.25ex}{e^{j2\pi d\left( {N_{x,T} - 1} \right) {{\sin (\theta) \cos (\phi)}} }}} \big]^T \hspace{-0.25ex}\otimes \notag \\
 & \hspace{-0.5ex}\frac{1}{\hspace{-0.25ex}\sqrt{N_{y,T}}\hspace{-0.25ex}}\big[\hspace{-0.25ex} {1\hspace{-0.25ex},{e^{\hspace{-0.25ex}j2\pi d  {{\sin (\hspace{-0.25ex}\theta\hspace{-0.25ex}) \sin (\hspace{-0.25ex}\phi\hspace{-0.25ex})}} }}\hspace{-0.35ex},\hspace{-0.25ex} \cdots\hspace{-0.35ex},\hspace{-0.25ex}{e^{\hspace{-0.25ex}j2\pi d\left( {N_{y,T} - 1} \right) {{\sin (\hspace{-0.25ex}\theta\hspace{-0.25ex}) \sin (\hspace{-0.25ex}\phi\hspace{-0.25ex})}} }}} \hspace{-0.25ex}\big]^T.
\end{align}
To reduce the RF chains utilization while covering the complete AoD and AoA supports, we define the quantized angle-pairs as: ${{\lambda _{x,b}^{n(t)}} \hspace{-0.5ex}=\hspace{-0.5ex} -1 + \frac{2n-1}{{{N_{x,T}}}}}$ for $n = 1, \cdots,{N_{x,T}}$ and ${{\lambda _{y,b}^{k(t)}} = -1 + \frac{2k-1}{{{N_{y,T}}}}}$ for $k = 1, \cdots,{N_{y,T}}$. Here, $N_{x,T}$ and $N_{y,T}$ are the antenna elements along x and y axis, respectively. Then, the quantized angle-pairs inside the AoD support satisfying \eqref{eq_AoD_Supp} are obtained as:  
\begin{equation}\label{eq_AOD_quantized_angles} 
	( \hspace{-0.3ex}{\lambda_{x,b}^{n(t)},\lambda _{y,b}^{k(t)}} \hspace{-0.25ex}) \hspace{-0.6ex} ~\big |\hspace{-0.5ex} \Big.~ {\sin (\theta)\hspace{-0.3ex}\cos (\phi)} \hspace{-0.5ex} \in \hspace{-0.3ex}{\boldsymbol{\lambda }}_{x,b}^{n(t)},\hspace{-0.3ex}{\sin (\theta)\sin (\phi)}\hspace{-0.5ex} \in\hspace{-0.3ex} {\boldsymbol{\lambda }}_{y,b}^{k(t)}, 
\end{equation}
where ${\boldsymbol{\lambda }}_{x,b}^{n(t)} \hspace{-0.5ex}=\hspace{-0.5ex} \big[ {\lambda _{x,b}^{n(t)} \hspace{-0.5ex}-\hspace{-0.5ex} \frac{1}{{{N_{x,T}}}},\lambda _{x,b}^{n(t)} \hspace{-0.5ex}+\hspace{-0.5ex} \frac{1}{{{N_{y,T}}}}} \big]$ is the boundary of $\lambda _{x,b}^{n(t)}$, and 
${\boldsymbol{\lambda }}_{y,b}^{k(t)} \hspace{-0.5ex}=\hspace{-0.5ex} \big[ {\lambda _{y,b}^{k(t)} \hspace{-0.5ex}-\hspace{-0.5ex} \frac{1}{{{N_{y,T}}}},\lambda _{y,b}^{k(t)} \hspace{-0.5ex}+\hspace{-0.5ex} \frac{1}{{{N_{y,T}}}}} \big]$ is the boundary of $\lambda _{y,b}^{k(t)}$. By using \eqref{eq_AoD_Supp} and \eqref{eq_AOD_quantized_angles}, ${\mathbf{F}}_b$ can be written in the form of transmit steering vector as:
%\begin{equation}\label{eq:eq_TX_RF}
%\mbox{$\squeezespaces{0.1}	{{\bf{F}}_b} \hspace{-0.5ex}=\hspace{-0.5ex} \big[ \hspace{-0.25ex}{{\bf{e}}_{b}^{(t)}\big(\hspace{-0.35ex} {\lambda ^{n_1 (t)}_{x,b} \hspace{-0.3ex},\lambda ^{k_1 (t)}_{y,b}} \hspace{-0.5ex}\big), \hspace{-0.25ex} \hdots,\hspace{-0.25ex} {\bf{e}}_{b}^{(t)}\big(\hspace{-0.25ex} {\lambda ^{n_{N_{{RF}_b}}\hspace{-0.3ex}(t)}_{x,b},\lambda ^{k_{N_{{RF}_b}}\hspace{-0.3ex}(t)}_{y,b}} \hspace{-0.5ex}\big)}\hspace{-0.25ex} \big]\hspace{-0.5ex} \in \hspace{-0.2ex}\mathbb{C}^{N_T \times N_{{RF}_b}}$}
%\end{equation} 
\begin{equation}\label{eq:eq_TX_RF}
	\begin{split}
&\hspace{-1.75ex}{{\bf{F}}_b} = \big[{\bf{e}}_{b}^{(t)}\big( {\lambda ^{n_1 (t)}_{x,b},\lambda ^{k_1 (t)}_{y,b}} \big),  \cdots, \\
&\hspace{9ex} \cdots, {\bf{e}}_{b}^{(t)}\big(\hspace{-0.25ex} {\lambda ^{n_{N_{{RF}_b}}\hspace{-0.3ex}(t)}_{x,b},\lambda ^{k_{N_{{RF}_b}}\hspace{-0.3ex}(t)}_{y,b}} \hspace{-0.25ex}\big)\hspace{-0.25ex} \big]\hspace{-0.9ex} \in \hspace{-0.2ex}\mathbb{C}^{N_T \times N_{{RF}_b}}\hspace{-0.35ex}.
	\end{split}
\end{equation}
Following similar approach, we can design the UAV receive RF stage via receive steering vector as follows: 
\begin{align}
&{\bf{e}}_{u}^{(r)}\hspace{-0.5ex}\left( {{\theta, \phi}} \right) \hspace{-0.5ex} = \notag \\ 
&\hspace{-0.75ex} \frac{1}{\sqrt{N_{x,r}}}[\hspace{-0.25ex} {1,{e^{- j2\pi d  {{\sin (\theta) \cos (\phi)}} }}\hspace{-0.25ex},\hspace{-0.25ex} \cdots\hspace{-0.25ex},\hspace{-0.25ex}{e^{ -j2\pi d\left( {N_{x,r} - 1} \right) {{\sin (\theta) \cos (\phi)}} }}}\hspace{-0.25ex}]^T \hspace{-0.4ex}\otimes \notag \\
& \hspace{-0.5ex}\frac{1}{\sqrt{N_{y,r}}\hspace{-0.25ex}}[\hspace{-0.25ex} {1\hspace{-0.25ex},{e^{\hspace{-0.25ex} -j2\pi d  {{\sin (\hspace{-0.25ex}\theta\hspace{-0.25ex}) \sin (\hspace{-0.25ex}\phi\hspace{-0.25ex})}} }}\hspace{-0.25ex}, \hspace{-0.35ex}\cdots\hspace{-0.35ex},\hspace{-0.25ex}{e^{\hspace{-0.25ex} -j2\pi d\left(\hspace{-0.25ex} {N_{y,r} \hspace{-0.25ex}-\hspace{-0.25ex} 1} \hspace{-0.25ex}\right) {{\sin (\hspace{-0.25ex}\theta\hspace{-0.25ex}) \sin (\hspace{-0.25ex}\phi\hspace{-0.25ex})}} }}}\hspace{-0.25ex}]^T.
\end{align}
\noindent Using quantized angle-pairs as ${{\lambda _{x,u}^{n(r)}} \hspace{-0.5ex}=\hspace{-0.5ex} -1 + \frac{2n-1}{{{N_{x,r}}}}}$ for $n = 1, \cdots,{N_{x,r}}$ and ${{\lambda _{y,u}^{k(r)}} = -1 + \frac{2k-1}{{{N_{y,r}}}}}$ for $k = 1, \cdots,{N_{y,r}}$, the UAV receive RF stage can be formulated as: \vspace{-0.5ex}
%\begin{equation}\label{eq:eq_RX_RF}
%\mbox{$\squeezespaces{0.1}	{{\bf{F}}_{u,r}} \hspace{-0.5ex}=\hspace{-0.5ex} \big[ \hspace{-0.25ex}{{\bf{e}}_{u}^{(r)\hspace{-0.5ex}}\big(\hspace{-0.35ex} {\lambda ^{n_1 (r)}_{x,u}\hspace{-0.3ex},\lambda ^{k_1 (r)}_{y,u}} \hspace{-0.5ex}\big), \hspace{-0.25ex} \hdots,\hspace{-0.25ex} {\bf{e}}_{u}^{(r)\hspace{-0.4ex}}\big(\hspace{-0.25ex} {\lambda^{n_{N_{{RF}_u}}\hspace{-0.5ex}(r)}_{x,u} \hspace{-0.5ex},\hspace{-0.5ex}\lambda ^{k_{N_{{RF}_u}}\hspace{-0.5ex}(r)}_{y,u}} \hspace{-0.65ex}\big)}\hspace{-0.25ex} \big]^{T\hspace{-0.5ex}}\hspace{-0.9ex} \in \hspace{-0.2ex}\mathbb{C}^{N_{{RF}_u}\hspace{-0.2ex} \times\hspace{-0.2ex} N_r} $} \vspace{-1ex}
%\end{equation} 
\begin{equation}\label{eq:eq_RX_RF}
	\begin{split}
&\hspace{-1.5ex}{{\bf{F}}_{u,r}} = \big[ {\bf{e}}_{u}^{(r)}\big( {\lambda ^{n_1 (r)}_{x,u},\lambda ^{k_1 (r)}_{y,u}} \big), \cdots, \\
&\hspace{9ex}	\hdots,\hspace{-0.25ex} {\bf{e}}_{u}^{(r)\hspace{-0.4ex}}\big(\hspace{-0.25ex} {\lambda^{n_{N_{{RF}_u}}\hspace{-0.5ex}(r)}_{x,u} \hspace{-0.25ex},\hspace{-0.25ex}\lambda ^{k_{N_{{RF}_u}}\hspace{-0.25ex}(r)}_{y,u}} \hspace{-0.25ex}\big)\hspace{-0.25ex} \big]^{T\hspace{-0.5ex}}\hspace{-0.9ex} \in \hspace{-0.2ex}\mathbb{C}^{N_{{RF}_u}\hspace{-0.2ex} \times\hspace{-0.2ex} N_r}\hspace{-0.35ex}, \vspace{-1ex}
	\end{split}
\end{equation}
where the quantized angle-pairs to provide AoA support while satisfying \eqref{eq_AoA_Supp} are obtained as: 	\vspace{-0.5ex}
\begin{equation}\label{eq_AOA_quantized_angles} 
\mbox{$\squeezespaces{0.4}( \hspace{-0.5ex}{\lambda _{x,u}^{n(r)},\lambda _{y,u}^{k(r)}} \hspace{-0.25ex}) \hspace{-1.15ex} ~\big |\hspace{-0.5ex} \Big.~ {\sin (\theta)\cos (\phi)} \hspace{-0.5ex} \in {\boldsymbol{\lambda }}_{x,u}^{n(r)},{\sin (\theta)\sin (\phi)}\hspace{-0.5ex} \in {\boldsymbol{\lambda }}_{y,u}^{k(r)},$} \vspace{-1ex}
\end{equation}
where ${\boldsymbol{\lambda }}_{x,u}^{n(r)} \hspace{-0.5ex}=\hspace{-0.5ex} \big[ {\lambda _{x,u}^{n(r)} \hspace{-0.5ex}-\hspace{-0.5ex} \frac{1}{{{N_{x,r}}}},\lambda _{x,u}^{n(r)} \hspace{-0.5ex}+\hspace{-0.5ex} \frac{1}{{{N_{x,r}}}}} \big]$ is the boundary of $\lambda _{x,u}^{n(r)}$, and 
${\boldsymbol{\lambda }}_{y,u}^{k(r)} \hspace{-0.5ex}=\hspace{-0.5ex} \big[ {\lambda _{y,u}^{k(r)} \hspace{-0.5ex}-\hspace{-0.5ex} \frac{1}{{{N_{y,r}}}},\lambda _{y,u}^{k(r)} \hspace{-0.5ex}+\hspace{-0.5ex} \frac{1}{{{N_{y,r}}}}} \big]$ is the boundary of $\lambda _{y,u}^{k(r)}$. After the design of $\mathbf{F}_b, \mathbf{F}_{u,r}$ RF stages, the UAV transmit RF stage is designed to support $K$ IoT users, which are clustered into $G$ groups based on their AoD information. Here, each group $g$ contains $K_g$ number of IoT users such that $K = \sum_{g=1}^G K_g$. The index $g_k = \sum_{g'=1}^{g-1} K_{g'} + k$ is used to denote the $k^{th}$ IoT user in group $g$. According to the user groups, we design $G$ different sub-blocks for the UAV transmit RF stage as $\mathbf{F}_{u,t}=\left[\mathbf{F}_{{u,t,1}}, \mathbf{F}_{{u,t,2}}, \cdots, \mathbf{F}_{{u,t,G}}\right] \in \mathbb{C}^{N_t \times N_{{RF}_u}}$, where $\mathbf{F}_{{u,t,g}} \in \mathbb{C}^{N_t \times N_{{RF}_{u,g}}}$ denotes the RF beamfomer for group $g$ such that $N_{{RF}_u}=\sum_{g=1}^G N_{{RF}_{u,g}}$. Then, $\mathbfcal{H}_2 = \mathbf{H}_2 \mathbf{F}_{u,t} \in \mathbb{C}^{K \times N_{{RF}_u}}$ is the reduced-size effective channel matrix seen from UAV transmit baseband stage $\mathbf{B}_{u,t}$. By defining $\mathbf{H}_2=\left[\mathbf{H}_{2,1}^T, \cdots, \mathbf{H}_{2,G}^T\right]^T \in \mathbb{C}^{K \times N_t}$, the effective channel matrix for second link can be expressed as: \vspace{-1ex}
\begin{equation}
	\mathbfcal{H}_2\hspace{-0.3ex}=\hspace{-0.6ex}\left[\hspace{-1ex}\begin{array}{cccc}
	\mathbf{H}_{2,1} \mathbf{F}_{u,t,1} & \mathbf{H}_{2,1} \mathbf{F}_{u,t,2} & \ldots & \mathbf{H}_{2,1} \mathbf{F}_{u,t,G} \\
		\mathbf{H}_{2,2} \mathbf{F}_{u,t,1} & \mathbf{H}_{2,2} \mathbf{F}_{u,t,2} & \ldots & \mathbf{H}_{2,2} \mathbf{F}_{u,t,G} \\
		\vdots & \vdots & \ddots & \vdots \\
		\mathbf{H}_{2,G} \mathbf{F}_{u,t,1} & \mathbf{H}_{2,G} \mathbf{F}_{u,t,2} & \ldots & \mathbf{H}_{2,G} \mathbf{F}_{u,t,G}
	\end{array}\hspace{-1ex}\right]\hspace{-0.3ex}, \label{eq:effective_channel_2}  \vspace{-0.5ex}
\end{equation}
where the diagonal block-matrix $\mathbfcal{H}_{2,g}=\mathbf{H}_{2,g} \mathbf{F}_{u,t,g}=\mathbf{Z}_{2,g} \mathbf{A}_{2,g} \mathbf{F}_{u,t}$ is the effective channel matrix for group $g$ and the off-diagonal block-matrix $\mathbfcal{H}_{2,\hat{g}}=\mathbf{H}_{2,g} \mathbf{F}_{u,t,g}=\mathbf{Z}_{2,\hat{g}} \mathbf{A}_{2,\hat{g}} \mathbf{F}_{u,t}$ is the effective interference channel matrix between groups $g$ and $\hat{g}, \forall \hat{g} \neq g$. The RF beamformer matrices are designed to eliminate inter-group interference as: \vspace{-0.5ex}
\begin{equation} \vspace{-0.5ex}
	\mathbf{A}_{2,\hat{g}} \mathbf{F}_{u,t,g} \approx 0, \hspace{1ex}\forall \hat{g} \neq g \hspace{1ex}\text { and }\hspace{1ex} \hat{g}, g=1, \cdots, G. 
\end{equation}
To design $\mathbf{F}_{u,t}$, which can satisfy the above zero condition, the columns of $\mathbf{F}_{u,t,g}$ should belong to the intersection of the null spaces of $\mathbf{A}_{2,\hat{g}}$, i.e., Span $\left(\mathbf{F}_{u,t,g}\right) \subset \cap_{\hat{g} \neq g} \operatorname{Null}\left(\mathbf{A}_{2,\hat{g}}\right)$. Moreover, in order to maximize the beamforming gain, the columns of $\mathbf{F}_{u,t,g}$ should belong to the subspace spanned by $\mathbf{A}_{2,g}$, i.e., $\operatorname{Span}\left(\mathbf{F}_{u,t,g}\right) \subset \operatorname{Span}\left(\mathbf{A}_{2,g}\right)$. Thus, the intersection of $\operatorname{Span}\left(\mathbf{A}_{2,g}\right)$ and $\operatorname{Null}\left(\mathbf{A}_{2,\hat{g}}\right), \forall \hat{g} \neq g$, should not be empty to obtain the RF beamformer matrix satisfying the above conditions. Similar to the design $\mathbf{F}_b$, the AoD support of the group $g$ can be expressed as the union of AoD supports for all IoT user in the corresponding group as:
\begin{align}
\mbox{$\squeezespaces{0.4}	\textrm{AoD}_g = \big\lbrace {\sin ( \theta  ){{\left[ {\cos ( \phi  ),\sin ( \phi )} \right]}}} \big| {\theta} \in \bm{\theta}_{2,g}, {\phi} \in \bm{\phi}_{2.g} \big\rbrace, $}
\end{align}
where $\squeezespaces{0.2}\bm{\theta}_{2,g} =\left[\theta_{2,g}^{\min}, \theta_{2,g}^{\max}\right]=\left[\min_{{2,g}_k,\hat{g}} \theta_{{2,g}_k,\hat{g}}, \max_{{2,g}_k,\hat{g}} \theta_{{2,g}_k,\hat{g}}\right]$ is the elevation angle support for group $g$ and $\bm{\phi}_{2.g}=$ $\left[\phi_{2,g}^{\min}, \phi_{2,g}^{\max}\right]=\left[\min_{{2,g}_k,\hat{g}} \phi_{{2,g}_k,\hat{g}}, \max_{{2,g}_k,\hat{g}} \phi_{{2,g}_k,\hat{g}}\right]$ is the azimuth angle support for group $g$. To achieve Span $\left(\mathbf{F}_{u,t,g}\right) \subset \operatorname{Span}\left(\mathbf{A}_{2,g}\right)$, the columns of the UAV transmit RF beamformer for the group $g$ can be constructed as: \vspace{-0.5ex}
	\begin{equation}
	\mathbf{F}_{u,t,g}=\{\mathbf{e}_{u}^{(t)}\left(\varphi, \vartheta\right) \big|\left(\varphi, \vartheta \right) \in \mathrm{AoD}_g\},
	\end{equation}
where $\varphi = \sin\theta \cos\phi$ and $\vartheta = \sin\theta \sin\phi$ represent the beam directions for the UAV transmit steering vector for uniform rectangular array (URA), which is defined as follows: 
\begin{equation}
\mathbf{e}_{u}^{(t)}(\varphi, \vartheta) = \mathbf{e}_{x,u}^{(t)}(\varphi) \otimes \mathbf{e}_{y,u}^{(t)}(\vartheta).
\end{equation} 
Then, we can construct the UAV transmit RF beamformer for each group $g$ based on the quantized angle-pairs as: \vspace{-0.5ex}
\begin{equation}\label{eq:eq_TX_RF_UAV}
	\begin{split}
&\hspace{-1.5ex}	\mathbf{F}_{u,t,g} = \big[{\bf{e}}_{u}^{(t)}( {\lambda ^{n_1 (t)}_{x,u },\lambda ^{k_1 (t)}_{y,u}}),  \cdots, \\
&\hspace{7ex}\cdots, {\bf{e}}_{u}^{(t)}(\hspace{-0.25ex} {\lambda ^{n_{N_{{RF}_{u,g}}}\hspace{-0.25ex}(t)\hspace{-0.25ex}}_{x,u}\hspace{-0.9ex},\hspace{-0.25ex}\lambda ^{k_{N_{{RF}_{u,g}}} \hspace{-0.25ex}(t)}_{y,u}} \hspace{-0.25ex})\hspace{-0.25ex} \big]\hspace{-0.75ex} \in \hspace{-0.5ex}\mathbb{C}^{\hspace{-0.25ex} N_t \hspace{-0.25ex} \times N_{{RF}_{u,g}}\hspace{-0.25ex}}, \vspace{-0.5ex}
	\end{split}
\end{equation} 
where ${\boldsymbol{\lambda }}_{x,u}^{n_i(t)} \hspace{-0.5ex}=\hspace{-0.5ex} \big[ {\lambda _{x,u}^{n_i(t)} \hspace{-0.5ex}-\hspace{-0.5ex} \frac{1}{{{N_{x,t}}}},\lambda _{x,u}^{n_i(t)} \hspace{-0.5ex}+\hspace{-0.5ex} \frac{1}{{{N_{x,t}}}}} \big]$ is the boundary of $\lambda _{x,u}^{n_i(t)}$, and 
${\boldsymbol{\lambda }}_{y,u}^{k_i(t)} \hspace{-0.5ex}=\hspace{-0.5ex} \big[ {\lambda _{y,u}^{k_i(t)} \hspace{-0.5ex}-\hspace{-0.5ex} \frac{1}{{{N_{y,t}}}},\lambda _{y,u}^{k_i(t)} \hspace{-0.5ex}+\hspace{-0.5ex} \frac{1}{{{N_{y,t}}}}} \big]$ is the boundary of $\lambda _{y,u}^{k_i(t)}$. The BS and UAV RF-stage designs do not require the fast time-varying instantaneous CSI and it is only based on the slow time-varying AoD/AoA information. Particularly, the design of BS and UAV RF beamformers require only four angular parameters, which are the mean of elevation and azimuth AoD (AoA) and their angular spread.
\vspace{-1em}
\subsection{BB-Stage Design}
After designing the transmit and receive RF beamformers for BS and UAV, we design the BB stages $\mathbf{B}_b$ and $\mathbf{B}_{u,r}$ based on the effective channel matrix $\mathbfcal{H}_1$ (as given in (\ref{eq:effective_first_channel})). By using singular value decomposition (SVD), we can write:
\begin{align} 
	\mathbfcal{H}_1 = \mathbf{U}_1 \mathbf{\Sigma}_1 \mathbf{V}^H_1, \vspace{-1ex}
\end{align}
where $\mathbf{U}_1$ $\in$ $\mathbb{C}^{{N_{{RF}_u}} \times rank(\mathbfcal{H}_1)}$ and $\mathbf{V}_1$ $\in$ $\mathbb{C}^{{N_{{RF}_b}} \times rank(\mathbfcal{H}_1)}$ are tall unitary matrices and $\mathbf{\Sigma}_1$ is the diagonal matrix with singular values in the decreasing order such that $\mathbf{\Sigma}_1 = \textrm{diag}(\sigma_1^2,\cdots,\sigma_{rank(\mathbfcal{H}_1)}^2)$ $\in$ $\mathbb{C}^{rank(\mathbfcal{H}_1) \times rank(\mathbfcal{H}_1)}$. Assuming rank$(\mathbfcal{H}_1) \geq K$, $\mathbf{V}_1$ can be partitioned as $\mathbf{V}_1 = [\mathbf{V}_{1,1}, \mathbf{V}_{1,2}]$ with $\mathbf{V}_{1,1}$ $\in$ $\mathbb{C}^{N_{{RF}_b} \times K}$. Then, the BB stages $\mathbf{B}_b$ and $\mathbf{B}_{u,r}$ for BS and UAV can be obtained as follows \cite{Asil_low_DACs}:	\vspace{-0.5ex}
\begin{equation} \label{eq:BB_BS}
	\mathbf{B}_b = \sqrt{\frac{P_T}{K}}\mathbf{V}_1 \in \mathbb{C}^{N_{{RF}_b} \times K}, 
\end{equation}
\begin{equation}
	\mathbf{B}_{u,r} = \mathbf{U}_1^H \in \mathbb{C}^{K \times N_{{RF}_u}}. \label{eq:BB_1st_link} \vspace{-1ex}
\end{equation}
Similarly, the reduced-size effective CSI $\bm{\mathcal{H}}_2$ given in (\ref{eq:effective_channel_2}) is employed for the UAV transmit BB stage design. We consider joint-group-processing (JGP) technique as designed in \cite{koc2020Access,mahmood2021energy,mobeen_3D}. The design of BB stage $\mathbf{B}_{u,t}$ not only reduces the intra-group interference but also mitigate the residual inter-group interference remaining after RF beamformer design. By applying the regularized zero-forcing (RZF) technique, $\mathbf{B}_{u,t}$ is defined as follows: 	
\begin{equation}
	\mathbf{B}_{u,t} = (\mathbfcal{H}_2^H \mathbfcal{H}_2 +\beta N_{{RF}_u}\mathbf{I}_{N_{{RF}_u}})^{-1}\mathbfcal{H}_2^H \in \mathbb{C}^{N_{{RF}_u} \times K}, \label{eq:BB_precoder}  \vspace{-1ex}
\end{equation}
where $\beta =\frac{\sigma^2}{P_T}$ is the regularization parameter and \squeezespaces{0.1}$\mathbf{I}_{N_{{RF}_u}}$ $\in$ $\mathbb{C}^{N_{{RF}_u} \times N_{{RF}_u}}$. The power constraint (i.e., $C_4$ in (\ref{eq:optimization})) will be adjusted by designing multi-user PA block $\mathbf{P}$ in Section III-C. 
\vspace{-2em}
\subsection{UAV Deployment and Multi-User Power Allocation}
After the formulation of RF beamformers $\mathbf{F}_b$, $\mathbf{F}_{u,r}$, $\mathbf{F}_{u,t}$ and BB stages $\mathbf{B}_{b}$, $\mathbf{B}_{u,t}$, $\mathbf{B}_{u,r}$, the optimization problem for maximum achievable rate can be formulated as follows:
\begin{gather}
	\begin{aligned}
		&\max_{\{\mathbf{P}, \mathbf{x}_o\}} \quad \textrm{R}_T(\mathbf{F}_b, \mathbf{B}_b, \mathbf{F}_{u,t}, \mathbf{B}_{u,t}, \mathbf{F}_{u,r}, \mathbf{B}_{u,r}, \mathbf{P}, \mathbf{x}_u) \\ 
		&\mbox{$\squeezespaces{0.3}\textrm{s.t.} \hspace{2ex} C_4: \hspace{1ex} \EX \{ \left\lVert \mathbf{s}_2 \right\rVert_2^2\} = \sum\nolimits_{k=1}^{K}p_k \mathbf{b}_{u,t,k}^H\mathbf{F}_{u,t}^H\mathbf{F}_{u,t}\mathbf{b}_{u,t,k}\leq P_T,$} \\
		&\quad \quad C_5: \hspace{1ex} p_k \geq 0, \hspace{1ex} \forall k,  \\
		&\mbox{$\squeezespaces{0.5} \quad \quad C_6: \hspace{1ex} \mathbf{x}_\mathrm{min} \leq \mathbf{x}_o \leq \mathbf{x}_\mathrm{max},$} 	\raisetag{0.7\baselineskip} \label{eq:optimization_reformulated}
	\end{aligned}   \vspace{-0.5ex}
\end{gather}
where $\textrm{R}_T$ is defined in (\ref{eq:optimization}). Even though the CM constraints for the RF beamformers and transmit power constraint for BS (i.e., $C_1$, $C_2$ and $C_3$ given in (\ref{eq:optimization})) are satisfied via the RF and BB stages developed in Section III-A and III-B, the updated optimization problem in (\ref{eq:optimization_reformulated}) is still non-convex due to the joint dependence of both the allocated powers $p_k$, and the UAV location $\mathbf{x}_o = [x_o, y_o]^T$ on the SINR expression in (\ref{eq:sinr}), which is used in the sum-rate $\mathrm{R}_2$ calculation as given in (\ref{eq:R_sum}). To overcome this challenge, we propose different PSO-based algorithmic solutions, which employ multiple agents, called particles, to explore the search space of objective function given in (\ref{eq:optimization_reformulated}). Initially, $M_p$ particles are randomly placed in search space, where each particle communicates with other particles to share their personal best solution and update the current global best solution for the objective function. The particles then move iteratively for $T$ iterations to reach the global optimum solution. Our PSO-based algorithms rely on two components, deterministic and stochastic, to guide the motion of the particles. The deterministic component utilizes knowledge from global and personal best solutions, while the stochastic component involves random movements. The proposed PSO-based algorithmic schemes are as follows: 1) optimal PA for fixed UAV location; 2) UAV positioning for equal PA; and 3) joint UAV positioning and optimal PA. 
\subsubsection{Optimal PA for Fixed UAV Location}
In this problem, we allocate optimal MU power values $\mathbf{P}$ for a fixed UAV location $\mathbf{x}_u$. Then, the $i^{th}$ particle at the $t^{th}$ iteration represents an instance of the PA matrix as follows:
\begin{equation}
\mathbf{P}_i^{(t)}=\operatorname{diag}(\sqrt{p_{1, i}^{(t)}}, \cdots, \sqrt{p_{K, i}^{(t)}}) \in \mathbb{R}^{K \times K}, \label{eq:P_PA} \vspace{-0.5ex}
\end{equation}
where, $i=1, \cdots, M_p$ and $t=0,1, \cdots, T$ are the particle index and iteration index, respectively. The objective function $R_{T}(\mathbf{F}_b, \mathbf{B}_b, \mathbf{F}_{u,t}, \mathbf{B}_{u,t}, \mathbf{F}_{u,r}, \mathbf{B}_{u,r}, \mathbf{P}_i^{(t)}, \mathbf{x}_u)$ is calculated after evaluating the personal best and communicating for the global best solutions for each $i^{th}$ particle. The particles move from $\mathbf{P}_i^{(t)}$ to $\mathbf{P}_i^{(t+1)}$ until the maximum number of iterations is reached. To satisfy the transmit power constraints $C_4$ and $C_5$ given in (\ref{eq:optimization_reformulated}), we define the following:
\begin{align}
\mbox{$\squeezespaces{0.5}	\kappa_i^{(t)} \hspace{-0.35ex}=\hspace{-0.35ex} \sqrt{\frac{P_T}{\EX \big\{ \left\lVert \mathbf{F}_{u,t}\mathbf{B}_{u,t}\hat{\mathbf{P}}_i^{(t)} \hat{\mathbf{d}} \right\rVert_2^2\big\}}} \overset{(a)}{=} \sqrt{\frac{P_T}{\sum\nolimits_{k=1}^K \hat{p}_{k,i}^{(t)} \mathbf{b}_{u,t,k}^H\mathbf{b}_{u,t,k}}}, $} \label{eq:kappa} \vspace{-0.5ex}
\end{align}
where (a) follows the unitary property of UAV transmit beamformer (i.e., $\mathbf{F}_{u,t}^H \mathbf{F}_{u,t}=\mathbf{I}_{N_{{RF}_u}})$. The normalized PA matrix is given as: $\hat{\mathbf{P}}_i^{(t)}=\operatorname{diag}(\sqrt{\hat{p}_{1, i}^{(t)}}, \cdots, \sqrt{\hat{p}_{K, i}^{(t)}}) \in \mathbb{R}^{K \times K}$ with $\hat{p}_{k, i}^{(t)} \in [0,1]$. Here, $\mathbf{P}_i^{(t)}$ satisfies the transmit power constraints for any $\hat{\mathbf{P}}_i^{(t)}$ by defining $\mathbf{P}_i^{(t)}=\kappa_i^{(t)} \mathbf{\hat{P}}_i^{(t)}$. The PSO-based optimal PA solution is defined by two variables: the position $\hat{\mathbf{P}}_i^{(t)} \in \mathbb{R}^{K \times K}$ and velocity $\mathbf{W}_i^{(t)} \in \mathbb{R}^{K \times K}$. Initially, the diagonal entries of $\hat{\mathbf{P}}_i^{(0)}$ are uniformly distributed over the range $[0,1]$, while the velocity is set as $\mathbf{W}_i^{(0)}=0$. During the iterations, the velocity and position matrices of $i^{th}$ particle are updated as follows: 
\begin{align}
\mbox{$\squeezespaces{0.1} \mathbf{W}_i^{(t+1)}\hspace{-0.95ex}=\gamma_1\hspace{-0.25ex} \mathbf{Y}_1^{(t)}\hspace{-0.5ex}(\hspace{-0.35ex}\hat{\mathbf{P}}_{\mathrm{best}}^{(t)}\hspace{-0.95ex}-\hat{\mathbf{P}}_i^{(t)}\hspace{-0.5ex})\hspace{-0.5ex}+\hspace{-0.5ex}\gamma_2\hspace{-0.25ex} \mathbf{Y}_2^{(t)}\hspace{-0.5ex}(\hspace{-0.35ex}\hat{\mathbf{P}}_{\mathrm{best},i}^{(t)}\hspace{-0.95ex}-\hat{\mathbf{P}}_i^{(t)}\hspace{-0.35ex})\hspace{-0.2ex}+\hspace{-0.25ex}\gamma_3^{(t)\hspace{-0.5ex}} \mathbf{W}_i^{(t)}\hspace{-0.95ex},$} \label{eq:velocity_PA}  \vspace{-2ex}
\end{align}
\begin{equation}
\mbox{$\squeezespaces{0.1} \hat{\mathbf{P}}_i^{(t+1)}=\operatorname{clip}(\hat{\mathbf{P}}_i^{(t)}+\operatorname{clip}(\mathbf{W}_i^{(t+1)},\left[w_{\text {min }}, w_{\text {max }}\right]),[0,1])$}, \label{eq:position_PA} 
\vspace{-2ex}
\end{equation}
where $\gamma_1$ and $\gamma_2$ are the learning parameters for the global best $\mathbf{P}_{b e s t}^{(t)}$ and the personal best $\mathbf{P}_{\mathrm{best},i}^{(t)}, \gamma_3^{(t)}=\mu-\frac{t}{T} (\mu_u - \mu_l)$ is the inertia parameters with the upper bound $\mu_u$ and lower bound $\mu_l$ for decreasing the velocity as the number of iterations increases, $\mathbf{Y}_1^{(t)}$ and $\mathbf{Y}_2^{(t)}$ are the random diagonal matrices with the uniformly distributed entries over $[0,1], w_{\min }$ and $w_{\max }$ denote the minimum and maximum acceptable velocity for the particles, respectively. Here, clip $(x,[a, b])=\max (a, \min (x, b))$ is used to prevent exceeding the maximum/minimum acceptable velocity and normalized power. Then, at $t^{th}$ iteration, the personal best of the $i^{th}$ particle and the global best over $M_p$ particles are respectively obtained as follows:
\begin{equation}
	\hat{\mathbf{P}}_{\mathrm{best},i}^{(t)} \hspace{-2ex}= \hspace{-2.5ex}\argmax_{\hat{\mathbf{P}}_{i}^{(t^*)}, \forall t^* = 0,1,\cdots, t\hspace{-2ex}} \hspace{-2.5ex}R_T\hspace{-0.4ex}(\hspace{-0.4ex}\mathbf{F}_b\hspace{-0.4ex},\hspace{-0.4ex} \mathbf{B}_b\hspace{-0.4ex}, \hspace{-0.4ex}\mathbf{F}_{u,t}\hspace{-0.4ex},\hspace{-0.4ex} \mathbf{B}_{u,t}\hspace{-0.4ex},\hspace{-0.4ex} \mathbf{F}_{u,r}\hspace{-0.4ex}, \hspace{-0.4ex}\mathbf{B}_{u,r}\hspace{-0.25ex},\hspace{-0.4ex} \kappa_{i}^{(t^*\hspace{-0.3ex})\hspace{-0.4ex}}\hat{\mathbf{P}}_{i}^{(t^*)\hspace{-0.5ex}}\hspace{-0.25ex},\hspace{-0.25ex} \mathbf{x}_u\hspace{-0.4ex}), \label{eq: personal_best} \vspace{-1ex}
	\end{equation}
\begin{equation} 
	\hat{\mathbf{P}}_{best}^{(t)} \hspace{-2ex}= \hspace{-2.5ex}\argmax_{\hat{\mathbf{P}}_{\mathrm{best},i}^{(t)}, \forall i = 0,1,\cdots, M_p\hspace{-2ex}} \hspace{-2.5ex}R_T\hspace{-0.4ex}(\hspace{-0.4ex}\mathbf{F}_b\hspace{-0.4ex},\hspace{-0.4ex} \mathbf{B}_b\hspace{-0.4ex}, \hspace{-0.4ex}\mathbf{F}_{u,t}\hspace{-0.4ex},\hspace{-0.4ex} \mathbf{B}_{u,t}\hspace{-0.4ex},\hspace{-0.4ex} \mathbf{F}_{u,r}\hspace{-0.4ex}, \hspace{-0.4ex}\mathbf{B}_{u,r}\hspace{-0.4ex},\hspace{-0.4ex} \kappa_{\mathrm{best},i}^{(t)\hspace{-0.4ex}}\hspace{-0.4ex}\hat{\mathbf{P}}_{\mathrm{best},i}^{(t)\hspace{-0.4ex}}\hspace{-0.4ex},\hspace{-0.4ex} \mathbf{x}_u\hspace{-0.6ex}), \label{eq:global_best}  \raisetag{1\baselineskip}	\vspace{-1ex}
\end{equation}
Finally, the multi-user PA matrix can be derived as follows:
\begin{equation}
\mathbf{P}=\kappa_{\mathrm{best}}^{(T)} \hat{\mathbf{P}}_{\mathrm{best}}^{(T)},
\end{equation}
where $\kappa_{\mathrm{best}}^{(t)}$ can be calculated by substituting $\hat{\mathbf{P}}_{\mathrm{best}}^{(T)}$ into (\ref{eq:kappa}). The summary of the proposed PSO-based optimal PA for fixed UAV location with HBF is outlined in Algorithm 1.
\setlength{\textfloatsep}{0pt}% Remove \textfloatsep
\begin{algorithm}[t!]\label{algo:1}
	\nonl \textbf{Input}: \hspace{-0.5ex} $M_p, T$, $(\theta, \phi)$, ($x_1$,$y_1$,$z_1$), ($x_u$,$y_u$,$z_u$). \\
	\nonl \textbf{Output}: \hspace{-0.5ex}$\mathbf{P}$, $\mathbf{F}_{b}, \mathbf{B}_{b}$, $\mathbf{F}_{u,t}, \mathbf{B}_{u,t}$, $\mathbf{F}_{u,r}, \mathbf{B}_{u,r}$. \\
	\SetAlgoLined 
	Formulate BS RF and BB stages using (\ref{eq:eq_TX_RF}), (\ref{eq:BB_BS}) \\
\squeezespaces{0.1}	Formulate UAV receive HBF stages using (\ref{eq:eq_RX_RF}), (\ref{eq:BB_1st_link}) \\
  	Construct UAV transmit HBF stages via (\ref{eq:eq_TX_RF_UAV}), (\ref{eq:BB_precoder}) \\
	\For{$i = 1:M_p$}{
		Initialize the velocity as $\mathbf{W}_i^{(0)} = \bf{0}$. \\
		Each entry of $\hat{\mathbf{P}}_{i}^{(0)}$ \hspace{-1ex}is uniformly distributed in $[0,1]$.\\
		Set the personal best $\hat{\mathbf{P}}_{\mathrm{best},i}^{(0)} = \hat{\mathbf{P}}_{i}^{(0)}$.
	}
	Find the global best $\hat{\mathbf{P}}_{\mathrm{best}}^{(0)}$ as in (\ref{eq:global_best}).\\
	\For {$t = 1:T$}{
		\For{$i = 1:M_p$}
		{
			Update the velocity $\mathbf{W}_i^{(t)}$ as in (\ref{eq:velocity_PA}).\\
			Update the position $\hat{\mathbf{P}}_{i}^{(t)}$ as in (\ref{eq:position_PA}).\\
			Find the personal best $\hat{\mathbf{P}}_{\mathrm{best},i}^{(t)}$ as in (\ref{eq: personal_best}).}
		Find the global best $\hat{\mathbf{P}}_{\mathrm{best}}^{(t)}$ as in (\ref{eq:global_best}).
	}
$\mathbf{P} = \kappa_{\mathrm{best}}^{(T)} \hat{\mathbf{P}}_{\mathrm{best}}^{(T)}$ \\
\caption{Proposed Joint HBF and Optimal PA for Fixed UAV Location (J-HBF-PSOPA-FL) Algorithm}
\end{algorithm}
\subsubsection{UAV Positioning for Equal PA}
In this problem, we optimize UAV location $\mathbf{x}_u = [x_u, y_u]^T$ in a given deployment span for fixed MU PA matrix $\mathbf{P}$, which is defined as:
\begin{equation}
	\mathbf{P} = \varepsilon \mathbf{I}_K \in \mathbb{C}^{K \times K}, \label{eq:EQ-PA}
\end{equation} 
where $\varepsilon$ is the normalization factor used to satisfy the power constraints $C_4$, $C_5$, and can be written as follows:
\begin{equation}
	\varepsilon = \sqrt{\frac{P_T}{\tr \EX \{ \hat{\mathbf{d}} \mathbfcal{H}_2^{H} \mathbf{T}^{H} \mathbf{F}_{u,t}^{H} \mathbf{F}_{u,t} \mathbf{T} \mathbfcal{H}_2\}}}, \label{eq:EQ_PA_constraint} 
\end{equation}  
where $\mathbf{T} = (\mathbfcal{H}_2^H \mathbfcal{H}_2 +\alpha N_{{RF}_u}\mathbf{I}_{N_{{RF}_u}})^{-1}$. Since the number of feasible UAV positions required to search space is $\big(\frac{x_{\mathrm{max}} - x_{\mathrm{min}}}{\Delta_x} \big) \times \big(\frac{y_{\mathrm{max}} - y_{\mathrm{min}}}{\Delta_y}\big)$, where $\Delta_x$ and $\Delta_y$ represents the search space resolution, hence, we propose a PSO-based algorithmic solution to find optimal UAV position while maximizing $R_T$. Following a similar approach as in Section III-C1, the $i^{th}$ particle at the $t^{th}$ iteration now represents an instance of the UAV location as:
\begin{equation}
	\mathbf{X}_i^{(t)} = [x_i^{(t)}, y_i^{(t)}]^T \in \mathbb{R}^{2}. \label{eq:Location_PSOL} 
\end{equation}
Here, the corresponding particle $i$ represents the candidate UAV position and calculates the objective function as $R_{T}(\mathbf{F}_b, \mathbf{B}_b, \mathbf{F}_{u,t}, \mathbf{B}_{u,t}, \mathbf{F}_{u,r}, \mathbf{B}_{u,r}, \mathbf{P}, \mathbf{X}_i^{(t)})$. Then, the position $\mathbf{X}_i^{(t)}$ and velocity $\mathbf{V}_i^{(t)}$ are updated as follows:
\begin{equation}
\mathbf{X}_i^{(t+1)}= \mathbf{X}_i^{(t)} + \mathbf{V}_i^{(t+1)}, \label{eq:position_PSOL} 
\end{equation}
\begin{equation}
\mbox{$\squeezespaces{0.1} \mathbf{V}_i^{(t+1)}\hspace{-0.95ex}=\gamma_1\hspace{-0.25ex} \mathbf{Y}_1^{(t)}\hspace{-0.75ex}(\hspace{-0.35ex}\hat{\mathbf{X}}_{\mathrm{best}}^{(t)}\hspace{-0.95ex}-\hat{\mathbf{X}}_i^{(t)}\hspace{-0.5ex})\hspace{-0.5ex}+\hspace{-0.5ex}\gamma_2\hspace{-0.25ex} \mathbf{Y}_2^{(t)}\hspace{-0.5ex}(\hspace{-0.35ex}\hat{\mathbf{X}}_{\mathrm{best},i}^{(t)}\hspace{-0.95ex}-\hat{\mathbf{X}}_i^{(t)}\hspace{-0.35ex})\hspace{-0.35ex}+\hspace{-0.5ex}\gamma_3^{(t)} \mathbf{V}_i^{(t)}\hspace{-0.35ex}$} \label{eq:velocity_PSOL} \vspace{-1ex} 
\end{equation}
Then, the personal and global best solutions for $i^{th}$ particle during $t^{th}$ iteration are obtained as follows: 
\begin{equation}
	\mathbf{X}_{\mathrm{best},i}^{(t)} \hspace{-2ex}= \hspace{-2.5ex}\argmax_{\mathbf{X}_{i}^{(t^*)}, \forall t^* = 0,1,\cdots, t\hspace{-2ex}} \hspace{-2.5ex}R_T\hspace{-0.4ex}(\hspace{-0.4ex}\mathbf{F}_b\hspace{-0.4ex},\hspace{-0.4ex} \mathbf{B}_b\hspace{-0.4ex}, \hspace{-0.4ex}\mathbf{F}_{u,t}\hspace{-0.4ex},\hspace{-0.4ex} \mathbf{B}_{u,t}\hspace{-0.4ex},\hspace{-0.4ex} \mathbf{F}_{u,r}\hspace{-0.4ex}, \hspace{-0.4ex}\mathbf{B}_{u,r}\hspace{-0.4ex},\hspace{-0.4ex} \mathbf{P}\hspace{-0.4ex},\hspace{-0.4ex} \mathbf{X}_i^{(t^*)}\hspace{-0.6ex}), \label{eq: personal_best_PSOL}  
\end{equation}
\begin{equation}
	\mathbf{X}_{best}^{(t)} \hspace{-1ex}= \hspace{-2.5ex}\argmax_{\mathbf{X}_{\mathrm{best},i}^{(t)}, \forall i = 0,1,\cdots, M_p\hspace{-2ex}} \hspace{-2.5ex}R_T\hspace{-0.4ex}(\hspace{-0.4ex}\mathbf{F}_b\hspace{-0.25ex},\hspace{-0.2ex} \mathbf{B}_b\hspace{-0.25ex}, \hspace{-0.25ex}\mathbf{F}_{u,t}\hspace{-0.25ex},\hspace{-0.25ex} \mathbf{B}_{u,t}\hspace{-0.2ex},\hspace{-0.25ex} \mathbf{F}_{u,r}\hspace{-0.25ex}, \hspace{-0.25ex}\mathbf{B}_{u,r}\hspace{-0.25ex},\hspace{-0.25ex} \hspace{-0.25ex}\mathbf{P}\hspace{-0.25ex},\hspace{-0.25ex} \mathbf{X}_{\mathrm{best},i}^{(t)\hspace{-0.25ex}}\hspace{-0.2ex}), \label{eq:global_best_PSOL}  \vspace{-2ex}
\end{equation}
After $T$ iterations, we update $\mathbf{x}_o$ $=$ $\mathbf{X}_{\mathrm{best}}^{(T)}$. The pseudo-code is given in Algorithm 2, which summarizes the proposed PSO-based UAV positioning for equal PA with HBF. 
\begin{algorithm}[t!]\label{algo:2}
	\nonl \textbf{Input}: \hspace{-0.5ex} $M_p, T$, $(\theta, \phi)$, ($x_1$,$y_1$,$z_1$), ($x_u$,$y_u$,$z_u$). \\
	\nonl \textbf{Output}: \hspace{-0.5ex}$ \mathbf{x}_o, \mathbf{P}$, $\mathbf{F}_{b}, \mathbf{B}_{b}$, $\mathbf{F}_{u,t}, \mathbf{B}_{u,t}$, $\mathbf{F}_{u,r}, \mathbf{B}_{u,r}$. \\
	\SetAlgoLined 
	Formulate BS RF and BB stages using (\ref{eq:eq_TX_RF}), (\ref{eq:BB_BS}) \\
	\squeezespaces{0.1}	Formulate UAV receive HBF stages using (\ref{eq:eq_RX_RF}), (\ref{eq:BB_1st_link}) \\
	Construct UAV transmit HBF stages via (\ref{eq:eq_TX_RF_UAV}), (\ref{eq:BB_precoder}) \\
	\For{$i = 1:M_p$}{
		Initialize the velocity as $\mathbf{V}_i^{(0)} = \bf{0}$. \\
		Each entry of $\mathbf{X}_{i}^{(0)}$ \hspace{-1ex}is uniformly distributed in $[0,1]$.\\
		Set the personal best $\mathbf{X}_{\mathrm{best},i}^{(0)} = \mathbf{X}_{i}^{(0)}$.
	}
	Find the global best $\mathbf{X}_{\mathrm{best}}^{(0)}$ as in (\ref{eq:global_best_PSOL}).\\
	\For {$t = 1:T$}{
		\For{$i = 1:M_p$}
		{
			Update the velocity $\mathbf{V}_i^{(t)}$ as in (\ref{eq:velocity_PSOL}).\\
			Update the position $\mathbf{X}_{i}^{(t)}$ as in (\ref{eq:position_PSOL}).\\
			Find the personal best $\mathbf{X}_{\mathrm{best},i}^{(t)}$ as in (\ref{eq: personal_best_PSOL}).}
		Find the global best $\mathbf{X}_{\mathrm{best}}^{(t)}$ as in (\ref{eq:global_best_PSOL}).
	}
$\mathbf{x}_o = \mathbf{X}_{\mathrm{best}}^{(T)}$ \\
Update $\mathbf{B}_b, \mathbf{B}_{u,t}, \mathbf{B}_{u,r}$ for $\mathbf{x}_o$.
	\caption{Proposed Joint HBF and UAV Deployment for Equal PA (J-HBF-PSOL-EQPA) Algorithm}
\end{algorithm}
\subsubsection{Joint UAV Positioning and Optimal PA}
In this problem, we jointly optimize UAV location $\mathbf{x}_u = [x_u, y_u]^T$ and $\mathbf{P}$, which are given by (\ref{eq:P_PA}) and (\ref{eq:Location_PSOL}), respectively. To solve this non-convex optimization problem, we propose a PSO-based algorithmic solution to optimize $\mathbf{x}_u$ and $\mathbf{P}$ whilst maximizing the total achievable rate $R_T$. Here, the $i^{th}$ particle at the $t^{th}$ iteration now represents an instance of the UAV location and multi-user PA matrix, which is given as follows:
\begin{equation}
\mbox{$\squeezespaces{0.1}	\mathbf{J}_{p,i}^{(t)} = [\mathbf{X}_i^{(t)}, \hat{\mathbf{P}}_i^{(t)}]^T = [x_i^{(t)}, y_i^{(t)},\sqrt{\hat{p}_{1,i}^{(t)}}\hspace{-0.5ex},\cdots,\hspace{-0.5ex} \sqrt{\hat{p}_{K,i}^{(t)}}]^T \in \mathbb{R}^{K+2}$}, \label{eq:Joint_PSO}
\end{equation}
where each particle $i$ represents the candidate UAV position and PA to $K$ IoT users, and calculates the objective function as $R_{T}(\mathbf{F}_b, \mathbf{B}_b, \mathbf{F}_{u,t}, \mathbf{B}_{u,t}, \mathbf{F}_{u,r}, \mathbf{B}_{u,r}, \kappa_i^{(t)}\hat{\mathbf{P}}_i^{(t)}, \mathbf{X}_i^{(t)})$. Here, $\hat{\mathbf{P}}_i^{(t)}=\operatorname{diag}(\sqrt{\hat{p}_{1, i}^{(t)}}, \cdots, \sqrt{\hat{p}_{K, i}^{(t)}}) \in \mathbb{R}^{K \times K}$ is the normalized PA matrix with $\hat{p}_{k, i}^{(t)} \in [0,1]$ and similar to Section III-C1, the transmit power constraints for any $\hat{\mathbf{P}}_i^{(t)}$ are satisfied by defining $\mathbf{P}_i^{(t)}=\kappa_i^{(t)} \mathbf{\hat{P}}_i^{(t)}$. Then, the position $\mathbf{J}_{p,i}^{(t)}$ and velocity $\mathbf{J}_{v,i}^{(t)}$ for $i^{th}$ particle during $t^{th}$ iteration are updated as follows:
\begin{equation}
	\mathbf{J}_{p,i}^{(t+1)}= \mathbf{J}_{p,i}^{(t)} + \mathbf{J}_{v,i}^{(t+1)}, \label{eq:position_PSOLPA} 
\end{equation}
\begin{equation}
	\mbox{$\squeezespaces{0.1} \mathbf{J}_{v,i}^{(t+1)}\hspace{-0.95ex}=\gamma_1\hspace{-0.25ex} \mathbf{Y}_1^{(t)}\hspace{-0.75ex}(\hspace{-0.35ex}\mathbf{J}_{p,\mathrm{best}}^{(t)}\hspace{-0.95ex}-\mathbf{J}_{p,i}^{(t)}\hspace{-0.25ex})\hspace{-0.25ex}+\hspace{-0.5ex}\gamma_2\hspace{-0.25ex} \mathbf{Y}_2^{(t)}\hspace{-0.5ex}(\hspace{-0.35ex}\mathbf{J}_{p,\mathrm{best},i}^{(t)}\hspace{-0.95ex}-\mathbf{J}_{p,i}^{(t)}\hspace{-0.35ex})\hspace{-0.35ex}+\hspace{-0.25ex}\gamma_3^{(t)} \mathbf{J}_{v,i}^{(t)}\hspace{-0.25ex}$}, \label{eq:velocity_PSOLPA}  \vspace{-1ex}
\end{equation}
Finally, the personal and global best solutions for $i^{th}$ particle during $t^{th}$ iteration are obtained as follows: \vspace{-0.5ex}
\begin{equation}
	\mathbf{J}_{p,\mathrm{best},i}^{(t)} \hspace{-2ex}= \hspace{-2.5ex}\argmax_{\mathbf{J}_{p,i}^{(t^*)}, \forall t^* = 0,1,\cdots, t\hspace{-2ex}} \hspace{-2.5ex}R_T\hspace{-0.3ex}(\hspace{-0.4ex}\mathbf{F}_b\hspace{-0.4ex},\hspace{-0.4ex} \mathbf{B}_b\hspace{-0.4ex}, \hspace{-0.4ex}\mathbf{F}_{u,t}\hspace{-0.4ex},\hspace{-0.4ex} \mathbf{B}_{u,t}\hspace{-0.4ex},\hspace{-0.4ex} \mathbf{F}_{u,r}\hspace{-0.4ex}, \hspace{-0.4ex}\mathbf{B}_{u,r}\hspace{-0.4ex},\hspace{-0.25ex} \kappa_{i}^{(t^*\hspace{-0.25ex})\hspace{-0.25ex}}\hspace{-0.4ex}\hat{\mathbf{P}}_i^{(t^*\hspace{-0.25ex})\hspace{-0.25ex}}\hspace{-0.4ex},\hspace{-0.4ex} \mathbf{X}_i^{\hspace{-0.25ex}(t^*)}\hspace{-0.6ex}), \label{eq: personal_best_PSOLPA}  \vspace{-0.5ex}
\end{equation}
\begin{equation} 
	\mathbf{J}_{p,\mathrm{best}}^{(t)} \hspace{-1.5ex}= \hspace{-2ex}\argmax_{\mathbf{J}_{p,\mathrm{best},i}^{(t)}, \forall i = 0,1,\cdots, M_p\hspace{-4ex}} \hspace{-2.5ex}R_T\hspace{-0.5ex}(\hspace{-0.4ex}\mathbf{F}_b\hspace{-0.4ex},\hspace{-0.4ex} \mathbf{B}_b\hspace{-0.4ex}, \hspace{-0.4ex}\mathbf{F}_{u,t}\hspace{-0.4ex},\hspace{-0.4ex} \mathbf{B}_{u,t}\hspace{-0.4ex},\hspace{-0.4ex} \mathbf{F}_{u,r}\hspace{-0.4ex}, \hspace{-0.4ex}\mathbf{B}_{u,r}\hspace{-0.4ex},\hspace{-0.25ex} \hspace{-0.4ex}\kappa_{\mathrm{best},i}^{(t)\hspace{-0.5ex}}\hspace{-0.25ex}\hat{\mathbf{P}}_{\mathrm{best,i}}^{(t)\hspace{-0.4ex}}\hspace{-0.6ex},\hspace{-0.6ex} \mathbf{X}_i^{\hspace{-0.4ex}(t)\hspace{-0.5ex}}\hspace{-0.25ex}), \label{eq:global_best_PSOLPA} 	
\end{equation}
After $T$ iterations, we update $\mathbf{x}_o = \mathbf{X}_{\mathrm{best}}^{(T)}$ and $\mathbf{P}=\kappa_{\mathrm{best}}^{(T)} \hat{\mathbf{P}}_{\mathrm{best}}^{(T)}$. Algorithm 3 gives the pseudo-code of the proposed PSO-based joint UAV positioning and optimal PA with HBF. 
\begin{algorithm}[t!]\label{algo:3}
	\nonl \textbf{Input}: \hspace{-0.5ex} $M_p, T$, $(\theta, \phi)$, ($x_1$,$y_1$,$z_1$), ($x_u$,$y_u$,$z_u$). \\
	\nonl \textbf{Output}: \hspace{-0.5ex}$ \mathbf{x}_o, \mathbf{P}$, $\mathbf{F}_{b}, \mathbf{B}_{b}$, $\mathbf{F}_{u,t}, \mathbf{B}_{u,t}$, $\mathbf{F}_{u,r}, \mathbf{B}_{u,r}$. \\
	\SetAlgoLined 
	Formulate BS RF and BB stages using (\ref{eq:eq_TX_RF}), (\ref{eq:BB_BS}) \\
	\squeezespaces{0.1}	Formulate UAV receive HBF stages using (\ref{eq:eq_RX_RF}), (\ref{eq:BB_1st_link}) \\
	Construct UAV transmit HBF stages via (\ref{eq:eq_TX_RF_UAV}), (\ref{eq:BB_precoder}) \\
	\For{$i = 1:M_p$}{
		Initialize the velocity as $\mathbf{J}_{v,i}^{(0)} = \bf{0}$. \\
		Each entry of $\mathbf{J}_{p,i}^{(0)}$ \hspace{-1ex}is uniformly distributed in $[0,1]$.\\
		Set the personal best $\mathbf{J}_{p,\mathrm{best},i}^{(0)} = \mathbf{J}_{p,i}^{(0)}$.
	}
	Find the global best $\mathbf{J}_{p,\mathrm{best}}^{(0)}$ as in (\ref{eq:global_best_PSOLPA}).\\
	\For {$t = 1:T$}{
		\For{$i = 1:M_p$}
		{
			Update the velocity $\mathbf{J}_{v,i}^{(t)}$ as in (\ref{eq:velocity_PSOLPA}).\\
			Update the position $\mathbf{J}_{p,i}^{(t)}$ as in (\ref{eq:position_PSOLPA}).\\
			Find the personal best $\mathbf{J}_{p,\mathrm{best},i}^{(t)}$ as in (\ref{eq: personal_best_PSOLPA}).}
		Find the global best $\mathbf{J}_{p,\mathrm{best}}^{(t)}$ as in (\ref{eq:global_best_PSOLPA}).
	}
	$\mathbf{x}_o = \mathbf{X}_{\mathrm{best}}^{(T)}$, $\mathbf{P} = \kappa_{\mathrm{best}}^{(T)} \hat{\mathbf{P}}_{\mathrm{best}}^{(T)}$ \\
	Update $\mathbf{B}_b, \mathbf{B}_{u,t}, \mathbf{B}_{u,r}$ for $\mathbf{x}_o$.
	\caption{Proposed Joint HBF, UAV Location Optimization and PA (J-HBF-PSOLPA) Algorithm} 
\end{algorithm}
\vspace{-1ex}
\section{Low-Complexity DL-Based Joint Power Allocation and UAV Positioning}\label{sec:low-ps}
\begin{figure}[!t]
	\centering
	\captionsetup{justification=centering}
	\includegraphics[height= 4cm, width=1\columnwidth]{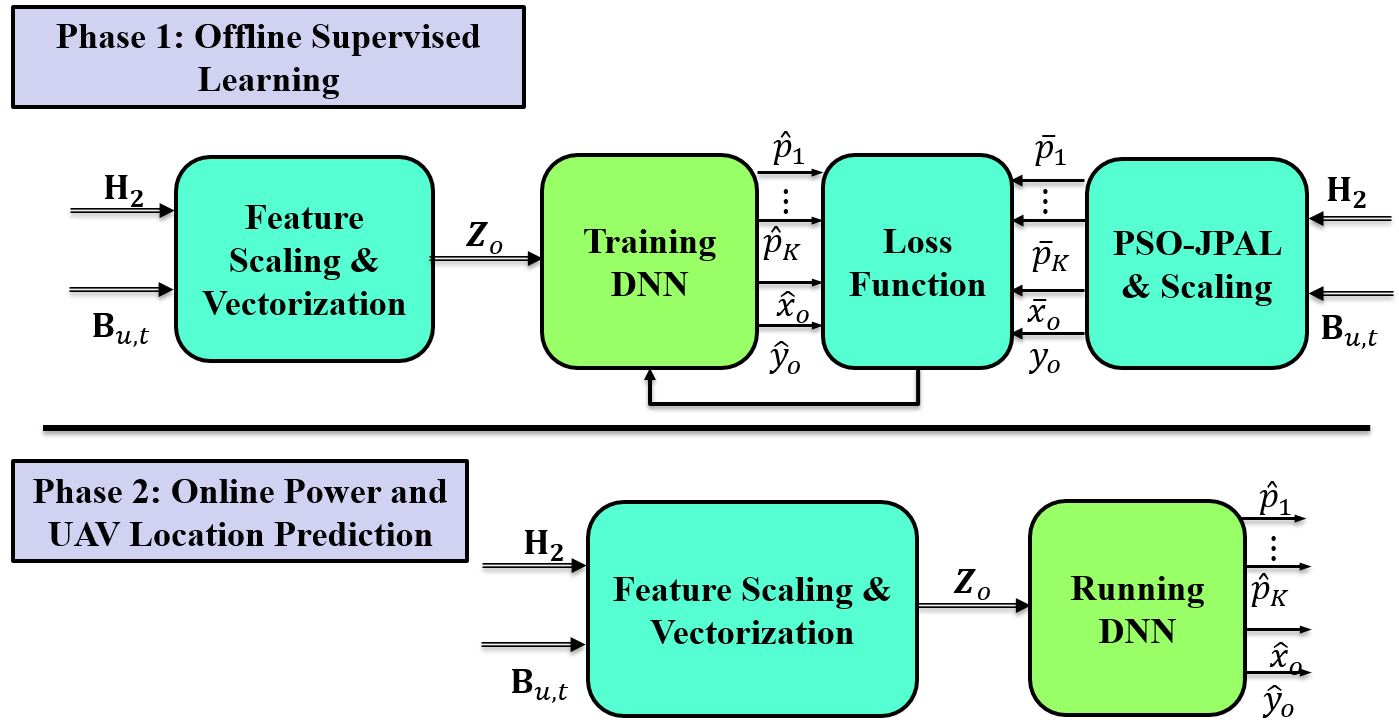}
		\vspace{-1em} 
	\caption{Block diagram of offline supervised learning and real-time prediction in J-HBF-DLLPA algorithm.}
	\label{fig:fig3}
\end{figure}

The proposed joint HBF, PA and UAV location schemes can achieve near-optimal capacity for a UAV-assisted MU-mMIMO IoT systems. Additionally, compared to the computationally expensive exhaustive search method, the proposed solutions offer higher computational efficiency. However, as the number of IoT users increases, the proposed PSO-based solutions require more iterations and longer run time, which may render them impractical for real-time online applications of UAV-assisted MU-MIMO IoT systems. To address this challenge, we propose a low-complexity DL-based algorithm, called J-HBF-DLLPA, which can achieve a near-optimal AR while maintaining a reasonable run time. The proposed algorithm has two phases, as illustrated in Fig. \ref{fig:fig3}: 1) Phase 1 applies the offline supervised learning via the optimal allocated power and UAV location values calculated by J-HBF-PSOLPA; and 2) Phase 2 runs the trained J-HBF-DLLPA algorithm for predicting the allocated powers and UAV location in the real-time online applications. Therefore, the remaining part of this section focuses on the deep neural network (DNN) architecture, loss functions, dataset preparation, and training process for the proposed low-complexity J-HBF-DLLPA algorithm.
\vspace{-1em}
\subsection{Proposed Deep Neural Network Architecture} 
We employ a fully-connected DNN architecture with four hidden layers as depicted in Fig. \ref{fig:fig4}, which aims to predict the optimal allocated powers for $K$ IoT users as well as UAV optimal location in a given deployment span. We consider $L_i$ neurons in each $i^{\text {th }}$ hidden layer (HL) with $i=1,\cdots,4$. As shown in Fig. \ref{fig:fig3}, the proposed J-HBF-DLLPA algorithm uses the effective channel matrix between UAV and $K$ IoT users $\mathbfcal{H}_2$ $\in$ $\mathrm{C}^{K \times N_{{RF}_u}}$ given in (\ref{eq:effective_channel_2}), and the UAV transmit BB stage $\mathbf{B}_{u,t}$ $\in \mathbb{C}^{N_{{RF}_u} \times K}$ given in (\ref{eq:BB_precoder}) as inputs, which are first subject to feature scaling and vectorization operations. Subsequently, the input layer feature vector is derived as:
\begin{align}
	\mathbf{z}_0=\left[\begin{array}{c}
		\omega_1 \mathbf{z}_{\tilde{\mathbf{h}}_{2,1}} \\
		\vdots \\
		\omega_1 \mathbf{z}_{\tilde{\mathbf{h}}_{2,K}} \\
		\omega_2 \mathbf{z}_{\mathbf{b}_{u,t,1}} \\
		\vdots \\
		\omega_2 \mathbf{z}_{\mathbf{b}_{u,t,K}} \\
		\omega_3 \mathbf{z}_{\mathbf{BB}_{u,t}} \\
		\omega_4 \mathbf{z}_{\mathbf{{BB,inv,u,t}}}
	\end{array}\right] \in \mathbb{R}^{L_0}, \label{eq:input_feature_vector}
\end{align}
where $L_0=(2N_t +2 N_{{RF}_u} + 2)K$ is the input feature size, $\mathbf{z}_{\tilde{\mathbf{h}}_{2,k}} = [\operatorname{Re}(\tilde{\mathbf{h}}_{2,k}^T), \operatorname{Im}(\tilde{\mathbf{h}}_{2,k}^T)]^T \in \mathbb{R}^{2N_t}$, $\mathbf{z}_{\mathbf{b}_{u,t,k}} = [\operatorname{Re}(\mathbf{b}_{u,t,k}^T), \operatorname{Im}(\mathbf{b}_{u,t,k}^T)]^T \in \mathbb{R}^{2 N_{{RF}_u}}$, $\mathbf{z}_{\mathbf{BB}_{u,t}^T} = [\mathbf{b}_{u,t,1}^H \mathbf{b}_{u,t,1},\cdots,\mathbf{b}_{u,t,K}^H \mathbf{b}_{u,t,K}]^T \in \mathbb{R}^{K}$ and $\mathbf{x}_{\mathbf{BB,inv,u,t}^T} = [\frac{1}{\mathbf{b}_{u,t,1}^H \mathbf{b}_{u,t,1}},\cdots,\frac{1}{\mathbf{b}_{u,t,K}^H \mathbf{b}_{u,t,K}}]^T \in \mathbb{R}^{K}$ are respectively the non-scaled input feature vectors for the effective channel, UAV transmit BB stage, the gain of each BB precoder vector and its inverse. By implementing the maximum absolute scaling \cite{Asil_DL}, the corresponding scaling coefficients are calculated as:
\begin{align}
	\omega_1 & =\max \left(|\mathbf{z}_{\tilde{\mathbf{h}}_{2,1}}^T|, \cdots,|\mathbf{z}_{\tilde{\mathbf{h}}_{2,K}}^T|\right)^{-1} \\
\omega_2 & =\max \left(|\mathbf{z}_{\tilde{\mathbf{b}}_{u,t,1}}^T|, \cdots,|\mathbf{z}_{\tilde{\mathbf{b}}_{u,t,K}}^T|\right)^{-1} \\
	\omega_3 & =\max \left(\mathbf{b}_{u,t,1}^H \mathbf{b}_{u,t,1}^H, \cdots, \mathbf{b}_{u,t,K}^H \mathbf{b}_{u,t,K}^H \right)^{-1} \\
\squeezespaces{0.1}	\omega_4 & =\max \left(\mathbf{b}_{u,t,1}^H \mathbf{b}_{u,t,1}^H, \cdots, \mathbf{b}_{u,t,K}^H \mathbf{b}_{u,t,K}^H \right).
\end{align}
The proposed algorithm utilizes the maximum absolute scaling technique to scale the input feature vector between -1 and 1 (i.e., $\mathbf{z}_0 \in(-1,1])$, which prevents certain features from dominating the learning process. In the offline supervised learning process, the optimal power allocation and UAV location values are calculated as the output labels via the computationally expensive J-HBF-PSOLPA algorithm. Similar to the input features, we also apply the maximum absolute scaling to the optimal allocated powers and UAV location as follows:
\begin{align}
	\bar{x}_o&=\frac{x_o}{\max(x_o, y_o)} \in[0,1] \label{eq:UAV_normalized_x} \\
	\bar{y}_o&=\frac{y_o}{\max(x_o, y_o)} \in[0,1] \label{eq:UAV_normalized_y} 
	\end{align}
	\begin{align}
\bar{p}_k&=\frac{p_k^{\mathrm{opt}}}{\max(p_1^{\mathrm{opt}}, \cdots, p_K^{\mathrm{opt}})} \in[0,1]. \label{eq:UAV_power}
\end{align}
To perform non-linear operations, we adopt the rectified linear unit (ReLU) as the activation function in the hidden layers (i.e., $f_r(z)=\max (0, z)$. Therefore, using the input feature vector of $\mathbf{z}_0$ given in (\ref{eq:input_feature_vector}), the output of $i^{th}$ hidden layer is computed as $\mathbf{z}_i=f_r(\mathbf{U}_{i-1} \mathbf{z}_{i-1}+\mathbf{b}_{i-1}) \in \mathbb{R}^{L_i}$, where $\mathbf{U}_{i-1} \in \mathbb{R}^{L_i \times L_{i-1}}$ and $\mathbf{b}_{i-1} \in \mathbb{R}^{L_i}$ are the weight matrix and bias vector, respectively. To ensure that the predicted output values are between $0$ and $1$, we apply the sigmoid function at the output layer (i.e., $f_\sigma(z)=\frac{1}{1+e^{-\infty}}$). Thus, the predicted power and UAV location values via the DNN architecture are written as follows:
\begin{figure}[!t]
	\centering
	\captionsetup{justification=centering}
	\includegraphics[height= 4cm, width=1\columnwidth]{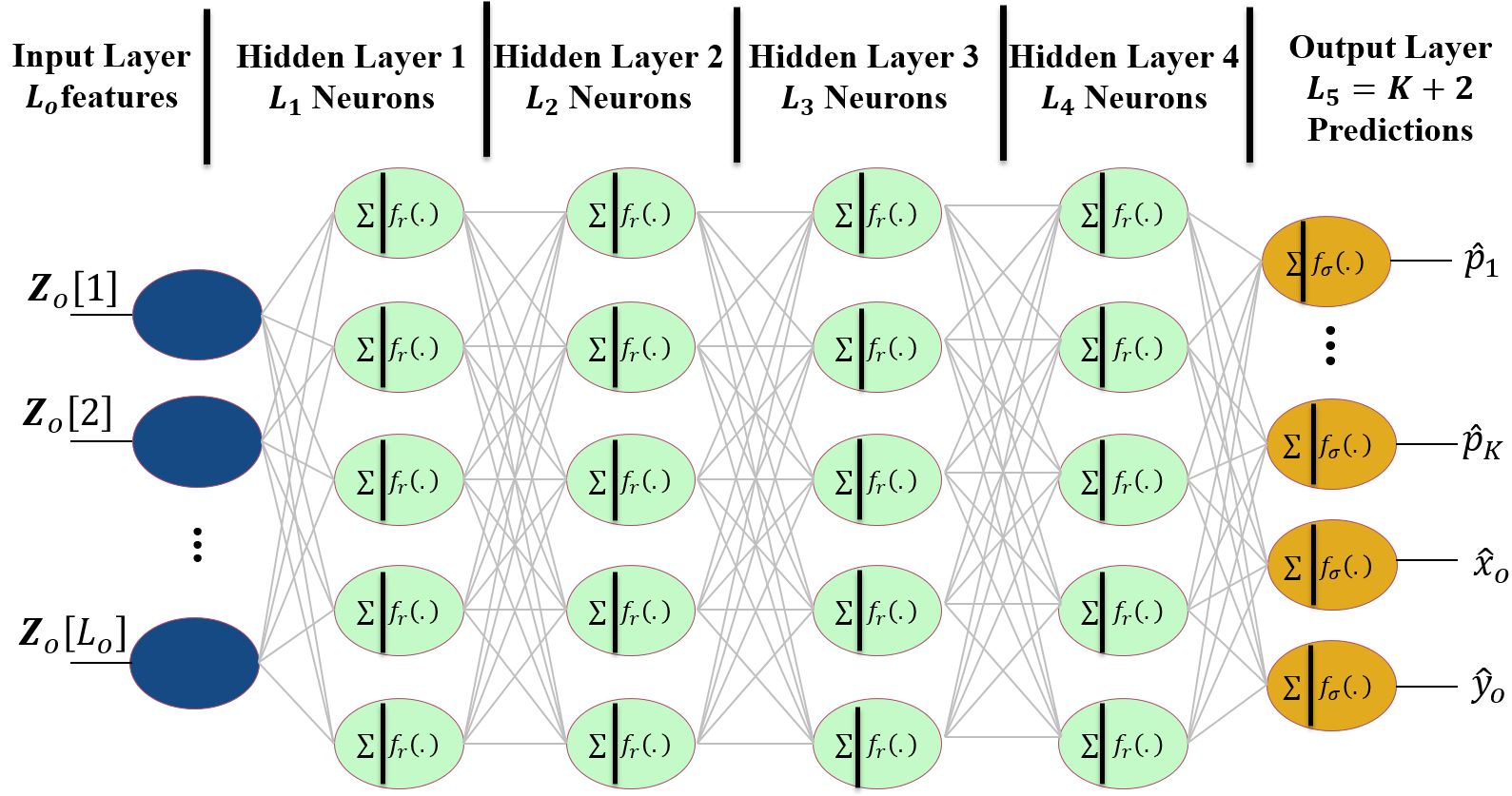} 
	\vspace{-2ex}
	\caption{Deep neural network architecture for J-HBF-DLLPA algorithm.}
	\label{fig:fig4}
\end{figure} 
\begin{align}
	& {\left[\hat{p}_1, \hat{p}_2, \cdots, \hat{p}_K, \hat{x}_o, \hat{y}_o\right]} \notag \\
	& =f_\sigma(\mathbf{U}_4 \mathbf{z}_4+\mathbf{b}_4) \notag \\
	& \mbox{$\squeezespaces{0.1}=f_\sigma(\mathbf{U}_4 f_r(\mathbf{U}_3 f_r(\hspace{-0.25ex}\mathbf{U}_2 f_r(\hspace{-0.25ex}\mathbf{U}_1 f_r(\hspace{-0.25ex}\mathbf{U}_0 \mathbf{x}_0+\mathbf{b}_0\hspace{-0.25ex})+\mathbf{b}_1\hspace{-0.25ex})+\mathbf{b}_2\hspace{-0.25ex})+\mathbf{b}_3)+\mathbf{b}_4\hspace{-0.25ex}).$} \raisetag{1.7\baselineskip}
\end{align}
\subsection{Loss Functions}
We here consider two loss functions by using the predicted and optimal power values: 1) mean square error (MSE); and 2) mean absolute error (MAE). When there are $S_r$ network realizations in the dataset, the MSE loss function is given by:
\begin{equation}
\mathcal{L}_{\mathrm{MSE}}\hspace{-0.1ex}=\hspace{-0.1ex}\frac{1}{S_r K}\hspace{-0.65ex} \sum_{i=1}^{S_r} \hspace{-0.35ex}\sum_{k=1}^K\hspace{-0.35ex}(\hspace{-0.35ex}\bar{p}_{k, i}\hspace{-0.35ex}-\hspace{-0.35ex}\hat{p}_{k, i})^2 \hspace{-0.5ex} + \hspace{-0.5ex}\frac{1}{S_r} \hspace{-0.5ex}\sum_{i=1}^{S_r} (\hspace{-0.35ex}\bar{x}_{o}\hspace{-0.35ex}-\hspace{-0.35ex}\hat{x}_{o})^2\hspace{-0.5ex} +\hspace{-0.5ex} \frac{1}{S_r} \hspace{-0.5ex}\sum_{i=1}^{S_r}\hspace{-0.35ex} (\hspace{-0.35ex}\bar{y}_{o}\hspace{-0.35ex}-\hspace{-0.35ex}\hat{y}_{o})^2.  \label{eq:loss_function_MSE}
\end{equation}
Similarly, the MAE loss function is written as:
\begin{equation}
\mathcal{L}_{\mathrm{MAE}}\hspace{-0.1ex}=\hspace{-0.1ex}\frac{1}{S_r K}\hspace{-0.65ex} \sum_{i=1}^{S_r} \hspace{-0.35ex}\sum_{k=1}^K\hspace{-0.15ex}|\hspace{-0.15ex}\bar{p}_{k, i}\hspace{-0.35ex}-\hspace{-0.35ex}\hat{p}_{k, i}| \hspace{-0.5ex} + \hspace{-0.5ex}\frac{1}{S_r} \hspace{-0.5ex}\sum_{i=1}^{S_r} |\hspace{-0.35ex}\bar{x}_{o}\hspace{-0.35ex}-\hspace{-0.35ex}\hat{x}_{o}|\hspace{-0.5ex} +\hspace{-0.5ex} \frac{1}{S_r} \hspace{-0.5ex}\sum_{i=1}^{S_r}\hspace{-0.35ex} |\hspace{-0.35ex}\bar{y}_{o}\hspace{-0.35ex}-\hspace{-0.35ex}\hat{y}_{o}|.  \label{eq:loss_function_MAE}
\end{equation}
The weight matrices $\mathbf{U}_i$ and bias vectors $\mathbf{b}_i$ of the DNN architecture are updated by back-propagating the gradients of the loss function from the output layer to the input layer. This helps in reducing the loss and accurately predicting the optimal allocated power and UAV location values, and thus, maximizing the total AR as expressed in (\ref{eq:optimization_reformulated}).
\vspace{-1em}
\subsection{Data Generation \& Training Process}
We generated a dataset of $S_r=100.000=10^5$ network realizations to train the proposed DNN architecture, as shown in Fig. \ref{fig:fig3}. For each realization, we randomly varied the path gains, AoD parameters, and UE location to generate the channel vector for each UE as given in (\ref{eq:second_channel}). The corresponding optimal allocated powers and UAV location are calculated via the J-HBF-PSOLPA algorithm (Section III-C3) and stored in the dataset. The total available dataset was split into 80$\%$ for training and 20$\%$ for validation for the offline supervised learning process (i.e., Phase $1$). After completing the offline supervised learning process, the online power allocation and UAV location (i.e., Phase 2) is tested with a purely new test dataset. The DNN architecture for the proposed J-HBF-DLLPA algorithm is implemented using the open-source DL libraries in TensorFlow \cite{Tensor_flow}.
\vspace{-1ex}
\section{Illustrative Results}
In this section, we present the Monte-Carlo simulation results based on the proposed algorithmic solutions. Table II outlines the simulation setup based on the 3D micro-cell scenario \cite{koc2020Access} for the results discussed hereafter. Moreover, for PSO-based results, we use $M_p = 20$, $\gamma_1 = \gamma_2 = 2$ and $\gamma_3 = 1.1$. 
 \begin{table}[!t]
	\centering
	\caption{Simulation Parameters \cite{koc2020Access}\vspace{-2ex}}
	\resizebox{\columnwidth}{!}{
		\begin{tabular}{|c|c|c|c|}
			\hline
			\multicolumn{2}{|c|}{Number of antennas}	& \multicolumn{2}{c|}{$(N_T, N_t, N_r) = 144$} \\ \hline 	
			Number of paths & Path loss exponent & $L = 10$ &  $3.6$ \\ \hline 	
			Frequency &  Channel Bandwidth & 28 GHz & 100 MHz  \\ \hline
			Noise PSD & Reference Path Loss $\alpha$ & $-174$ dBm/Hz & 61.34 dB \\ \hline 	
			BS/Gateway height &  UAV height & 10 m & 20 m  \\ \hline			
			User groups & \# of users per group & $G = 1$ or $2$ & $K_g = \frac{K}{G}$ \\ \hline 	
			UAV x-axis range & UAV y-axis range & $[x_{min}, x_{max}] = \left[0,100\right] m$ & $[y_{min}, y_{max}] = \left[0,100\right] m$ \\ \hline
			Mean AAoD/AAoA ($1^{st}$ link) & Mean AAoD ($2^{nd}$ link) & $120^{\circ}$  & $\phi_{g} = 21^{\circ} + 120^{\circ} (g-1)$  \\ \hline
			Mean EAoD/EAoA ($1^{st}$ link) & Mean EAoD ($2^{nd}$ link) & $60^{\circ}$  & $\theta_{g} = 60^{\circ}$  \\ \hline  	
			Azimuth/Elevation Angle Spread & \# of network realizations & $\pm 10^{\circ}$ & 2000 \\ \hline	
			Minimum horizontal distance $\tau_{h,\mathrm{min}}$ & Maximum horizontal distance $\tau_{h,\mathrm{min}}$ & $0$ & $100$ [m]\\ \hline	
		\end{tabular} 
	} 	\label{tab:tab1} \vspace{-2em}
\end{table}
\begin{figure}[!t] 
	% 	\centering
	\subfloat[\label{fig:fig5a}]{% 
		\includegraphics[width =0.485\columnwidth,height=4cm]{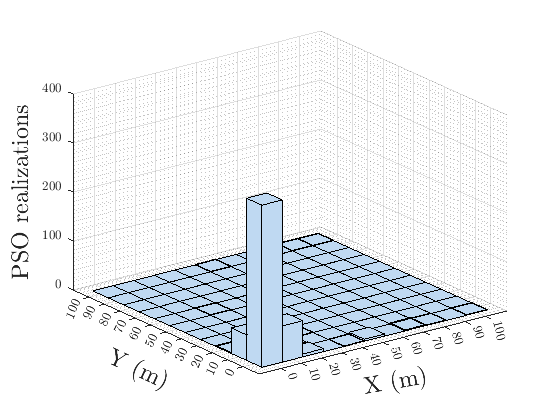} 
	} \hfil 
	\subfloat[\label{fig:fig5b}]{% 
		\includegraphics[width =0.485\columnwidth, height= 4cm]{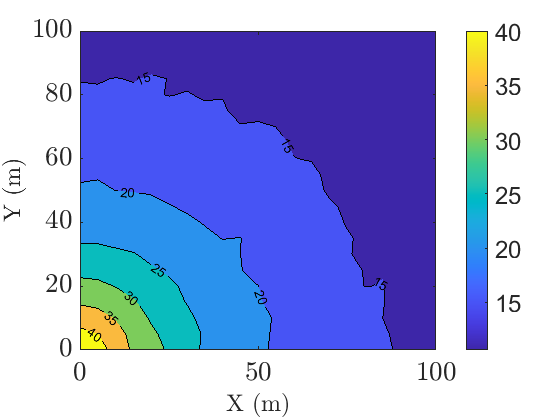} 
	} \vspace{-2ex}
	\caption{Comparison of AR $\mathrm{R}_1$. (a) Number of realizations for best UAV location using J-HBF-PSOL-EQPA. (b) Contour plot using exhaustive search.}
	\label{fig:fig5}
\end{figure}
\begin{figure}[!t]
	\vspace{-2.5ex}
	\centering
	\captionsetup{justification=centering}	
	\includegraphics[scale = 0.5]{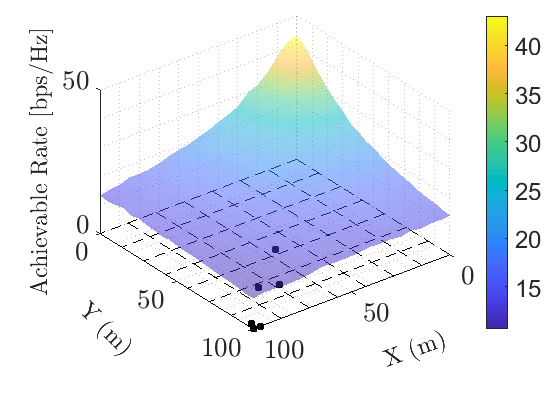} 
 \vspace{-3ex}
	\caption{Achievable rate $\mathrm{R}_1$ vs. $(x-y)$-coordinates at $P_T = 40$ dBm.}
	\label{fig:fig6}
\end{figure}
\vspace{-1ex}
\subsection{Spectral Efficiency}
In this section, we compare the spectral efficiency of the proposed PSO-based algorithmic solutions for a UAV-assisted MU-mMIMO IoT systems. To achieve this, we first analyze the rate of each link individually for a fixed UAV location and transmit power $P_T = 20$ dBm. Specifically, we consider that the BS is located at $(x_b,y_b,z_b) = (0,0,10)$, the UAV is deployed at fixed location $(x_u,y_u,z_u) = (50,50,20)$ and the IoT users are randomly distributed and located at a far distance from BS ($x_k, y_k$) $\in$ $[50,100]$. We then compare the optimal UAV location for maximizing the capacity of the first link using two schemes: 1) exhaustive search; and 2) the proposed PSO-based UAV deployment and equal PA (J-HBF-PSOL-EQPA) over a given deployment span of $[(x_{\mathrm{min}}, y_{\mathrm{min}}), (x_{\mathrm{max}}, y_{\mathrm{max}})] = [0,100]$. The results show that the optimal location for the UAV to maximize the capacity of the first link is close to the BS, as demonstrated in Fig. \ref{fig:fig5}(b). Furthermore, the proposed J-HBF-PSOL-EQPA can find the global optimal solution for almost 95$\%$ of the realizations, as shown in \ref{fig:fig5}(a). Fig. \ref{fig:fig6} plots the rate of first link $R_1$ using J-HBF-PSOL-EQPA versus different 2-D UAV locations, which shows similar performance to exhaustive search solution presented in Fig. \ref{fig:fig5}(b).\par 
\begin{figure}[!t] 
	% 	\centering
	\subfloat[\label{fig:fig7a}]{% 
		\includegraphics[width =0.485\columnwidth,height=4cm]{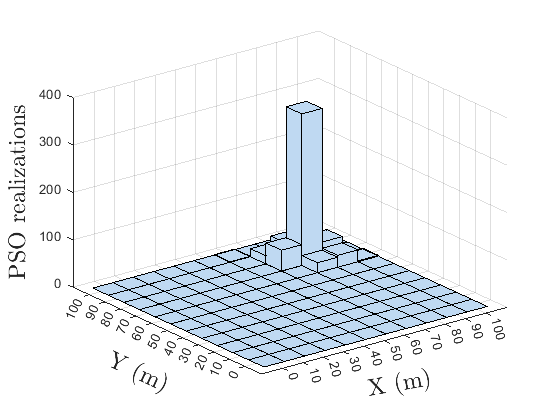} 
	} \hfil 
	\subfloat[\label{fig:fig7b}]{% 
		\includegraphics[width =0.485\columnwidth, height= 4cm]{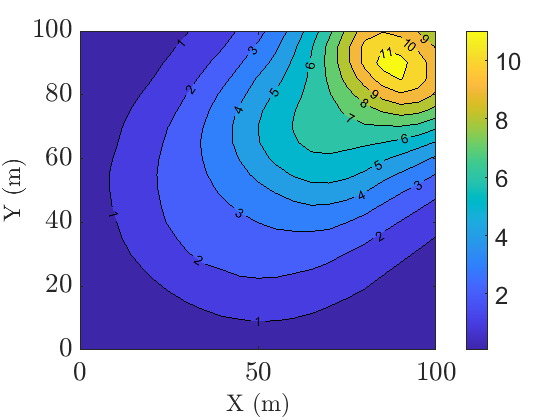} 
	} \vspace{-2ex}
	\caption{Comparison of AR $\mathrm{R}_2$. (a) Number of realizations for best UAV location using J-HBF-PSOL-EQPA. (b) Contour plot using exhaustive search.}
	\label{fig:fig7}
\end{figure}
\begin{figure}[!t]
	\centering
	\captionsetup{justification=centering}	
	\includegraphics[scale = 0.5]{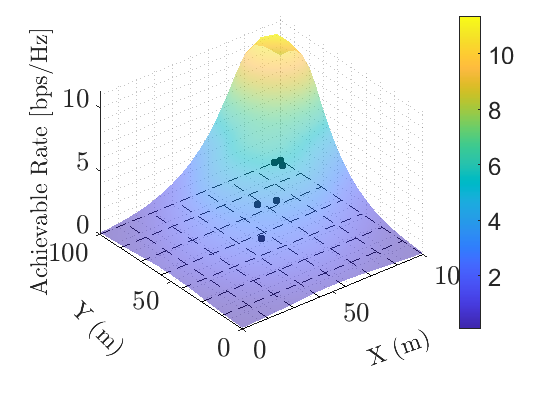} 
 \vspace{-3ex}
	\caption{Achievable rate $\mathrm{R}_2$ vs. $(x-y)$-coordinates at $P_T = 40$ dBm.}
	\label{fig:fig8}
\end{figure}
\begin{figure}[!t]
	\centering
	\captionsetup{justification=centering}	
	\includegraphics[height=6cm, width=1\columnwidth]{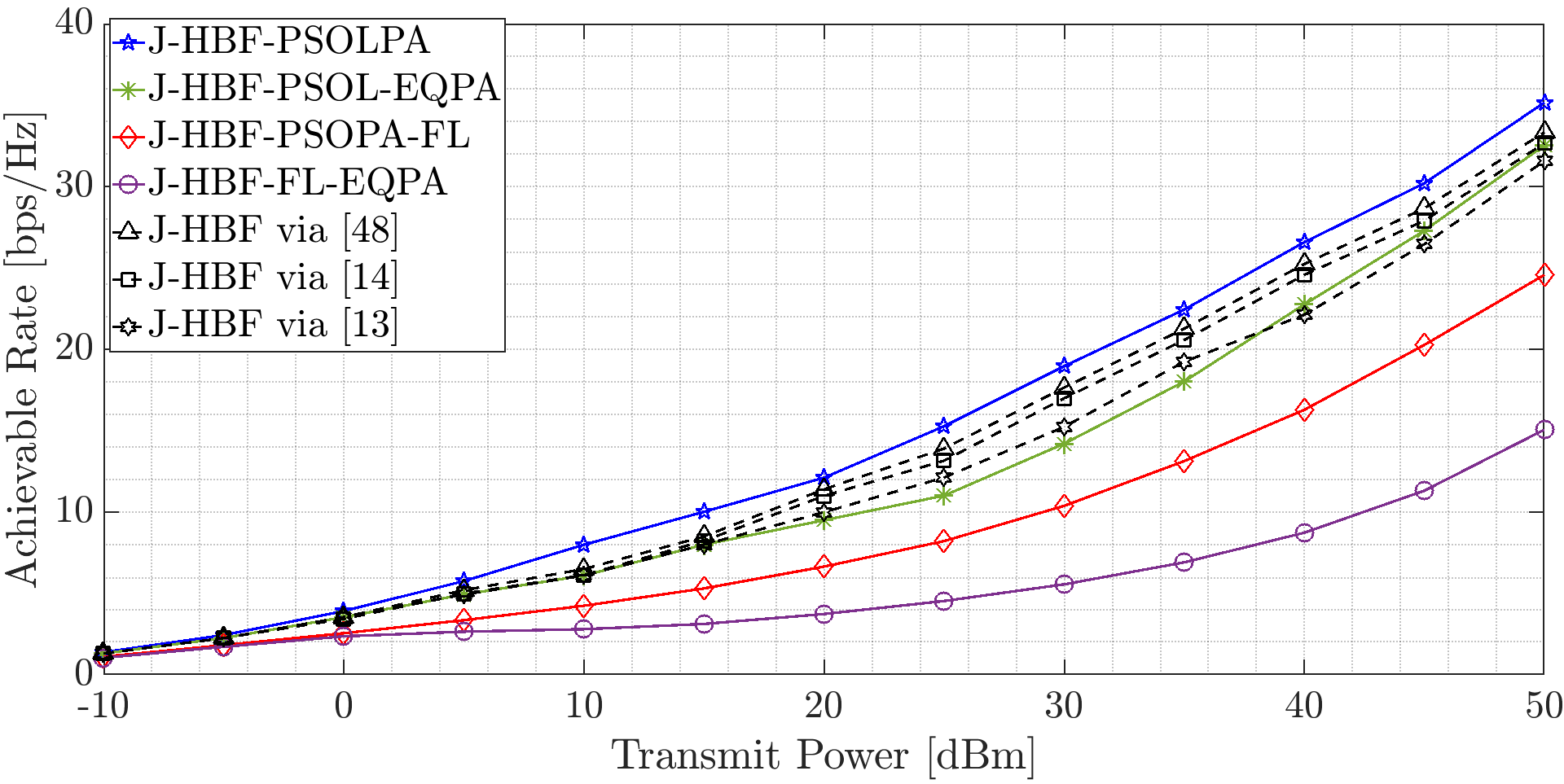} 
	\caption{Total AR $\mathrm{R}_T$ vs. $P_T$ for different proposed algorithmic schemes.}
	\label{fig:fig9}
		\vspace{-1em}
\end{figure}

Fig. \ref{fig:fig7} analyzes the total rate of second link $\mathrm{R}_2$ for exhaustive search and J-HBF-PSOL-EQPA. Due to the randomness in the placement of the IoT users, finding a single optimal UAV location that covers all IoT users while minimizing interference is a challenging task. However, the proposed J-HBF-PSOL-EQPA can find optimal UAV placement close to global solution as depicted in Fig. \ref{fig:fig7}(a). Fig. \ref{fig:fig8} plots the AR of second link versus different UAV 2-D locations for $P_T = 20$ dBm, which shows that UAV placement by the proposed J-HBF-PSOL-EQPA can give higher AR when compared to UAV deployed at some fixed location. Fig. \ref{fig:fig9} compares the AR versus transmit power $P_T$ of the proposed HBF solution for four cases: 1) PSO-based UAV location and PSO-based PA (J-HBF-PSOLPA); 2) PSO-based UAV location and equal PA (J-HBF-PSOL-EQPA); 3) fixed UAV location and PSO-based PA (J-HBF-PSOPA-FL); and 4) fixed UAV location and equal PA (FL-EQPA). The results show that all proposed algorithmic schemes can increase the total capacity of a UAV-assisted MU-mMIMO IoT systems when compared to the FL-EQPA case. Comparing the single optimizations (i.e., PSO location-EQ PA and fixed location-PSO PA), we can see that optimizing UAV location only can provide a higher AR than allocating optimal power to multiple IoT users. However, in a highly dynamic environment, where the IoT users are far located and randomly distributed, the joint optimization of UAV location and power allocation can achieve higher spectral efficiency (i.e., $\mathrm{R}_{\mathrm{PSOL-PSOPA}}$ $\geq$ $\mathrm{R}_{\mathrm{PSOL-EQPA}}$ $\gg$ $\mathrm{R}_{\mathrm{FL-PSOPA}}$). Moreover, compared to FL-EQPA case at $P_T = 40$ dBm, the proposed J-HBF-PSOPA-FL, J-HBF-PSOL-EQPA, and J-HBF-PSOLPA schemes can increase the total AR by $77\%$, $155\%$, and $200\%$, respectively. We also compare the performance with the existing HBF solutions. For instance, compared to HBF schemes presented in \cite{sohrabi2016hybrid, HBF_PE-AltMin}, the proposed J-HBF-PSOLPA can provide a higher AR. Similarly, compared to the iterative successive approximation (ISA) algorithmic solution in \cite{HBF_Relay_MIMO}, which requires full CSI, the proposed J-HBF-PSOLPA can provide better performance with reduced CSI overhead size.
\vspace{-1em}
\subsection{Buffer-Aided Transmission}
Section V-A of this study presents a comparison of different proposed PSO-based algorithms for conventional relaying in UAV-assisted wireless systems, i.e., UAV receives the data transmitted by BS in first time slot, decode the data, and then forwards it to multiple IoT users in the second time slot. This pre-scheduled approach may not perform well in UAV-assisted wireless systems because the channel qualities (i.e., $\mathbf{H}_1$ and $\mathbf{H}_2$) can vary significantly with time, preventing the UAV relay from exploiting the best transmitting and receiving channels. In addition, unlike static relays, UAVs can fly close to the base station, store the data in a buffer, and then fly close to multiple IoT users. The results in Section V-A are based on a fixed optimized location for the UAV relay, which can increase the rate of one link but may degrade the rate for other link \footnote{Since $\mathrm{R}_2$ is the minimum of ($\mathrm{R}_1, \mathrm{R}_2$), therefore UAV is deployed close to multiple users to maximize $\mathrm{R}_2$, however, it increases the pathloss for first link, which results in slight rate degradation for $\mathrm{R}_1$ (Fig. \ref{fig:fig5} - \ref{fig:fig8}).}. To fully exploit the potential of a mobile relay and to further maximize the capacity of a UAV-assisted MU-mMIMO IoT systems, we propose a buffer-aided UAV relay that can store the data in a buffer while transitioning from one optimal location for link $1$ to an optimal location for link $2$. By considering two possible locations for UAV, we can maximize the SINR for each link, leading to a higher total rate. \par  

\begin{figure}[!t]
	\centering
	\captionsetup{justification=centering}	
	\includegraphics[height=6cm, width=1\columnwidth]{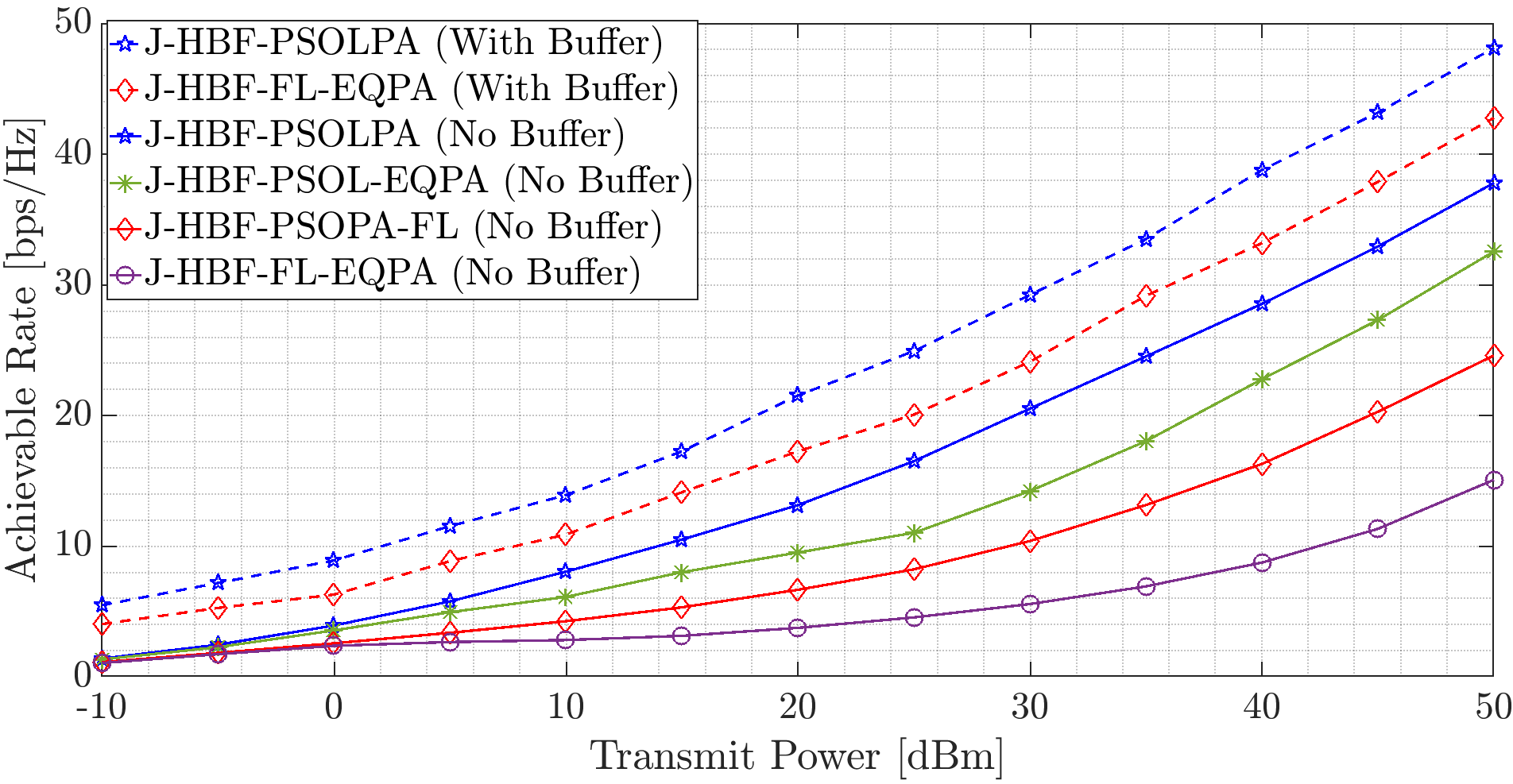} 
	\caption{Total AR $\mathrm{R}_T$ vs. $P_T$ for buffer-aided transmissions.}
	\label{fig:fig10}
\end{figure}
\begin{figure}[!t]
	\centering
	\captionsetup{justification=centering}	
	\includegraphics[height=6cm, width=1\columnwidth]{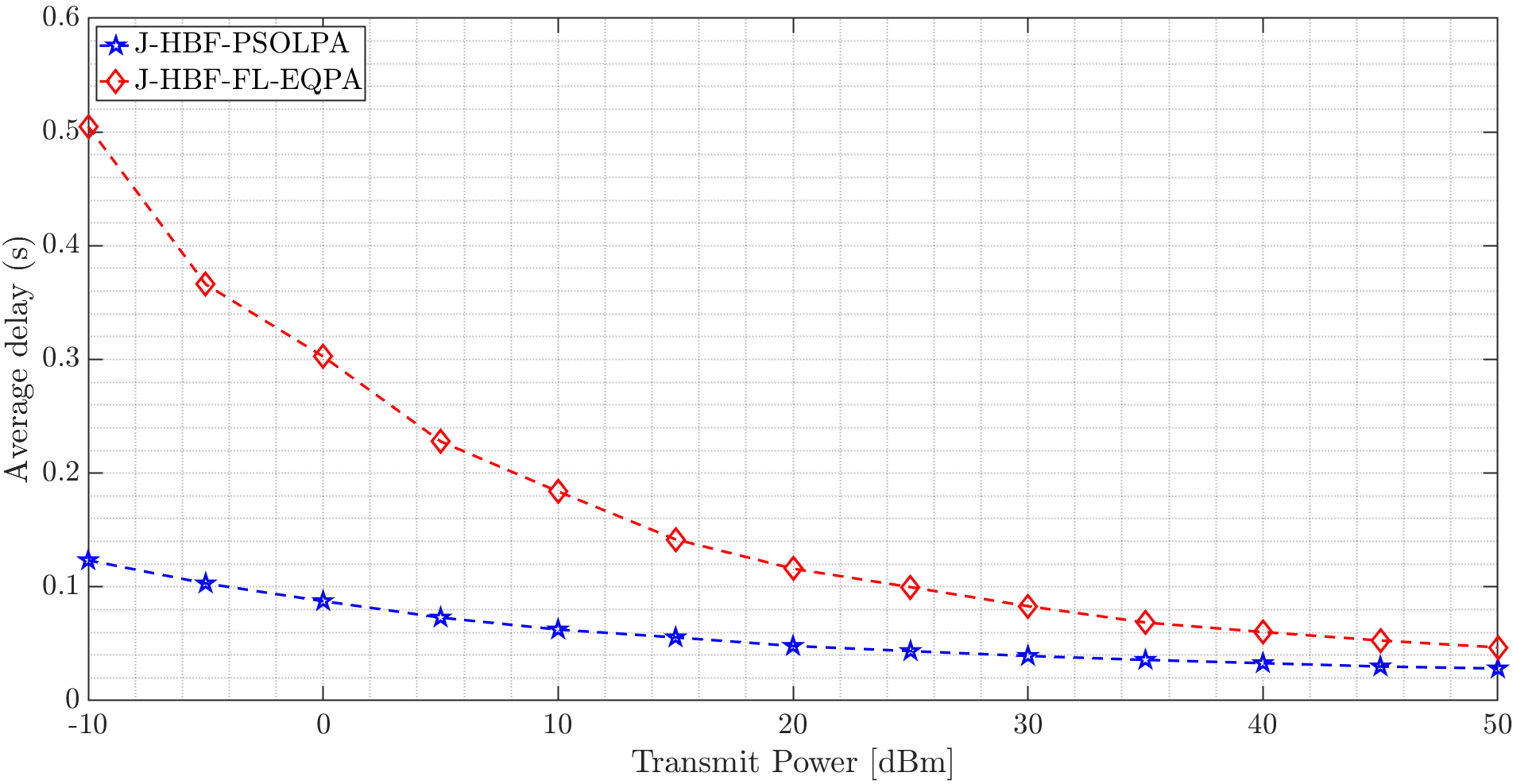} 
	\caption{Average delay $D$ vs. transmit power $P_T$.}
	\label{fig:fig11}
\end{figure}
\begin{figure}[!t]
	\centering
	\captionsetup{justification=centering}	
	\includegraphics[height=6cm, width=1\columnwidth]{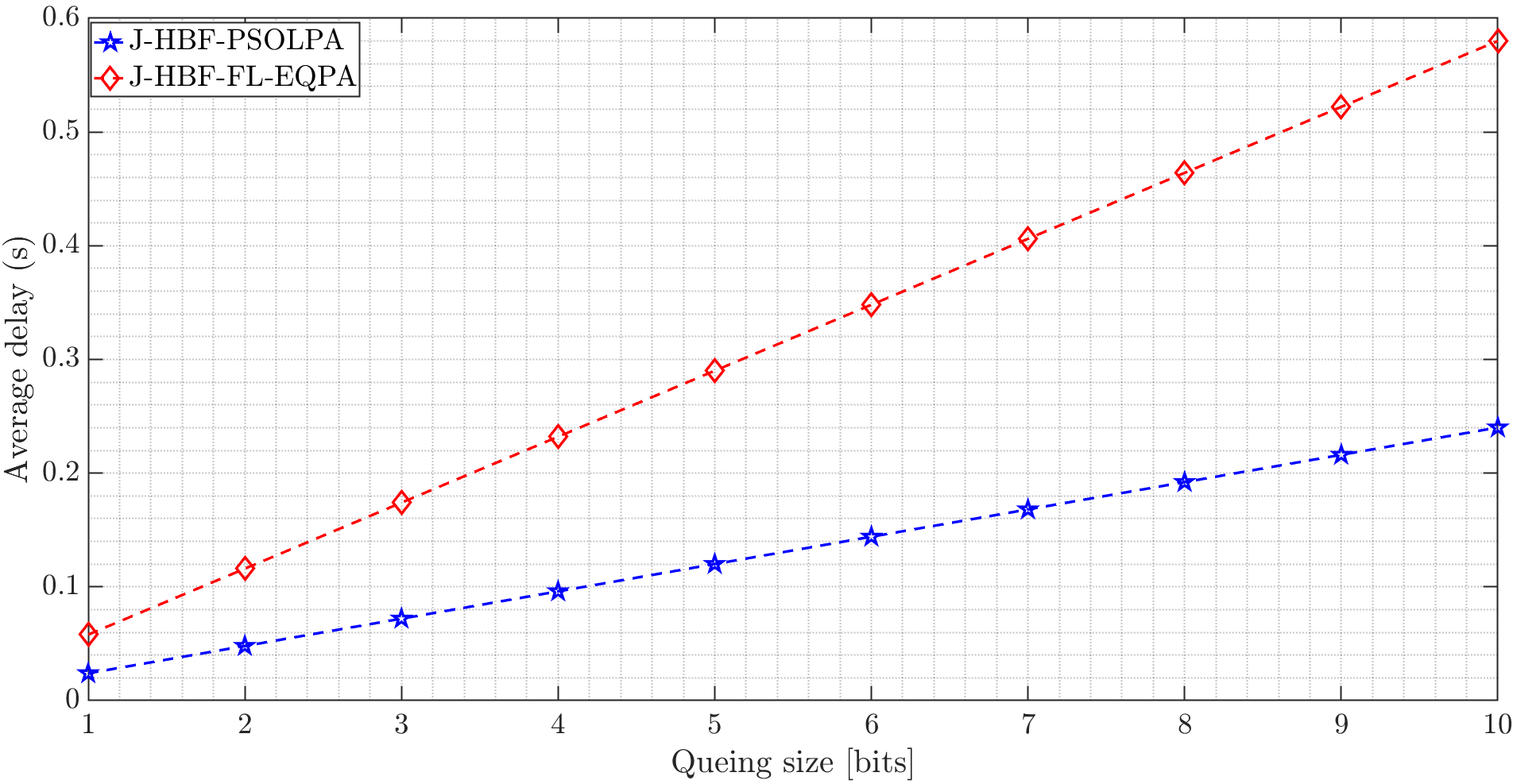} 
	\caption{Average delay $D$ vs. queuing size $Q$.}
	\label{fig:fig12}
	\vspace{-2ex}
\end{figure}
\begin{figure}[!t]
	\centering
	\captionsetup{justification=centering}	
	\includegraphics[height=6cm, width=1\columnwidth]{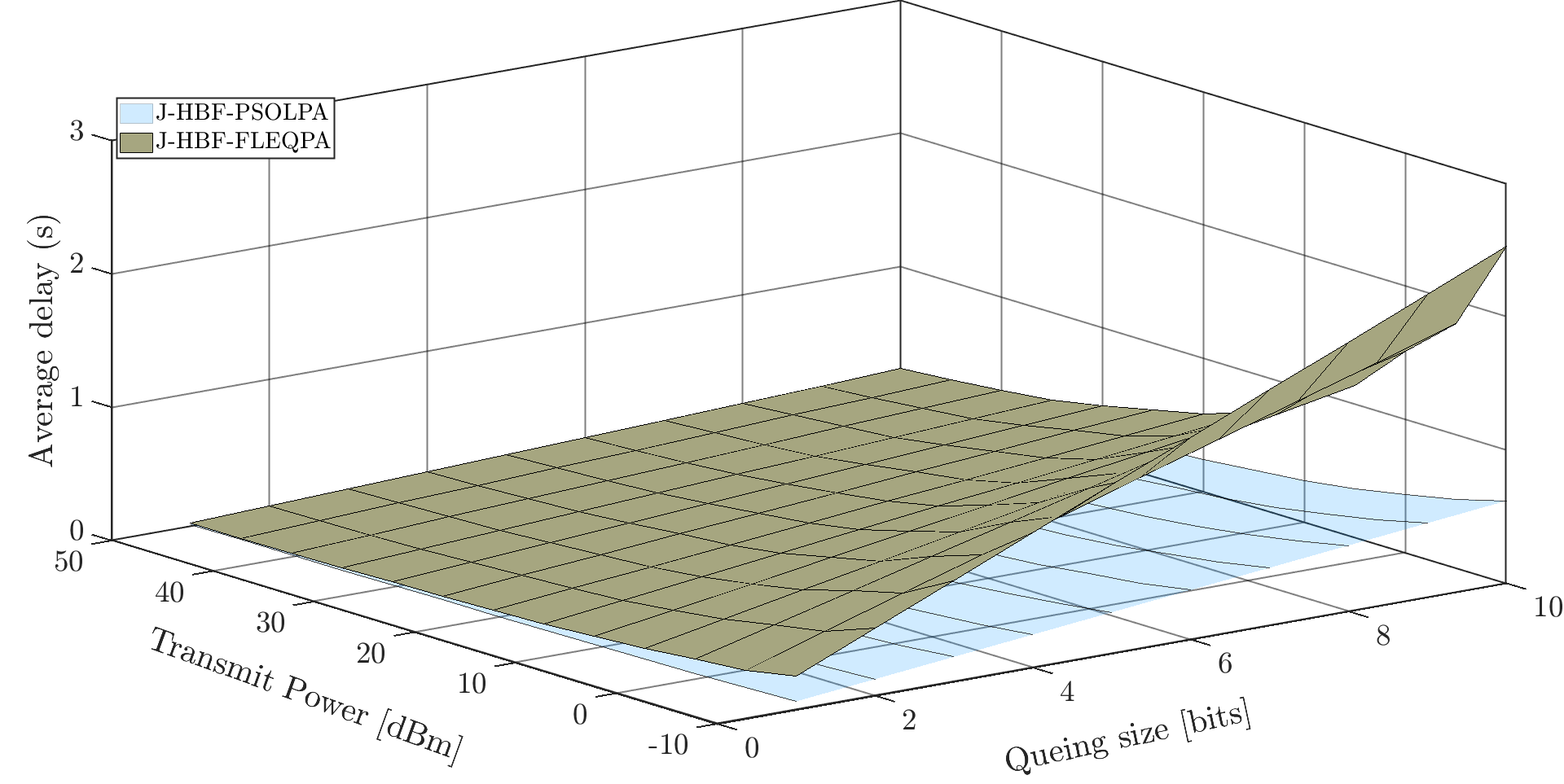} 
	\caption{Average delay $D$ vs. transmit power $P_T$ and queuing size $Q$.}
	\label{fig:fig13}
\end{figure}

In Fig. \ref{fig:fig10}, we compare AR versus transmit power for UAVs with buffer and without buffer. For the buffer-aided UAV, we consider two scenarios for delay-unconstrained applications. In FL-EQPA, we consider UAV is deployed at fixed location ($x_u, y _u = 50,50$) and multiple IoT users have equal PA. However, the UAV does not transmit in consecutive time slots. In the second scenario, J-HBF-PSOLPA, we consider two different UAV locations (close to BS for link $1$ and close to multiple IoT users for link $2$) and do not use pre-scheduled transmissions. The results show that using buffers at UAV can significantly increase the capacity (e.g., from $10$ bps/Hz for FL-EQPA (without buffer) to $62$ bps/Hz for J-HBF-PSOLPA (with buffer) at $P_T = 40$ dBm, which represents more than five fold increase in capacity). It is important to note that we assume a buffer size as $B = \infty$, and thus, it represents a performance upper bound for a UAV-assisted MU-mMIMO IoT system in a delay-constrained transmissions. For delay-unconstrained transmissions, the average delay tends to $\infty$ as $B \rightarrow \infty$. However, with the simple heuristic modifications proposed in this subsection, the proposed algorithmic solutions for delay-unconstrained transmission can also be employed for delay-constrained transmission at the expense of a small performance degradation due to the delay constraint. Let $D$ denote the waiting time (delay) experienced for a bit transmitted from the BS to the multiple IoT users via UAV relay equipped with buffer and having queuing size $Q$ bits. Then, for the HD UAV-assisted MU-mMIMO IoT systems, we use Little's Law to relate $D$ to $Q$. Let $\lambda = \min(\mathrm{R}_1, \mathrm{R}_2)$ be the arrival rate, then we can express $D$ as\cite{little_law}: 
\begin{equation}
	D = \frac{Q}{\min(\mathrm{R}_1, \mathrm{R}_2)}. \label{eq:delay}
\end{equation}
We assume a first-in-first-out (FIFO) queuing mechanism for the UAV buffer, where the UAV receives data from the BS at a rate of $\mathrm{R}_1$, and transmits it to the IoT users at rate of $\mathrm{R}_2$ incurring a delay $D$ due to queuing bits $Q$ of buffer \footnote{It is important to note that the arrival rate $\mathrm{R}_1$ is greater than the departure rate $\mathrm{R}_2$ for the given system (i.e., $\mathrm{R}_1 > \mathrm{R}_2$), the stability of the system is still guaranteed due to the assumption of $\min(\mathrm{R}_1, \mathrm{R}_2)$ being used as the arrival rate. This ensures that the overall arrival rate is always less than or equal to the departure rate, which is necessary for the stability of the system.}. Fig. \ref{fig:fig11} plots the average delay versus transmit power $P_T$ for the cases of: 1) FL-EQPA; and 2) J-HBF-PSOLPA. It can be seen that for a fixed queuing size of $Q = 2$ bits, increasing $P_T$ can result in reduced average delay due to an increased AR for both cases. However, the proposed J-HBF-PSOLPA can reduce the average delay by approximately $50\%$ when compared to FL-EQPA. In Fig. \ref{fig:fig12}, we plot average delay versus queuing size $Q$ for a fixed transmit power $P_T = 20$ dBm. As expected, higher queuing size results in increased delay. However, by utilizing the proposed J-HBF-PSOLPA, we can decrease $D$ by more than $140\%$ (e.g., at $Q = 8$ bits, $D$ can be reduced from $0.48$s to $0.2$s). Finally, we compare the average delay versus $P_T$ and $Q$ for FL-EQPA and J-HBF-PSOLPA in Fig. \ref{fig:fig13}. Our results show that the proposed J-HBF-PSOLPA can significantly reduce the average delay for any combination of $P_T$ and $Q$. Thus, it can be applied to both delay-unconstrained and delay-constrained applications in UAV-assisted MU-mMIMO IoT systems.
\begin{table}[!t]
	\centering
	\caption{DNN Parameters \vspace{-2ex}}
	\resizebox{\columnwidth}{!}{
		\begin{tabular}{|c|c|c|c|}
			\hline
			\multicolumn{2}{|c|}{Regularizer}	& \multicolumn{2}{c|}{L2} \\ \hline 	
			$1^{st}$ hidden layer size & $2^{nd}$ hidden layer size & $L_1 = 1024$ &  $L_2 = 512$ \\ \hline
			$3^{rd}$ hidden layer size & $4^{th}$ hidden layer size & $L_3 = 256$ & $L_4 = 128$ \\ \hline
			Optimizer & Learning Rate & ADAM & 0.001 \\ \hline
			Dataset Size & Test Data & $S_r = 100,000$ & $S_t = 1,000$ \\ \hline
			Batch Size & Epoch Size & $32$ & $15$ \\ \hline
		\end{tabular} 
	} 	\label{tab:tab2}
\end{table}
\vspace{-2ex}
\subsection{Low-Complexity DL-Based Performance}
\begin{figure}[!t] 
	% 	\centering
	\subfloat[\label{fig:fig14a}]{% 
		\includegraphics[width =0.485\columnwidth,height=4cm]{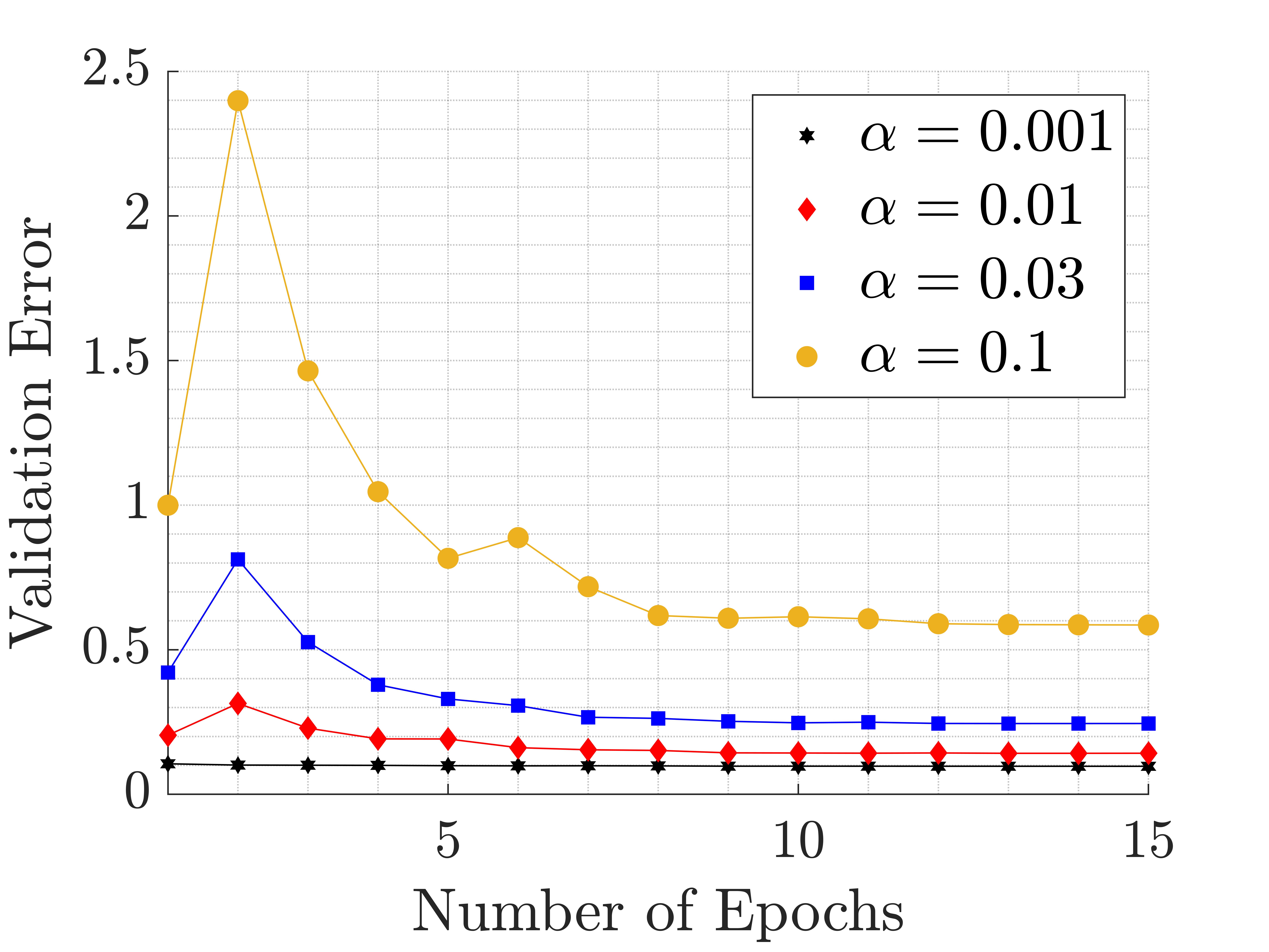} 
	} \hfil 
	\subfloat[\label{fig:fig14b}]{% 
		\includegraphics[width =0.485\columnwidth, height= 4cm]{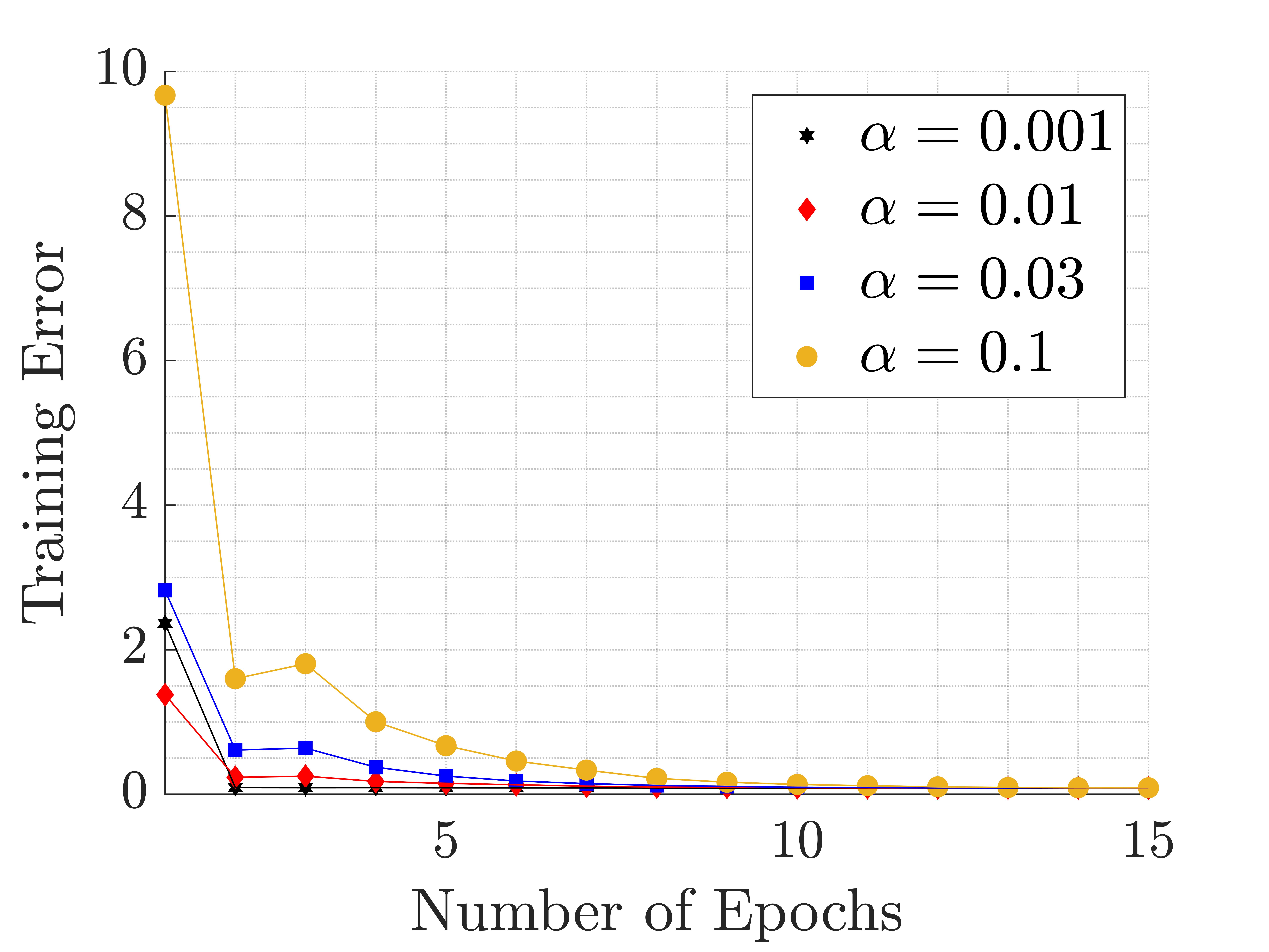} 
	} \vspace{-1ex}
	\caption{MSE loss vs. epoch for different learning rates. (a) Validation error. (b) Training error.}
	\label{fig:fig14}
	\vspace{-2em}
\end{figure}

\begin{figure}[!t] 
	% 	\centering
	\subfloat[\label{fig:fig15a}]{% 
		\includegraphics[width =0.485\columnwidth,height=4cm]{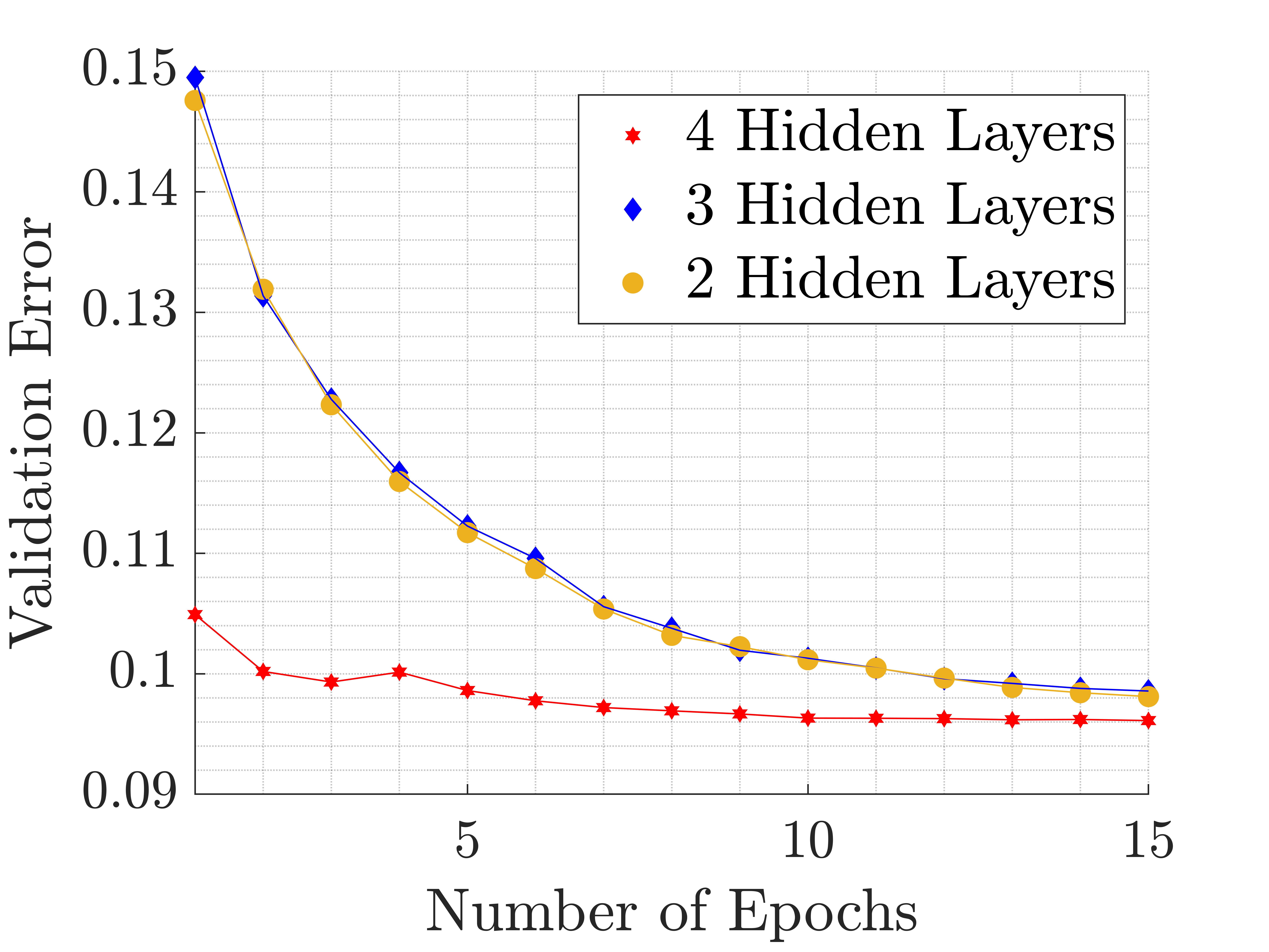} 
	} \hfil 
	\subfloat[\label{fig:fig15b}]{% 
		\includegraphics[width =0.485\columnwidth, height= 4cm]{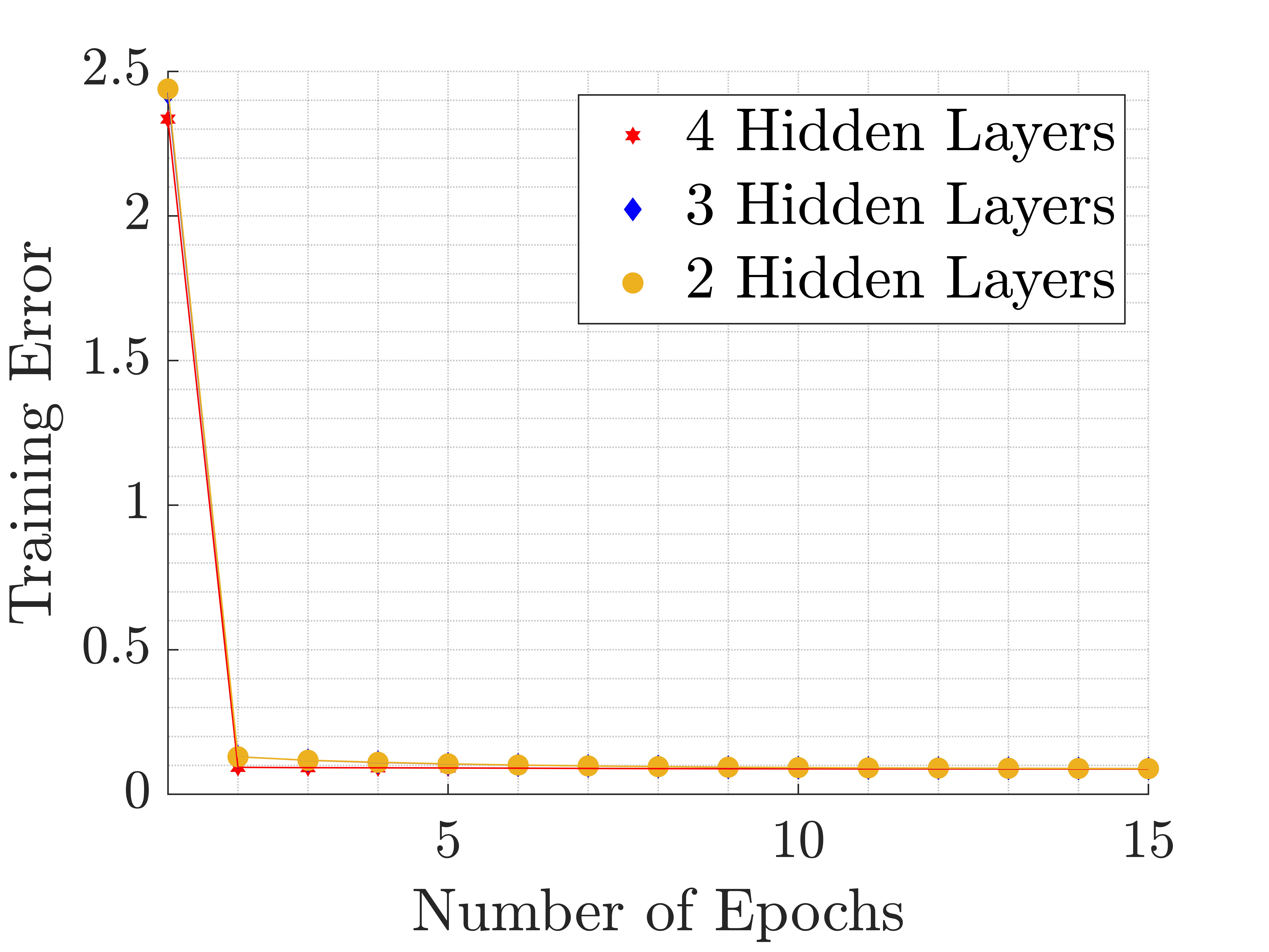} 
	} 
	\caption{MSE loss vs. epoch for different hidden layers. (a) Validation error. (b) Training error.}
	\label{fig:fig15}
\end{figure}

In this section, we present the results obtained using the proposed low-complexity J-HBF-DLLPA algorithm and compare its performance with that of J-HBF-PSOLPA. Table \ref{tab:tab2} outlines the hyper-parameters for the proposed DNN architecture, which is given in Fig. \ref{fig:fig4}. Fig. \ref{fig:fig14} exhibits the MSE performance for training and validation datasets under varying learning rates $\alpha = \{0.001, 0.01, 0.03, 0.1\}$ over a total of $15$ epochs. The larger values of $\alpha$, such as $0.1$ and $0.03$ result in higher validation error, which requires a significant number of iterations to converge. In contrast, smaller values of $\alpha = 0.01$ and $0.001$ can provide lower MSE for validation dataset, even with a few epochs. Notably, $\mathrm{MSE}_{\alpha = 0.001} < \mathrm{MSE}_{\alpha = 0.01} < \mathrm{MSE}_{\alpha = 0.03} \ll \mathrm{MSE}_{\alpha = 0.1}$, which emphasizes the importance of appropriately choosing the learning rate. Similarly, Fig. \ref{fig:fig15} compares the MSE performance for training and validation datasets with different number of hidden layers (HL) in DNN architecture. We evaluate the error for $15$ epoch for HL $= \{2,3,4\}$. It can be seen that DNN with $2$ or $3$ HL can achieve low MSE. However, it may require a large number of epochs to converge. By using HL $= 4$, we can achieve a better MSE performance for validation dataset ($\mathrm{MSE}_{\mathrm{HL} = 4} < \mathrm{MSE}_{\mathrm{HL} = 3} < \mathrm{MSE}_{\mathrm{HL} = 2}$). \par
\begin{figure}[!t] 
	% 	\centering
	\subfloat[\label{fig:fig16a}]{% 
		\includegraphics[width =0.485\columnwidth,height=4cm]{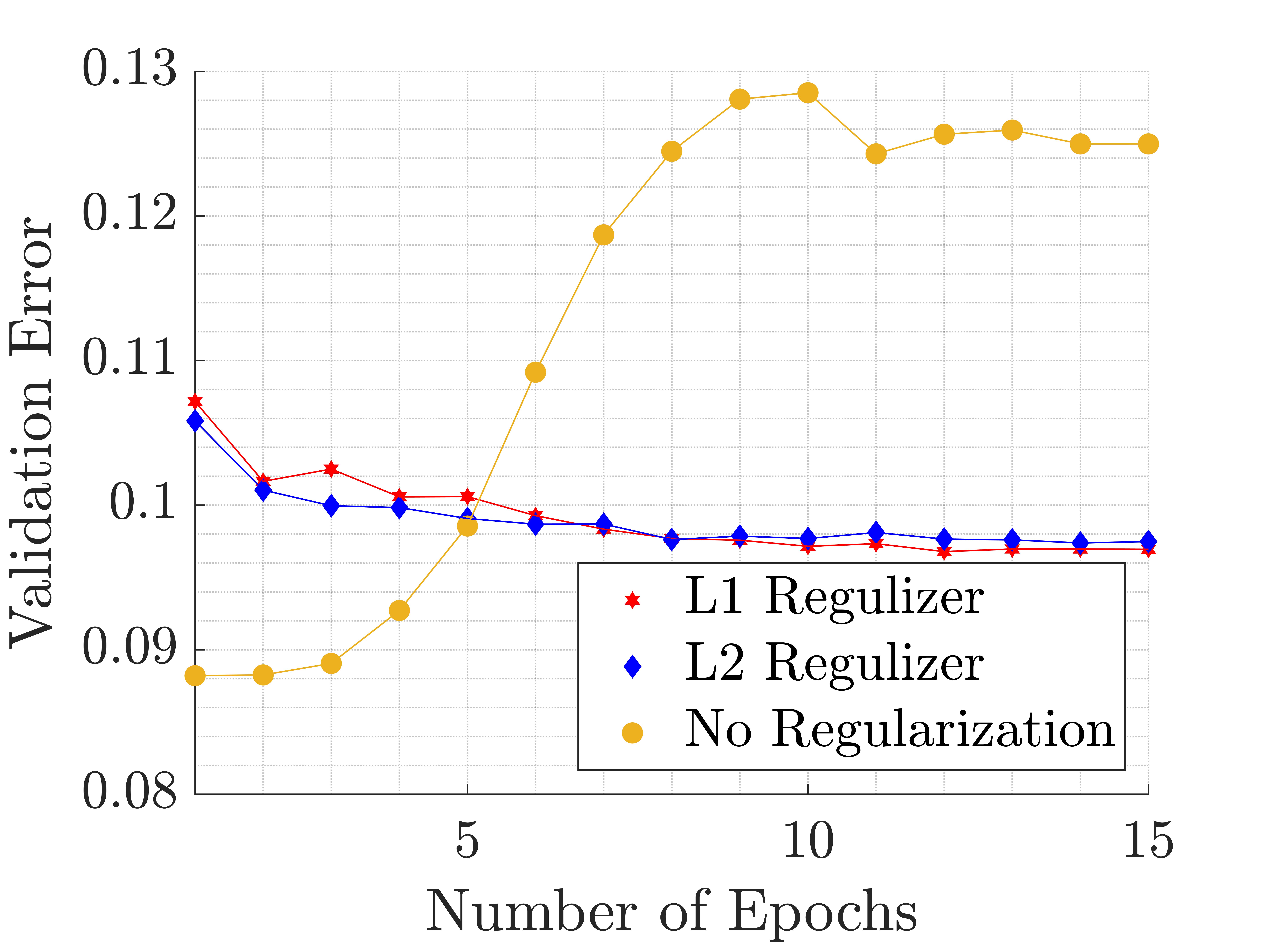} 
	} \hfil 
	\subfloat[\label{fig:fig16b}]{% 
		\includegraphics[width =0.485\columnwidth, height= 4cm]{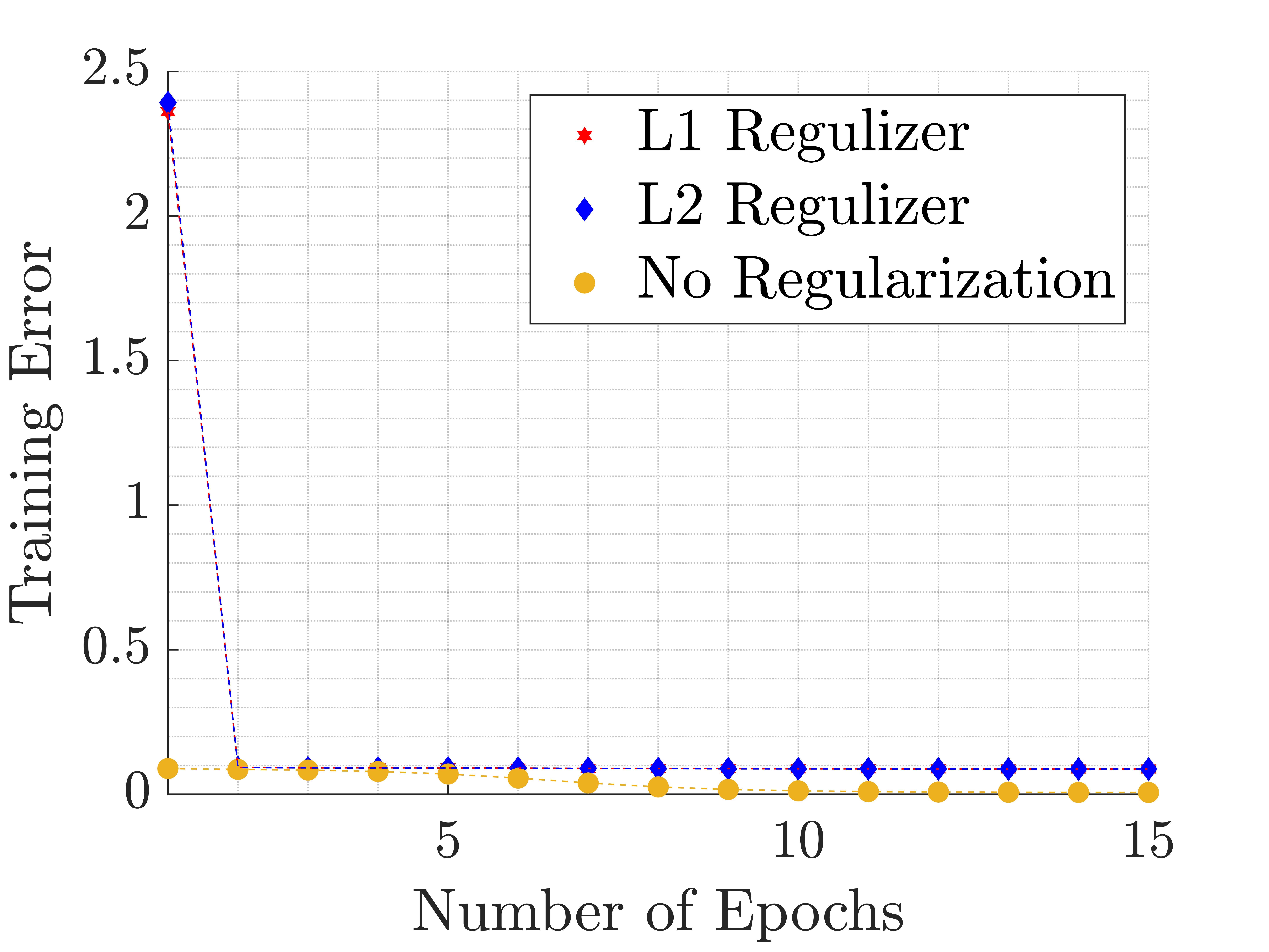} 
	} 
	\caption{MSE loss vs. epoch for regularization. (a) Validation error. (b) Training error.}
	\label{fig:fig16}
		\vspace{-2em}
\end{figure}

\begin{figure}[!t] 
	% 	\centering
	\subfloat[\label{fig:fig17a}]{% 
		\includegraphics[width =0.485\columnwidth,height=4cm]{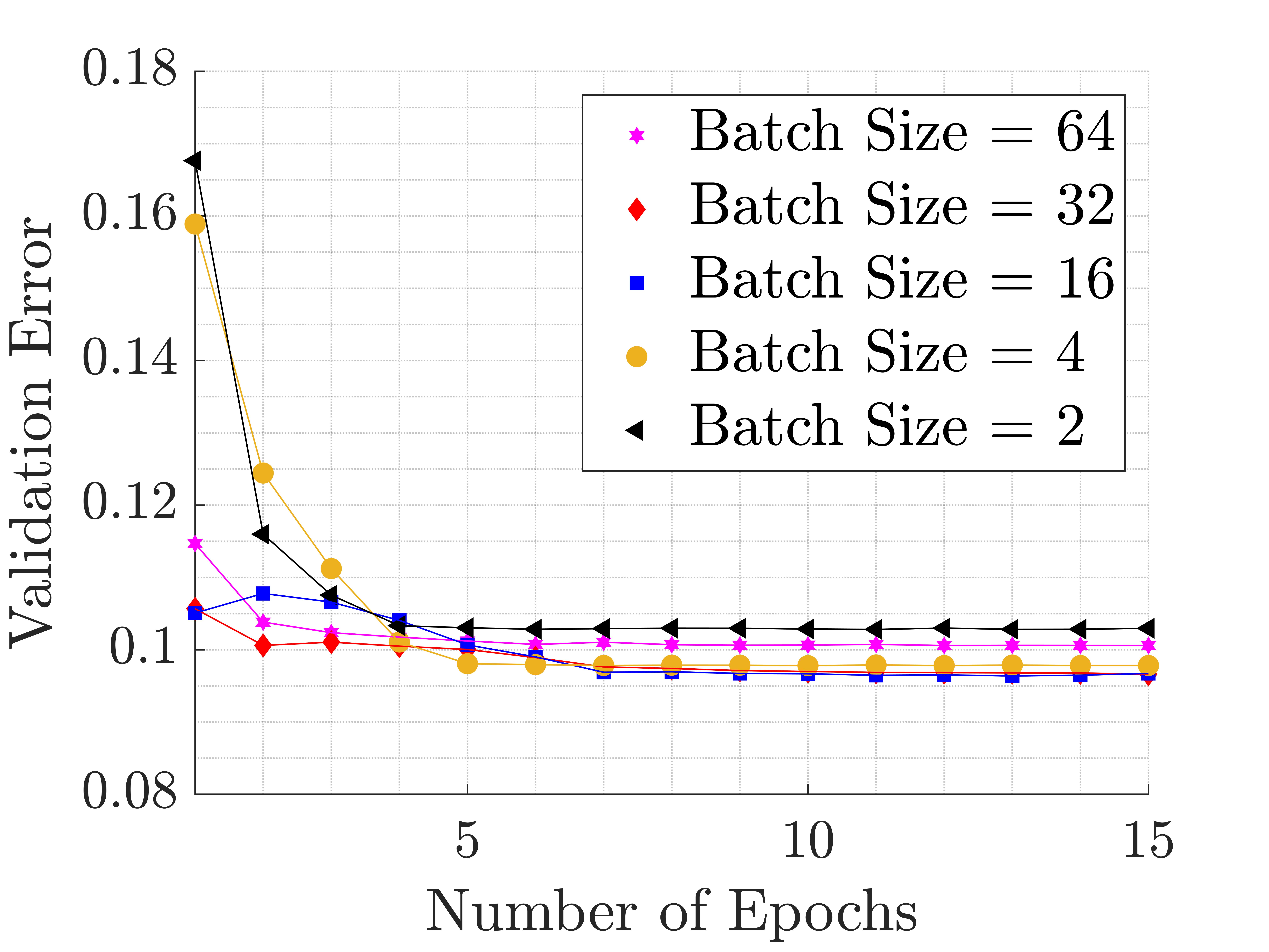} 
	} \hfil 
	\subfloat[\label{fig:fig17b}]{% 
		\includegraphics[width =0.485\columnwidth, height= 4cm]{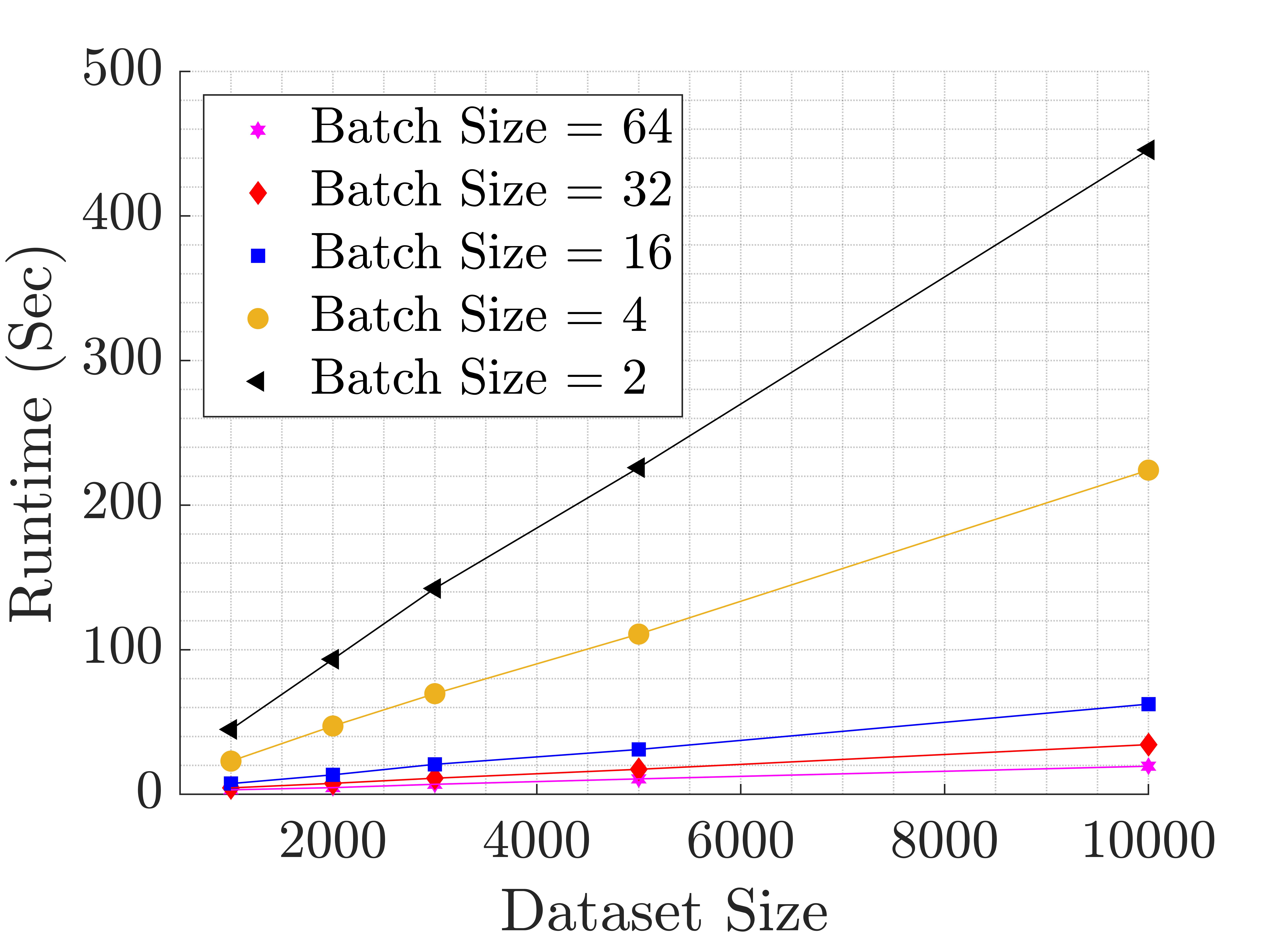} 
	} 
	\caption{MSE loss vs. epoch for different batch sizes. (a) Validation error. (b) Runtime.}
	\label{fig:fig17}
\end{figure}

 
To avoid over-fitting of the training data, our proposed DNN architecture incorporates regularization techniques. In Fig. \ref{fig:fig16}, we compare the MSE performance of the training and validation datasets for the following cases: 1) no regularization; 2) L$1$ regularization; and 3) L$2$ regularization. The use of either L$1$ or L$2$ regularization techniques prevent over-fitting, and can predict the power and location values close to optimal solution on test data \footnote{When there is no regularization, then the DNN architecture gives very low MSE for training dataset but its performance on validation dataset degrades after a few epochs.}. Another important factor in optimizing the DNN architecture is selecting an appropriate batch size ($B_S$), which can have a significant impact on memory usage, convergence speed, training stability, and model generalization. In Fig. \ref{fig:fig17}(a), we compare the MSE performance of the validation dataset for $B_S$ $= \{2, 4, 16, 32, 64\}$. Our results indicate that $B_S = 4, 16,$ or $32$ can provide lower MSE, but the runtime for $B_S=4$ and $16$ can be significantly longer than BS$= 32$, especially for larger datasets with $100,000$ realizations. Therefore, we use $B_S = 32$, which can provide a lower MSE with low computational complexity. \par 
Fig. \ref{fig:fig18} compares the total AR of J-HBF-PSOLPA with low-complexity DL solutions, namely, J-HBF-DLLPA with MSE, and J-HBF-DLLPA with MAE, for $G=1$, $K=3$ at $P_T = 20$ dBm. We provide the AR performance evaluations for training, validation, and test datasets. For benchmark comparison, we compare the proposed DL-based solutions with J-HBF-PSOLPA and FL-EQPA. The numerical results show that both DL-based solutions can achieve AR close to PSO-based solution for all datasets and outperform FL-EQPA. For instance, J-HBF-DLLPA with MSE and MAE can provide the total AR of $14.3$ and $14.2$ bps/Hz, which is $98.6\%$ and $97.4\%$ of the total capacity achieved by J-HBF-PSOLPA, indicating a performance difference of only $1.3\%$ and $2.06\%$, respectively. Furthermore, compared to FL-EQPA, both DL-based solutions show more than two-fold increase in AR. 
\vspace{-1em}
\subsection{Complexity Analysis}
In this section, we compare the computational complexity of several proposed solutions for UAV-assisted MU-mMIMO IoT systems. Specifically, we compare: 1) J-HBF-PSOPA-FL; 2) J-HBF-PSOL-EQPA; 3) J-HBF-PSOLPA; and 4) J-HBF-DLLPA. In Fig. \ref{fig:fig19}, we present the runtime results for different numbers of IoT users for $G=2$, where $K = \{2,4,6,8\}$. We provide the runtime for $1000$ network realizations for test data.  It is noteworthy to mention that the offline trained DNN architecture for J-HBF-DLLPA algorithm run on MATLAB\footnote {For the MATLAB runtime results, we implement both PSO-based J-HBF-PSOLPA and DL-based J-HBF-DLLPA via a PC with Intel Core(TM) i7-4770 CPU @ 3.4 GHz and 32 GB RAM.}. The proposed J-HBF-DLLPA outperforms the computational complex J-HBF-PSOLPA, J-HBF-PSOL-EQPA, and J-HBF-PSOPA-FL algorithms by significantly reducing the runtime. For example, for a small number of IoT users ($K = 2$),  J-HBF-PSOLPA, J-HBF-PSOL-EQPA, and J-HBF-PSOPA-FL take approximately $489$, $234$, and $376$ sec to predict $K+2$ values, whereas with J-HBF-DLLPA, only about $2$ sec are enough to predict the optimal UAV location and power allocation values. This means that the proposed J-HBF-DLLPA with MSE requires only about $0.4\%-0.8\%$ of the runtime of different PSO-based algorithmic solutions without impacting the total AR performance. When the number of IoT users are increased to $K=8$, then J-HBF-PSOLPA, J-HBF-PSOL-EQPA, and J-HBF-PSOPA-FL can take around $1725$, $976$, and $1453$ sec, representing an increase of approximately $252\%$, $317\%$, and $287\%$, respectively, whereas with the proposed J-HBF-DLLPA, the predictions take only about $2.2$ sec, which is $\approx 0.127\%$ of runtime of J-HBF-PSOLPA. Then, we consider the computational complexity of the proposed angular-based HBF design. For the RF beamformers $\mathbf{F}_b$ and $\mathbf{F}_{u,r}$ require $\mathcal{O}(N_T)$ and $\mathcal{O}(N_r)$ operations, respectively. Meanwhile, for $\mathbf{F}_{u,t}$, the complexity is $\mathcal{O}(G N_t)$ to serve $G$ different IoT user groups. For BB stages $\mathbf{B}_{b}$ and $\mathbf{B}_{u,r}$, we require $\mathcal{O}(N_{{RF}_b}^3)$ and $\mathcal{O}(N_{{RF}_u}^3)$ operations, respectively. For $\mathbf{B}_{u,t}$, the complexity is $\mathcal{O}(N_{{RF}_u}^3 + K N_{{RF}_u}^2)$, where $\mathcal{O}(K N_{{RF}_u}^2)$ is for computation of $\mathbfcal{H}_2^H\mathbfcal{H}$ and $\mathcal{O}(N_{{RF}_u}^3)$ is for matrix inversion. Overall, the HBF design for BS and UAV requires $\mathcal{O}(N_T + N_r + G N_t + N_{{RF}_b}^3 + N_{{RF}_u}^3 + K N_{{RF}_u}^2)$ operations. Let us consider the J-HBF-PSOPA-FL algorithm, which requires $\mathcal{O}(M_p)$ to initialize particles and $\mathcal{O}(K^2 M_p)$ operations to find $K$ global solutions. Then, for a total of $T$ PSO iterations, we need $\mathcal{O}(T M_p(N_{{RF}_u}^3 + K N_{{RF}_u}^2))$. Ignoring the small computational terms (for instance, matrix multiplication, matrix inversion), the overall computational complexity of J-HBF-PSOPA-FL is $\mathcal{O}(T M_p(N_{{RF}_u}^3 + K N_{{RF}_u}^2))$. Similarly, the complexities of J-HBF-PSOL-EQPA and J-HBF-PSOLPA are computed as $\mathcal{O}(T M_p(N_T + N_r + G N_t + N_{{RF}_b}^3 + N_{{RF}_u}^3 + K N_{{RF}_u}^2)$ and $\mathcal{O}(T M_p(N_T + N_r + G N_t + N_{{RF}_b}^3 + (N_{{RF}_u}^3 + K N_{{RF}_u}^2)^2)$, respectively. Now, let us discuss the proposed DNN architecture with the hyper-parameters provided in Table \ref{tab:tab2}. The complexity of the training phase can be calculated as follows: 
\begin{figure}[!t]
	\centering
	\captionsetup{justification=centering}	
	\includegraphics[height=6cm, width=1\columnwidth]{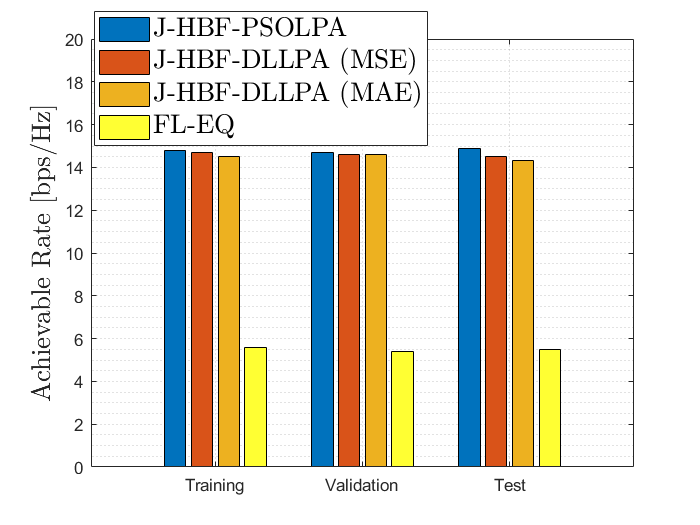} 
	\caption{Total AR evaluation on training, validation and test dataset for J-HBF-PSOLPA and J-HBF-DLLPA.}
	\label{fig:fig18}
\end{figure}
\begin{itemize}
	\item Forward Pass: 1) the \textit{input layer} has $(2N_t +2 N_{{RF}u} + 2)K$ neurons, and computing the output of the input layer requires $\mathcal{O}((2N_t +2 N{{RF}_u} + 2)K)$ operations; 2) there are four HL with $1024, 512, 256$ and $128$ neurons, respectively, and the total complexity of computing the output of each HL can be computed as follows: $\mathcal{O}(1024)$ (layer 1), $\mathcal{O}(512)$ (layer 2), $\mathcal{O}(256)$ (layer 3), $\mathcal{O}(128)$ (Layer 4). Therefore, the total complexity of computing the output of all HL is $\mathcal{O}(1024+512+256+128) = \mathcal{O}(1920)$; 3) the \textit{output layer} has $K+2$ neurons, so the complexity of final layer is $\mathcal{O}(K+2)$. The overall complexity of the forward pass is $\mathcal{O}((2N_t +2 N_{{RF}u} + 2)K + 3840 + K+2) = \mathcal{O}((2N_t +2 N{{RF}_u} + 2)K)$.
	\item Backward Pass: The backward propagation involves computing gradients of the loss function with respect to the weights and biases of the DNN. The overall complexity of the backward pass is $\mathcal{O}(N_{\text{params}})$, where $N_{\text{params}}$ is the total number of parameters in the DNN \footnote{The total number of parameters in the DNN can be calculated as follows: (i) input layer to HL $1: (2N_t +2 N_{{RF}_u} + 2)K \times 1024$, (ii) HL $1$ to HL $2: 1024 \times 512$, (iii) HL $2$ to HL $3: 512 \times 256$, (iv) HL $3$ to HL $4: 256 \times 128$, (v) HL $4$ to output layer: $128$}.
	\item Assuming a batch size of 32 and 15 epochs, the total number of forward and backward passes would be $N_{\text{batch}} \cdot N_{\text{epochs}} = 480$. Therefore, the total computational complexity of the DNN during training phase can be approximated as $\mathcal{O}(N_{\text{params}}N_{\text{batch}}N_{\text{epochs}}) \approx \mathcal{O}(3.42 \cdot 10^{11})$.
\end{itemize}
\begin{figure}[!t]
	\centering
	\captionsetup{justification=centering}	
	\includegraphics[height=6.5cm, width=1\columnwidth]{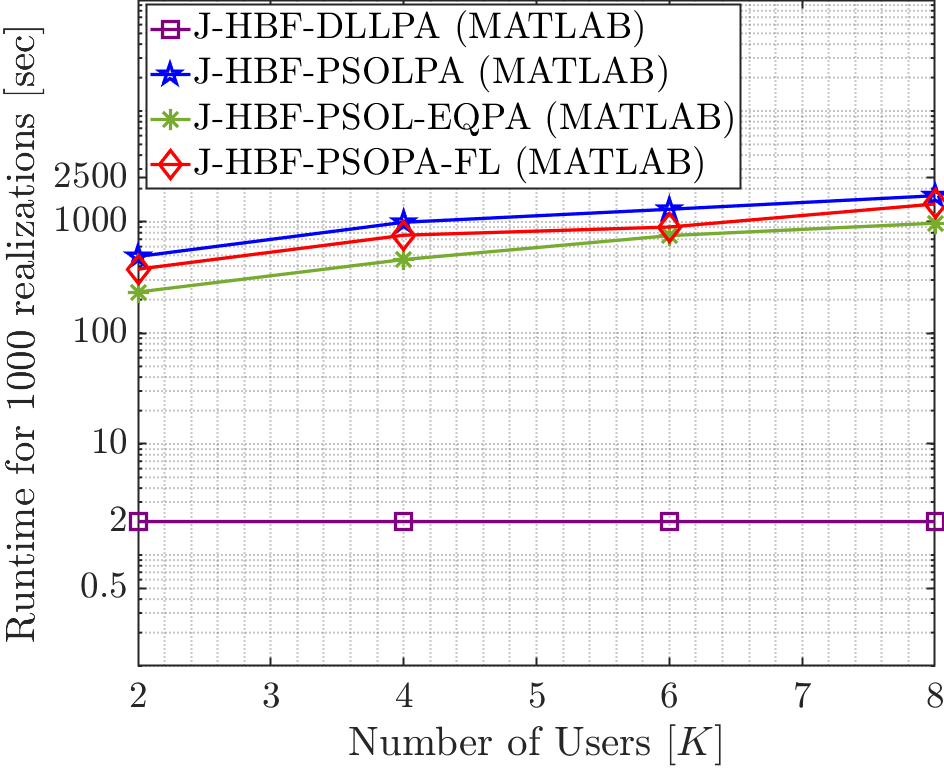} 
	\vspace{-2ex}
	\caption{Runtime comparison of proposed HBF solutions for different numbers of users.}
	\label{fig:fig19}
\end{figure}
Compared to different proposed solutions using PSO, the DNN architecture has a higher computational complexity during training phase, however, once fully trained, the proposed DNN architecture requires only $\mathcal{O}(1)$ operations for predicting $K+2$ output values for a input feature vector of size $(2N_t + 2N_{{RF}_u} + 2)K$. This is because the complexity is proportional to the number of layers and neurons in each layer, which is a constant value for a trained DNN. Thus, the proposed J-HBF-DLLPA offers a low-complexity solution for real-time applications in UAV-assisted MU-mMIMO IoT systems.
\vspace{-1ex}
\section{Conclusions}
In this paper, a UAV-assisted MU-mMIMO IoT communications system has been considered, where the UAV operates as a decode-and-forward relay between BS and multiple IoT users. The problem of jointly designing the hybrid beamforming stages for BS and UAV together with UAV deployment and optimal power allocation to multiple users is taken into consideration. For this challenging non-convex problem, we have proposed three different PSO-based algorithmic solutions to optimize the UAV location, and the power allocated to IoT users, which are not directly accessible to BS. Then, based on the optimized UAV location, and power allocation, the hybrid beamforming stages for BS and UAV transmit and receive are sequentially updated for a maximum total achievable rate. In particular, the RF stages are designed using the angular location of nodes, reducing the number of RF chains, while the BB stages are designed using reduced-dimension effective channel, which significantly reduces MU interference among IoT nodes. Then, a deep learning-based low-complexity joint hybrid beamforming, UAV location optimization and power allocation scheme (J-HBF-DLLPA) has been proposed for maximizing the achievable rate. The results illustrate that the proposed PSO-based solutions can significantly enhance the capacity of a UAV-assisted MU-mMIMO IoT system as well as reduce the average delay for delay-constrained transmissions. Additionally, the proposed J-HBF-DLLPA can closely approach the capacity of PSO-based solutions and greatly reducing the runtime by $99\%$, which makes the implementation suitable for real-time online applications in UAV-assisted MU-mMIMO IoT systems.

\balance
\bibliographystyle{IEEEtran}
\bibliography{references}

\end{document}